\documentclass[11pt]{article}

\usepackage[utf8]{inputenc}
\usepackage{amsmath}
\usepackage{amsfonts}
\usepackage{amssymb}
\usepackage{amsthm}
\usepackage{hyperref}
\usepackage{tikz}
\usepackage{graphicx}
\usepackage{cleveref}
\usepackage{natbib}
\usepackage{enumitem}
\usepackage{booktabs}
\usepackage{multirow}
\usepackage{multicol}
\usepackage{array}
\usepackage{float}
\usepackage{caption}

\usepackage{titletoc}

\hypersetup{
	colorlinks=true,
	linkcolor=black,
	citecolor=black}

\usetikzlibrary{positioning, calc, shapes.geometric, shapes.multipart, shapes, arrows.meta, arrows, 	decorations.markings, external, trees}

\tikzstyle{Arrow} = [
thick,
decoration={
	markings,
	mark=at position 0.99 with {  % fix bug with tikz arrow head interference with decoration
		\arrow[thick]{latex}
	}
},
shorten >= 3pt, preaction = {decorate}
]
\tikzstyle{BiArrow} = [
thick,
decoration={
	markings,
	mark=at position 0.99 with {  % fix bug with tikz arrow head interference with decoration
		\arrow[thick]{latex}
	}
},
shorten >= 3pt, preaction = {decorate}
]
\tikzstyle{RedArrow} = [
red, 
line width = 0.5mm,
decoration={
	markings,
	mark=at position 0.99 with {
		\arrow[line width = 0.5mm,red]{latex}
	}
},
shorten >= 3pt, preaction = {decorate}
]

\tikzstyle{BlueArrow} = [
blue, 
line width = 0.5mm,
decoration={
	markings,
	mark=at position 0.99 with {
		\arrow[line width = 0.5mm,blue]{latex}
	}
},
shorten >= 3pt, preaction = {decorate}
]

\tikzstyle{Dashed} = [
thick,dashed,
decoration={
	markings,
	mark=at position 0.99 with {  % fix bug with tikz arrow head interference with decoration
		\arrow[thick]{latex}
	}
},
shorten >= 3pt, preaction = {decorate}
]

\tikzset{
    BiArrow/.style={
        thick,
        decoration={
            markings,
            mark=at position 0.99 with {
                \arrow[thick]{latex}
            },
            mark=at position 0 with {
                \arrow[thick]{latex}
            }
        },
        shorten >= 3pt,
        preaction = {decorate}
    }
}

\AfterEndEnvironment{remark}{\noindent\ignorespaces}
\newtheorem{assumption}{Assumption}
\AfterEndEnvironment{assumption}{\noindent\ignorespaces}
\newtheorem{proposition}{Proposition}
\AfterEndEnvironment{proposition}{\noindent\ignorespaces}

\AfterEndEnvironment{theorem}{\noindent\ignorespaces}
\newtheorem{corollary}{Corollary}
\AfterEndEnvironment{corollary}{\noindent\ignorespaces}

\AfterEndEnvironment{condition}{\noindent\ignorespaces}

\AfterEndEnvironment{lemma}{\noindent\ignorespaces}

\theoremstyle{definition}
\newtheorem{exmp}{Example}
\AfterEndEnvironment{exmp}{\noindent\ignorespaces}

\AfterEndEnvironment{design}{\noindent\ignorespaces}

\AfterEndEnvironment{defi}{\noindent\ignorespaces}

\def \bz {\boldsymbol{z}}
\def \bZ {\boldsymbol{Z}}
\def \bA {\boldsymbol{A}}

\def \bX {\boldsymbol{X}}

\def \bG {\boldsymbol{G}}
\def \bE {\boldsymbol{E}}
\def \bV {\boldsymbol{V}}

\def \bD {\boldsymbol{D}}
\def \cR {\mathcal{R}}

\def \EX {\mathbb{E}}

\def \bbeta {\boldsymbol{\beta}}

\usepackage[status=draft,inline,index]{fixme}
\fxsetup{theme=color,margin=false,mode=multiuser}
\FXRegisterAuthor{bw}{bwe}{BW}

\title{\textbf{Sensitivity analysis for contamination in egocentric-network randomized trials with interference}}

\author{Bar Weinstein and Daniel Nevo\thanks{barwein@tauex.tau.ac.il, danielnevo@tauex.tau.ac.il. 
}  
\vspace{.1cm} \\
Department  of Statistics and Operations Research, Tel Aviv University
}
\date{\today}

\begin{document}
\maketitle

\begin{abstract}
     Egocentric-Network Randomized Trials (ENRTs) are increasingly used to estimate causal effects under interference when measuring complete sociocentric network data is infeasible. 
    ENRTs rely on egocentric network sampling, where a set of egos is first sampled, and each ego recruits a subset of its neighbors as alters. Treatments are then randomized across egos. 
    While the observed ego-networks are disjoint by design, the underlying population network may contain edges connecting them, leading to contamination.
    Under a design-based framework, we show that the Horvitz-Thompson estimators of direct and indirect effects are biased whenever contamination is present. 
    To address this, we derive bias-corrected estimators and propose a novel sensitivity analysis framework based on sensitivity parameters representing the probability or expected number of missing edges. This framework is implemented via both grid sensitivity analysis and probabilistic bias analysis, providing researchers with a flexible tool to assess the robustness of the causal estimators to contamination. 
    We apply our methodology to the HIV Prevention Trials Network 037 study, finding that ignoring contamination may lead to underestimation of indirect effects and overestimation of direct effects.
\end{abstract}

\noindent%

\section{Introduction}
\label{sec:intro}
In causal inference and design of experiments, it is commonly assumed that there is no interference between units, that is, the treatment assigned to one unit does not affect the outcomes of other units \citep{Cox1958}.
However, researchers are often interested in settings where 
the treatment of interest can spill over from treated units to untreated ones,
leading to interference between units.
For example, in infectious disease epidemiology,
 vaccinating some individuals may protect unvaccinated individuals \citep{Halloran1991}.
 % The term interference encompasses a wide range of transmission mechanisms,
 % including social influence, information diffusion, contagion, and infectious disease spread.

 To estimate causal effects under interference, 
researchers often rely on detailed information about the network structure
that encodes the relationships between units.
Such data is typically assumed to be on a complete \textit{sociocentric network}, 
where all units in the network and their
connections are observed \citep{Hudgens2008, aronow_estimating_2017, Forastiere2021,Tchetgen2020,Ogburn2024}.
This finite population can be composed of a collection of disjoint clusters or a single connected network. However, collecting complete sociocentric network data is often logistically challenging and expensive, limiting the application of network-based causal inference methods or resorting to sub-networks sampled from the population network. 

A common sampling design is \emph{egocentric network sampling} \citep{marsden2002egocentric,Perry2018},
where a set of egos (also known as indexes) are sampled from the population,
    and each ego recruits a subset of its neighbors in the population network as alters (non-indexes).
The observed network is then assumed to be composed of disjoint ego-networks comprising each ego and its recruited alters.
This design is attractive due to its logistical feasibility and accessibility.
Such egocentric sampling can be combined with randomized experiments,
where treatments are randomly assigned to the recruited egos, 
resulting in an \emph{egocentric-network randomized trial} (ENRT) \citep{Buchanan2018,Buchanan_2024,Chao2023,Fang2023}.
ENRTs are increasingly used in various studies, 
such as HIV prevention \citep{Latkin2009, Latkin2013, Booth2016, Miller2018}, 
substance addiction \citep{Sherman2009}, 
and mental health \citep{Asmus2017, Desrosiers2020}.

However, while most methods assume that the ego-networks are disjoint, alters may be connected to multiple egos in the population network, and egos may be connected to other egos, resulting in contamination between ego-networks.
Thus, the egocentric sampling process may lead to missing edges in the observed network,
    resulting in misspecification of the network interference structure \citep{Weinstein2026}.
Previous analyses of ENRTs typically assumed that the ego-networks
are \emph{non-overlapping} in the population network \citep{Buchanan2018,Buchanan_2024,Chao2023,Fang2023}.
That is, alters are connected only to one ego, and egos are not connected to other egos.
Such an assumption may not hold in practice and be difficult to verify with the observed data alone, absent additional validation studies or reference data.

In this paper, we relax the non-overlapping networks assumption and allow for the possibility of overlapping ego-networks.
Under a design-based inference paradigm, where the only source of randomness is the treatment assignment, we show that the Horvitz-Thompson (HT) estimators of the indirect and direct effects based on the observed ENRT data are biased whenever there is contamination between ego-networks.
We develop a sensitivity analysis framework to assess the impact of such contamination on the causal estimates, and we couple it with bias-corrected estimators of the indirect and direct effects.

These bias-corrected estimators rely on sensitivity parameters that encode 
the probability or expected number of missing alter--ego and ego--ego edges in the population network.
% These parameters can be specified using covariates and prior knowledge about suspected contamination. 
Crucially, because we treat the underlying population network as fixed but partially unobserved, these edge probabilities do not imply a stochastic superpopulation model of network generation. Rather, they quantify the researcher's epistemic uncertainty regarding latent parts of the interference network.
For the direct effect on egos, the framework requires an additional sensitivity parameter capturing the ratio of the sample average direct effect among egos exposed to at least one treated neighbor to that of unexposed egos.
We provide practical guidelines for specifying these parameters under various scenarios using covariates and prior knowledge.
Furthermore, while our framework is designed to evaluate robustness across various contamination scenarios, we also demonstrate how these sensitivity parameters can be directly calibrated using internal validation data, such as recall surveys in follow-up. This approach reduces the reliance on pure sensitivity analysis and can guide data collection strategies in future trials.

To improve asymptotic precision without compromising design-based validity, we augment these bias-corrected estimators with working outcome regression models. We further develop a conditional two-fold cross-fitting procedure tailored to the ENRT design. 
The overall framework is implemented using either grid sensitivity analysis \citep{rosenbaum1983assessing,greenland1996basic}
 or a probabilistic bias analysis \citep{Greenland2005,Lash2014,Fox2021}.
% Finally, we illustrate the proposed methodology through comprehensive simulations and a re-analysis of the HIV Prevention Trials Network (HPTN) 037 study.

% In the first, bias-corrected estimators are computed over a grid of sensitivity parameter values,
% while in the second, prior distributions are specified for the sensitivity parameters,
%     and posterior distributions for the causal estimates are obtained via Monte Carlo.

The HPTN 037 study \citep{Latkin2009} serves as a primary motivating example for our work. HPTN 037 was an ENRT that aimed to evaluate the efficacy of a peer education intervention to reduce HIV risk behaviors among people who inject drugs. 
In the trial, treated egos received training on the prevention of risky behaviors and were encouraged to share this information with their alters.
While previous analyses of this study \citep{Latkin2009, Buchanan2018, aroke2023, Chao2023} assumed that ego-networks are disjoint, contamination remains a significant concern.
Specifically, alters might inject drugs or have sexual relations with egos from other ego-networks, or egos might be connected to each other.
In this paper, we relax the assumption of disjoint ego-networks and assess the robustness of the causal estimates to contamination between the ego-networks.
Our analysis suggests that ignoring contamination may lead to underestimation of the indirect effect and overestimation of the direct effect.

Our work relates to several strands of literature.
First, it contributes to the growing literature on causal inference under interference 
where the true network structure is latent and only proxies are observed by the researchers \citep{Toulis2013,Li2021,Hardy2019,boucher2021estimating,weinstein2025}.
Second, it adds to the literature on interference with sampled network data 
which has primarily focused on random node sampling or snowball sampling designs 
under linear-in-means parametric models \citep{chandrasekhar2011econometrics, Sewell2017, Yauck2022,marray2024}.
Third, it contributes to the literature on sensitivity analysis under interference,
which has focused on unmeasured confounding \citep{VanderWeele2014},
or on settings where interference depends on two networks, but only one of which is observed \citep{Egami2020}.
Finally, it relates to the literature on cross-cluster contamination in cluster-randomized trials,
which often recommend changing the clusters boundaries to reduce contamination \citep{Hayes2017},
or more recently, develop data-driven methods for clusters construction \citep{leung2025cluster}.

The rest of the paper is organized as follows.
In Section~\ref{sec:egocentric_rct}, we introduce the setup of ENRTs, define the causal estimands of interest, and show that the commonly used HT estimators based on the observed data are biased under contamination.
In Section~\ref{sec:sa}, we develop a sensitivity analysis framework for contamination between ego-networks, providing bias-corrected estimators of the indirect and direct effects.
In Section~\ref{sec:sim}, we illustrate the proposed framework via simulations.
In Section~\ref{sec:data}, we re-analyze the HPTN 037 study,
    assessing the robustness of the causal estimates to possible contamination.
Finally, we conclude with a discussion in Section~\ref{sec:discussion}.  
All proofs are provided in the Appendix.
The \textbf{R} package \texttt{ENRTsensitivity} that implements the proposed sensitivity analysis method 
is available at 
\url{https://github.com/barwein/ENRTsensitivity}.

\section{Egocentric-network randomized trials}
\label{sec:egocentric_rct}
\subsection{Setup and recruitment process}
\label{subsec:recruitment}
Let $\bG = (\bV,\bE)$ denote the population network,
where $\bV = \{1,\ldots,N\}$ is the set of units, and $\bE$ is the set of edges, assumed to be binary. The size of the population $N$ is assumed to be finite but unknown.
The population network $\bG$ can be described by its adjacency matrix $\bA$,
where $A_{ij} = 1$ if edge $(i,j) \in \bE$, and $0$ otherwise.
We assume that the population network is undirected and that there are no self-edges,
that is, $A_{ij} = A_{ji}$ and $A_{ii} = 0$ for all $i$. 

Units are recruited to the study via an \emph{egocentric sampling design}.
The sampling process consists of two stages. 
In the first stage, a set of egos (indexes) are sampled from the population.
Then, in the second stage, each ego recruits a subset of its neighbors in the population network as alters (non-indexes).
Formally, the recruitment process is as follows:
\begin{enumerate}
    \item Sample egos from the population.
    Denote the set of recruited egos by $\cR_e \subset \bV$, 
     and $n_e = \lvert \cR_e \rvert$ as the number of egos.
    We do not impose any restrictions on the sampling process of egos.
    \item Each ego recruits alters from its neighbors in the population network that were not recruited as egos in the first stage. 
    We assume that each alter is recruited by a single ego, resulting in $n_e$ disjoint ego-networks.
    This is often enforced in the recruitment process by researchers to avoid overlaps between the observed ego-networks, e.g., as was the case in the HPTN 037 study \citep{Latkin2009}. Denote the set of recruited alters by $\cR_a \subset \bV \setminus \cR_e$, and let $n_a = \lvert \cR_a \rvert$ be their number.
\end{enumerate}
The set of all recruited units is $\cR = \cR_e \cup \cR_a$, yielding
an egocentric sample of size $n = n_e + n_a$.
We assume that this sample is a strict subset of the population, i.e., $n < N$.
The observed network $\widetilde{\bG} = (\cR, \widetilde{\bE})$ 
is a subgraph of the population network $\bG$ composed of $n_e$ disjoint ego-networks.
The set of observed edges is $\widetilde{\bE} = 
\left\{(i,j) \in \bE : i \in \cR_e, j\in \cR_a , e(j) = i\right\}$,
where $e(j) \in \cR_e$ is the ego that recruited alter $j \in \cR_a$.
Clearly, $\widetilde{\bE} \subset \bE$.
Let $\widetilde{\bA}$ be the adjacency matrix of the observed network $\widetilde{\bG}$.

\begin{figure}[!hbtp]
    \centering
    \includegraphics[width=0.7\linewidth]{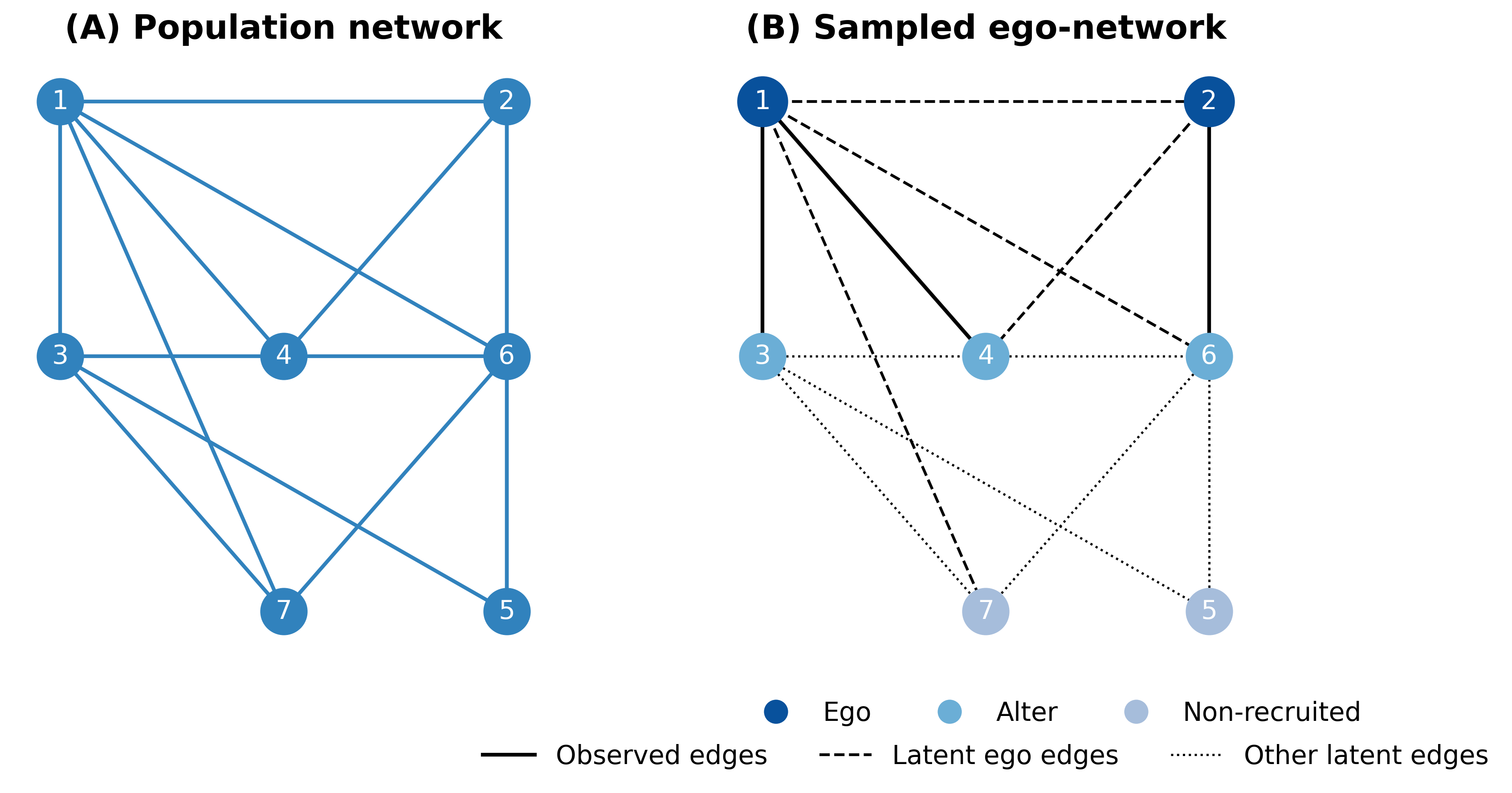
    }
    \caption{Illustration of egocentric sampling. 
    Panel~(A) shows the full $N=7$ population network $\bG$.
    Panel~(B) illustrates a single egocentric sample $\widetilde{\bG}$: each ego (nodes 1 and 2; $n_e=2$) reports only a subset of its true neighborhood as alters (nodes 3, 4, and 6; $n_a=3$).
    The observed network is composed of disjoint ego–networks (solid lines),
    while ego–ego and unreported ego–alter connections remain latent (dashed lines).
    All other edges are also unobserved (dotted lines).}
    \label{fig:egonetwork}
\end{figure}

Figure~\ref{fig:egonetwork} illustrates the egocentric sampling process in a small hypothetical population. Although the sampled egos are connected and some alters are connected to multiple egos in the population network, 
each ego reports only a subset of its true neighborhood, yielding an observed network made up of disjoint ego-networks.

\subsection{Experimental design, potential outcomes, and causal estimands}
\label{susec:doe_po_estimands}
In what follows, all assumptions, estimands, and estimators are conditioned on the recruited sample. Specifically, we consider in-sample causal effects only. Let $\bZ$ be the binary treatment assignment vector, where $Z_i = 1$ if unit $i$ is treated, and $Z_i=0$ otherwise.
We assume that treatments are randomly assigned to the recruited egos via Bernoulli randomization.
\begin{assumption}[Bernoulli experimental design]
\label{ass:bernoulli_design}
    Treatments are independently assigned only to recruited egos,
    $\Pr(Z_i = 1 \mid \cR) = p_z \mathbb{I}\{i \in \cR_e\}$ with a  known probability $p_z \in (0,1)$.
\end{assumption}
Denote the treatment space given the sample as $\mathcal{Z}_{\cR} = \left\{\bz \in \{0,1\}^N : \Pr(\bZ=\bz \mid \cR) > 0 \right\}$, where treatments of non-recruited units are assumed to be equal to zero with probability one. To avoid clutter, the conditioning on $\cR$ includes the specification of egos and alters sets as well. Furthermore, the distribution of $\bZ$ always refers to the conditional distribution of $\bZ$ given the recruitment $\cR$.

Let $Y_i(\bz)$ be the potential outcomes of each unit under treatment vector $\bz \in \mathcal{Z}_{\cR}$. We take a design-based approach, where the potential outcomes are assumed to be fixed, and the only source of randomness is the treatment assignment. 
Let $\bZ_{-i}$ denote the treatment assignment of all units except unit $i$.
We assume that $\bZ_{-i}$ affects the potential outcomes of unit $i$ only
 through the values of the following exposure mapping \citep{Manski2013,aronow_estimating_2017}.
\begin{assumption}[Exposure mapping]
\label{ass:expos_map}
    Let $F(\bZ_{-i},\bA) =\mathbb{I}\left\{\sum_{j\neq i} Z_jA_{ij} > 0\right\}$. For any $\bz, \bz' \in \mathcal{Z}_{\cR}$, if $z_i = z'_i$ and $F(\bz_{-i},\bA) = F(\bz'_{-i},\bA)$, then $Y_i(\bz) = Y_i(\bz')$. 
\end{assumption}
Denote the exposures in the true population network by $F_i = F(\bZ_{-i}, \bA)$.
The exposure $F_i$ indicates whether at least one of the neighbors of unit $i$ is treated.
Assumption~\ref{ass:expos_map} implies that the exposure mapping is correctly specified \citep{Saevje2024},
and that only treatments of neighbors affect the potential outcomes of unit $i$. Binary exposure mapping specifications are common in the ENRT literature \citep{Buchanan2018,Chao2023,Fang2023}.
For clarity, we present the main results under this binary specification, 
but extend our framework to a three-level exposure mapping specification distinguishing
between zero, one, or two-or-more treated neighbors in Appendix~\ref{apdx.sec:three_level_em}.
By Assumption~\ref{ass:expos_map}, we can express the potential outcomes as
$Y_i(\bZ)=Y_i(Z_i,F_i)$.
Consequently, each unit has four effective potential outcomes,
$Y_i(z,f)$ for $z,f\in\{0,1\}^2$, indicating whether unit $i$ is treated and whether at least one of its neighbors is treated.
We assume that the observed outcomes $Y_i, \; i \in \cR$ are generated from one of the potential outcomes.
\begin{assumption}[Consistency]
\label{ass:consis} For all $i \in \cR$, the observed outcome is given by
$$
Y_i = \sum_{z,f \in \{0,1\}^2}Y_i(z,f) \mathbb{I}\left\{Z_i=z, F_i=f\right\}.
$$    
\end{assumption}
In addition, we will also consider cases where covariates $\bX_i$ are observed for each recruited unit. We will clarify their role later on.

We consider two types of causal estimands, separate for egos and alters.
%  as common in the literature\footnote{There are interpretation issues with estimands such as $\EX\big[Y_i(1,0)-Y_i(0,1) \big]$, see \citet{Crawford2019}.}.
The first is the sample average indirect effect of treatment on the alters, 
\begin{equation}
    \label{eq:indirect_effect}
    IE = \frac{1}{n_a} \sum_{i\in\cR_a} \left[Y_i(0, 1) - Y_i(0,0)\right].
\end{equation}
The $IE$ is the average effect on alters of being exposed to at least one treated neighbor compared to not being exposed.
In Appendix~\ref{apdx_subsec:rr_estimand}, we also consider the indirect effect on the relative risk scale, for binary outcomes.
The second estimand is the sample average direct effect of treatment on the egos,
\begin{equation}
    \label{eq:directed_effect} 
    DE = 
    \frac{1}{n_e}\sum_{i\in\cR_e} \left[Y_i(1, 0) - Y_i(0,0)\right].
\end{equation}
The $DE$ is the average effect on egos of being treated compared to not being treated when not exposed to any treated neighbors.
Our explicit separation of causal estimands for egos and alters resolves interpretation issues of other estimands \citep{Crawford2019}.
Furthermore, by not conflating the two groups, we avoid imposing additional assumptions and structure, as the recruitment process described in Section~\ref{subsec:recruitment} may induce different selection mechanisms for egos and alters, leading to distinct potential outcome means across the two groups.

\subsection{Naive estimation and bias under contamination}
% \subsection{Estimation with the observed data}
The observed data consist of the treatment assignments $\bZ$, the outcomes $Y_i$, the observed network $\widetilde{\bA}$, and possibly covariates $\bX_i$ of the recruited units $i \in \cR$.
By Assumption~\ref{ass:consis}, the observed outcomes $Y_i$
are generated from the potential outcomes with exposures $F_i$ 
created by the \emph{population network} $\bA$.  However, the observed network $\widetilde{\bA}$ might not include all the relevant edges present in the population network $\bA$ 
due to the egocentric sampling process, resulting in a \emph{misspecified network interference structure} \citep{Weinstein2026}. Let $\widetilde{F}_i = F(\bZ_{-i}, \widetilde{\bA})$, be the observed exposures. Due to this network misspecification, the observed exposures $\widetilde{F}_i$ may differ from the true exposures $F_i$. Specifically, under Assumption~\ref{ass:expos_map}, only missing ego--ego and alter--ego edges induced by the sampling process can lead to incorrect observed exposures.

Consider, for example, the context of Figure~\ref{fig:egonetwork}.
If only ego 1 is treated, for alter 6 we observe exposure $\widetilde{F}_6=0$ since its ego in the observed network (node 2) is not treated. However,  its true exposure is $F_6=1$ due to its link to ego 1 in the population network. 
Similarly, under the same treatment assignment, we will observe for ego 2 exposure $\widetilde{F}_2=0$ while $F_2=1$ in the population network
since ego 2 is connected to ego 1.

Due to the ENRT design, the observed exposures of egos are $\widetilde{F}_i =0$ for all $i \in \cR_e$,
while the observed exposures of alters can be either zero or one, depending on their observed ego's treatment status.
Specifically, by Assumption~\ref{ass:bernoulli_design}, alters are exposed to a treated ego in the observed network with probability $\Pr(\widetilde{F}_i=1 \mid i \in \cR_a) = p_z$.

In practice, researchers estimate causal effects only with the observed exposures.
The most common estimator in design-based inference settings is the Horvitz-Thompson (HT) estimator.
For the indirect effect \eqref{eq:indirect_effect}, its form is
\begin{equation}
    \label{eq:IE_naive_HT}
    \widehat{IE} = 
    \frac{1}{n_a}
    \sum_{i \in \cR_a}
    \left[
    \frac{
    \mathbb{I}\{\widetilde{F}_i=1\}Y_i}{p_z} 
    -
    \frac{
    \mathbb{I}\{\widetilde{F}_i=0\}Y_i}{1 - p_z} 
    \right],
\end{equation}
and for the direct effect \eqref{eq:directed_effect},
\begin{equation}
    \label{eq:DE_naive_HT}
    \widehat{DE} = 
    \frac{1}{n_e}
    \sum_{i \in \cR_e}
    \left[
    \frac{
    \mathbb{I}\{Z_i=1\}Y_i}{p_z} 
    -
    \frac{
    \mathbb{I}\{Z_i=0\}Y_i}{1 - p_z} 
    \right].
\end{equation}
If some alter--ego or ego--ego edges are missing in the observed network $\widetilde{\bA}$,
% due to the egocentric sampling process,
% then for some units the observed exposures $\widetilde{F}_i$ may not agree with the true exposures $F_i$ under all possible treatment assignments $\mathcal{Z}_{\cR}$.
% In that case, 
both estimators \eqref{eq:IE_naive_HT} and \eqref{eq:DE_naive_HT} might be biased.
\begin{proposition}
    \label{prop:bias_naive_ht}
    Under ENRT design and Assumptions~\ref{ass:bernoulli_design}--\ref{ass:consis},
    \begin{equation*}
        \begin{aligned}
        \EX_{\bZ}\left[\widehat{IE} \right] 
        &=
        IE + 
        \frac{1}{n_a} 
    \sum_{i\in\cR_a}
    \frac{p_z -\pi_i^a}{1-p_z}
    \left[
    Y_i(0, 1) - Y_i(0,0)
    \right],
    \\
      \EX_{\bZ}\left[\widehat{DE} \right] 
        &=
        DE + 
        \frac{1}{n_e} 
    \sum_{i\in\cR_e}
    \pi_i^e
    \left\{
    \left[ Y_i(1, 1) - Y_i(0,1)\right]
    - \left[ Y_i(1,0) - Y_i(0,0) \right]
    \right\},
        \end{aligned}
    \end{equation*}
    where $\pi_i^a = \Pr(F_i=1\mid i\in \cR_a)$ and $\pi_i^e = \Pr(F_i=1 \mid i \in \cR_e)$ are the probabilities that an alter and an ego, respectively, are exposed to at least one treated neighbor in the population network $\bA$.
\end{proposition}
When all alters are connected only to their own ego, then $\pi_i^a=p_z$ for all $i\in \cR_a$, and $\widehat{IE}$ is an unbiased estimator of $IE$.
Otherwise, whenever some alters are connected to more than one ego in the population network,
 $\pi_i^a > p_z$, and the indirect effect estimator \eqref{eq:IE_naive_HT} based on
  the observed exposures can be biased. 
The magnitude of the bias depends on the number of egos each alter is connected to in the population network.
That is, the more egos an alter is connected to, the larger the gap between $\pi_i^a$ and $p_z$, and hence the larger the bias. 
In addition, we show in Appendix~\ref{apdx_subsec:direction_bias_to_null} that under unit-level monotonicity (in either direction) of the potential outcomes $Y_i(0,1)$ and $Y_i(0,0)$, the naive indirect estimator $\widehat{IE}$ is
biased towards the null of no indirect effect.

If an ego is not connected to any other egos in the population network, then $\pi_i^e=0$. If this is the case for all $i \in \cR_e$, then $\widehat{DE}$  is an unbiased estimator of \eqref{eq:directed_effect}. However, if at least one ego $i$ is connected to at least one other ego in the population network, then $\pi_i^e>0$, and the direct effect estimator $\widehat{DE}$ based on the observed data can be biased.
Similarly to alters, the magnitude of the bias depends on the number of ego--ego edges in the population network.
Furthermore, in Appendix~\ref{apdx_subsec:sign_pres}, we analyze the direction of the bias of $\widehat{DE}$ relative to the null of no direct effect. We demonstrate that the direction of the bias is governed by the ratio of the sample average direct effect among egos exposed to at least one treated neighbor to the effect among non-exposed egos.  
Generally, if the direct effects are larger, on average, in the presence of treated neighbors than in their absence, $\widehat{DE}$ is biased away from the null. Conversely, if the direct effects are smaller, on average, in the presence of treated neighbors, the estimator is biased towards the null.

\section{Sensitivity analysis for contamination}
\label{sec:sa}

Proposition~\ref{prop:bias_naive_ht} shows that the HT estimators based on the observed data are biased whenever there is contamination between the ego-networks.
We refer to contamination as the presence of alter--ego and ego--ego edges in the population network $\bA$,
that are missing from the observed network $\widetilde{\bA}$ due to the egocentric sampling process. 
Under such contamination, the ego-networks are not disjoint in the population network,
 leading to incorrect exposures when computed using the observed network.  
    
By the structure of the egocentric sampling and the exposure mapping specification (Assumption~\ref{ass:expos_map}), alters with observed exposure $\widetilde{F}_i=1$ will have correctly classified exposures. 
That is, $\widetilde{F}_i=1$ implies $F_i=1$ with probability one for $i \in \cR_a$.
However, alters with $\widetilde{F}_i=0$ can have a true exposure of $F_i=1$ or $F_i=0$,
depending on the structure of the population network $\bA$.
Therefore, the observed outcomes $Y_i$ for alters with $\widetilde{F}_i=0$ can be generated from either $Y_i(0,1)$ or $Y_i(0,0)$.

On the other hand, all egos have observed exposures of $\widetilde{F}_i=0$, 
as they are linked only to alters in their ego-network in the observed network $\widetilde{\bA}$.
However, the observed outcomes of egos that are assigned to treatment $Z_i=z$ can be generated from $Y_i(z,1)$ or $Y_i(z,0)$, implying that observed outcomes of egos are possibly linked to four different potential outcomes, compared to only two among the alters.

We develop a sensitivity analysis framework that enables researchers to assess the impact of possible contamination between ego-networks on the causal estimates.
In Sections~\ref{subsec:ie} and \ref{subsec:de}, we provide bias-corrected estimators of the indirect and direct effects 
that account for possible exposure misspecification due to the egocentric-network recruitment process. 
Furthermore, in Section~\ref{subsec:augmented_cf}, we show how to augment these estimators with an outcome model through a cross-fitting procedure tailored to the design-based framework taken in this paper, enabling researchers to integrate
covariates into the analysis to improve efficiency. Such model-assisted estimation is common in survey sampling \citep{Saerndal2003} and causal inference \citep{ding2024first}.

Building on Proposition \ref{prop:bias_naive_ht}, the bias-corrected indirect and direct effect estimators are functions of the exposure probabilities under the true population network $\pi_i^a$ and $\pi_i^e$, respectively.
The structure of the egocentric sampling design and our setup 
enables us to decompose $\pi_i^a$ and $\pi_i^e$ into basic
 components that are straightforward to compute given the specification
  of the presumed \emph{expected number} or the \emph{probability} of missing alter--ego and ego--ego edges.
 These components will form the basis of our sensitivity parameters, 
 enabling seamless integration of covariates and prior knowledge into the sensitivity analysis.
Crucially, we maintain the design-based framework throughout, treating the underlying population network $\bA$ as fixed but partially unobserved. Therefore, probability statements regarding latent edges do not reflect a stochastic superpopulation model or assumptions about a data-generating process. Rather, they quantify the researcher's epistemic uncertainty regarding the fixed, unobserved edges in the interference network.
 We provide multiple examples and practical guidelines on how to specify these parameters in Section~\ref{subsec:rho_param}.
 
For the direct effect, we require an additional sensitivity parameter to account for the complexity resulting from possibly four potential outcomes generating the observed outcomes. This parameter is the ratio of the sample average direct effect among egos exposed to at least one treated neighbor to the effect among non-exposed egos, as described in Section~\ref{subsec:de}.

While the primary focus of this paper is to assess the robustness of the estimates over plausible contamination scenarios, researchers can sometimes calibrate the sensitivity parameters from auxiliary information. In Section~\ref{subsec:int_validation_data}, we demonstrate how different types of internal validation data, such as recall surveys at follow-up, can be leveraged to estimate these parameters. This approach reduces the reliance on pure sensitivity analysis and can motivate data collection and design choices in future studies.

\subsection{Indirect effect}
\label{subsec:ie}
Define the bias-corrected indirect effect estimator as a function of the observed data and $\pi_i^a$,
\begin{equation}
    \label{eq:IE_corrected}
    \widehat{IE}_{adj} = 
    \frac{1}{n_a}
    \sum_{i \in \cR_a}
    \frac{1-p_z}{1-\pi_i^a}
    \left[
    \frac{
    \mathbb{I}\{\widetilde{F}_i=1\}Y_i}{p_z} 
    -
    \frac{
    \mathbb{I}\{\widetilde{F}_i=0\}Y_i}{1 - p_z} 
    \right].
\end{equation}
Given $\pi_i^a$, \eqref{eq:IE_corrected} is an unbiased estimator of the indirect effect \eqref{eq:indirect_effect}.
\begin{proposition}
\label{prop:ie_sa}
  Under ENRT design and Assumptions~\ref{ass:bernoulli_design}--\ref{ass:consis}, we have
  $\EX_{\bZ}\big[\widehat{IE}_{adj} \big] = IE$.
\end{proposition}
By the ENRT design and Assumption~\ref{ass:expos_map}, $\pi_i^a \geq p_z$, with equality if and only if there are no additional latent alter--ego edges for alter $i$.
Consequently, if all alters have the same exposure probability, $\pi^a_i = \pi^a$,
then $\widehat{IE}_{adj} = \frac{1-p_z}{1-\pi^a} \widehat{IE}$.
Therefore, $\text{sign}\big(\widehat{IE}_{adj}\big) = \text{sign}\big(\widehat{IE}\big)$, and $\widehat{IE}_{adj} \geq \widehat{IE}$ if $\widehat{IE} >0$ and vice versa otherwise.

We derive variance estimators for the bias-corrected estimator $\widehat{IE}_{adj}$. 
By the ENRT design, alters in the same ego-network all have the same treatment assignments and observed exposure status. Thus, we can calculate the variance by using the aggregated outcomes in each ego-network.
Let $r^a_i = \frac{1-p_z}{1-\pi_i^a}\left[\frac{
    \mathbb{I}\{\widetilde{F}_i=1\}Y_i}{p_z} 
    -
    \frac{
    \mathbb{I}\{\widetilde{F}_i=0\}Y_i}{1 - p_z}\right]$ such that $\widehat{IE}_{adj}=\frac{1}{n_a}\sum_{i\in \cR_a} r^a_i$. 
 The design-based variance estimator is \citep{ding2024first}
\begin{equation}
    \label{eq:ie_adj_var_esti}\widehat{V}_{\bZ}\left(\widehat{IE}_{adj}\right) = 
    \frac{1}{n_a^2}\sum_{i \in \cR_e}
    \left(T_i - \overline{T}\right)^2, 
\end{equation}
where $T_i =\sum_{j \in \cR_a : e(j)=i} 
    r^a_j $ and $\overline{T} = \frac{1}{n_e}\sum_{i \in \cR_e} T_i$ is the average over ego-networks. 
    In Appendix~\ref{apdx_subsec:var_estimation}, we derive the variance estimator and establish the asymptotic normality of the bias-corrected estimator $\widehat{IE}_{adj}$, ensuring that Wald-type confidence intervals using the variance estimator $\widehat{V}_{\bZ}\left(\widehat{IE}_{adj}\right)$ achieve a valid, albeit conservative, asymptotic coverage rate.

\subsection{Direct effect}
\label{subsec:de}

As the observed outcomes of egos are possibly generated from four different potential outcomes, compared to only two for the alters, bias correction for the direct effect estimator requires specifying the relation between the direct effect among non-exposed egos ($F_i=0$) and the effect among egos exposed to at least one treated neighbor ($F_i=1$).
Denote the direct effect of treatment on an ego $i \in \cR_e$  with exposure value $f \in \{0,1\}$ by $\Delta_i(f) = Y_i(1,f) - Y_i(0,f)$.
Let $\overline{\Delta}(f)= \frac{1}{n_e}\sum_{i \in \cR_e}\Delta_i(f)$ denote the sample average effect for $f \in \{0,1\}$. That is, with a slight abuse of notations, $\overline{\Delta}(0) = DE$ as defined in \eqref{eq:directed_effect}.
Let $\kappa$ be the ratio of the sample average direct effect among egos exposed to at least one treated neighbor $\overline{\Delta}(1)$ to the effect among non-exposed egos $\overline{\Delta}(0)$:
\begin{equation}
    \label{eq:kappa}
    \overline{\Delta}(1) = \kappa \overline{\Delta}(0).
\end{equation}
Under the design-based framework taken in the paper, where the potential outcomes are fixed, there is always a constant $\kappa$ satisfying \eqref{eq:kappa}.
The parameter $\kappa$ will serve as an additional sensitivity analysis parameter. 
We discuss the interpretation of $\kappa$ below.

Let $\overline{\pi}^e = \frac{1}{n_e}\sum_{i \in \cR_e} \pi_i^e$ be the sample average probability of exposure among the egos.
We define an inflation factor $u_e = n_e \left[1 + \overline{\pi}^e(\kappa-1) 
    \right]$ 
    that adjusts the effective sample size of egos based on the expected contamination and $\kappa$.
    Using this factor, we define the bias-corrected direct effect estimator as a function of the observed data, $\overline{\pi}^e$ and $\kappa$,
\begin{equation}
    \label{eq:DE_corrected}
    \widehat{DE}_{adj} =
    \frac{1}{1 + \overline{\pi}^e(\kappa-1)}
    \widehat{DE} = 
    \frac{1}{u_e}
    \sum_{i \in \cR_e}
    \left[
    \frac{
    \mathbb{I}\{Z_i=1\}Y_i}{p_z} 
    -
    \frac{
    \mathbb{I}\{Z_i=0\}Y_i}{1 - p_z} 
    \right]
    .
\end{equation}
An additional assumption is needed for $\widehat{DE}_{adj}$ to be unbiased, as we now formalize.
For two vectors $\boldsymbol{a},\boldsymbol{b} \in \mathbb{R}^m$, let the empirical covariance be $S_{\boldsymbol{a},\boldsymbol{b}} = \frac{1}{m}\sum_{i=1}^m (a_i -\overline{a})(b_i -\overline{b})$, where $\overline{a} = \frac{1}{m}\sum_i a_i$. To eliminate bias, we require that the empirical covariance between exposure probabilities $\pi_i^e$ and the direct effect among exposed egos $\Delta_i(1)$ is equal to the empirical covariance between $\pi_i^e$ and the direct effect among non-exposed egos $\Delta_i(0)$. Let $\boldsymbol{\pi}^e = \left(\pi_1^e,\ldots,\pi_{n_e}^e\right)$ and $\boldsymbol{\Delta}_i(f) = \left(\Delta_1(f),\ldots, \Delta_{n_e}(f)\right)$ be the vectors of exposure probabilities and direct effects at the unit-level of the egos.
\begin{assumption}
    \label{ass:cov_pi_de_egos}
    $S_{\boldsymbol{\pi}^e, \boldsymbol{\Delta}(1)}=S_{\boldsymbol{\pi}^e, \boldsymbol{\Delta}(0)}$. 
\end{assumption}
To provide a better understanding of the implications of Assumption~\ref{ass:cov_pi_de_egos}, consider the representation of the potential outcomes under a saturated  model
\begin{equation*}
    Y_i(z,f) = \beta_{0i} + \beta_{1i}z + \beta_{2i}f + \beta_{3i}zf.
\end{equation*} 
Under this representation, $\Delta_i(1) = \beta_{1i} + \beta_{3i}$, $\Delta_i(0) = \beta_{1i}$, and simple calculations show that Assumption~\ref{ass:cov_pi_de_egos} is equivalent to 
\begin{equation}
\label{eq:no_cov_pi_de}
    S_{\boldsymbol{\pi}^e, \boldsymbol{\beta}_{3}} = 
    \frac{1}{n_e}\sum_{i \in \cR_e}
    (\pi_i^e - \overline{\pi}^e)(\beta_{3i} - \overline{\beta_{3}}) =0.
\end{equation}
That is, the empirical covariance between the exposure probabilities $\pi_i^e$ and the treatment-exposure interaction terms $\beta_{3i}$ is zero.
Eq.~\eqref{eq:no_cov_pi_de} naturally holds if these interaction effects are homogeneous across all egos (i.e., $\beta_{3i} = \beta_3$ for some constant $\beta_3$) or if the exposure probabilities are homogeneous (i.e., $\pi_i^e=\pi^e$ for some constant $\pi^e \in [0,1]$). In practice, this assumption would only be violated if egos that are highly connected to other egos systematically exhibit different interaction responses to the intervention compared to less connected egos.

Given $\pi_i^e$ and $\kappa$, \eqref{eq:DE_corrected} is an unbiased estimator of the direct effect \eqref{eq:directed_effect}.
\begin{proposition}
\label{prop:de_sa}
    Under ENRT design and Assumptions~\ref{ass:bernoulli_design}--\ref{ass:cov_pi_de_egos}, we have
     $\EX_{\bZ}\left[\widehat{DE}_{adj} \right] = DE$.
\end{proposition}
% Assumption~\ref{ass:cov_pi_de_egos} is required to eliminate residual bias.
As described above, under homogeneous exposure probabilities, the empirical covariance requirement stated in Assumption~\ref{ass:cov_pi_de_egos} is trivially satisfied.
\begin{corollary}
    If exposure probabilities are homogeneous ($\pi_i^e = \pi^e$ for all $i\in\cR_e$), then $\EX_{\bZ}\left[\widehat{DE}_{adj} \right] = DE$. 
\end{corollary}
Reasoning about the range of plausible $\kappa$ values depends on the mechanism under study. 
For example, in vaccine efficacy studies, if treatment represents vaccines, 
it is reasonable to assume that the direct effect of vaccination is larger, on average, among egos with non-vaccinated neighbors compared to those with vaccinated neighbors (who benefit from some indirect protection), implying $\kappa < 1$.
On the other hand, in the HPTN 037 study, which we analyze in Section~\ref{sec:data}, the treatment is a behavioral intervention reliant on peer education. In such settings, researchers may believe the intervention to be mutually reinforcing. That is, the direct effect may be larger, on average, for egos who are also exposed to at least one treated ego compared to those who are not, which implies $\kappa > 1$.
While researchers must often reason about ranges for $\kappa$ based on subject-matter expertise, it is also possible to estimate $\kappa$ directly if appropriate internal validation data are collected, as we detail in Section \ref{subsec:int_validation_data}.

If $\pi_i^e=0$ or $\kappa=1$, the bias-corrected estimator $\widehat{DE}_{adj}$ reduces to the naive HT estimator $\widehat{DE}$.
Thus, $\widehat{DE}$ is unbiased if, for example, $\kappa=1$ and the probabilities $\pi_i^e$ are homogeneous. That is, $\widehat{DE}$ can be an unbiased estimator of $DE$ even if there is some contamination between the egos ($\pi^e>0$). See Appendix~\ref{apdx_subsec:proof_prop3} for more details.

% Moreover,
% if $\kappa > 1 - \frac{1}{\overline{\pi}^e}$,
% $\text{sign}\big(\widehat{DE}_{adj}\big) = \text{sign}\big(\widehat{DE}\big)$. As $\kappa > 1 - \frac{1}{\overline{\pi}^e}$ in most cases (for $\pi^e=0.01$ this implies $\kappa > -99$), the bias-corrected estimator generally preserves the direction of the naive estimator.
Furthermore,
if $\kappa > 1 - \frac{1}{\overline{\pi}^e}$ (e.g., for $\overline{\pi}^e=0.01$ this implies $\kappa > -99$),
$\text{sign}\big(\widehat{DE}_{adj}\big) = \text{sign}\big(\widehat{DE}\big)$, i.e., the bias-corrected estimator will have the same sign as the naive estimator.
In addition, if $\kappa>1$ or $\kappa \in (1 - \frac{1}{\overline{\pi}^e},1)$, the bias-corrected estimator $\widehat{DE}_{adj}$ is closer to or further away from, respectively, zero than the naive estimator $\widehat{DE}$.
See Appendix~\ref{apdx_subsec:sign_pres} for further details.

The requirement of Assumption~\ref{ass:cov_pi_de_egos} in Proposition~\ref{prop:de_sa} can be replaced with alternative formulations. We derive nonparametric bounds for $DE$ based on the observed data, $\pi_i^e$ and $\kappa$, enabling researchers to obtain valid intervals for $DE$ without Assumption~\ref{ass:cov_pi_de_egos}.
Furthermore, we show that replacing the average-level proportionality between $\overline{\Delta}(1)$ and $\overline{\Delta}(0)$, as expressed in \eqref{eq:kappa}, with a unit-level proportionality assumption between $\Delta_i(1)$ and $\Delta_i(0)$, produces a modified bias-corrected estimator that is an unbiased estimator of $DE$ without Assumption~\ref{ass:cov_pi_de_egos}. See Appendix~\ref{apdx_subsec:relaxing_ass4} for more details. 

Our simulation study (Section~\ref{sec:sim} and Appendix~\ref{apdx.sec:sim}) assessed the impacts of violating Assumption~\ref{ass:cov_pi_de_egos} and misspecifying the sensitivity parameter $\kappa$. We found that violations of Assumption~\ref{ass:cov_pi_de_egos} resulted in negligible bias for the corrected direct effect estimators, albeit with slight variance underestimation. The impact of misspecifying $\kappa$, however, depends on its direction. If $\kappa$ is misspecified in the wrong direction (e.g., specifying $\kappa < 1$ when the true $\kappa > 1$), the bias-corrected estimator can exhibit greater bias and lower coverage than the naive estimator. Conversely, when $\kappa$ is misspecified in magnitude but correctly oriented in direction, the bias-corrected estimator outperforms the naive estimator, maintaining smaller bias and higher coverage.
    
We derive a conservative variance estimator for the bias-corrected estimator $\widehat{DE}_{adj}$. 
Let $r^e_i = \frac{
    \mathbb{I}\{Z_i=1\}Y_i}{p_z} 
    -
    \frac{
    \mathbb{I}\{Z_i=0\}Y_i}{1 - p_z}$ 
    such that $\widehat{DE}_{adj}=
    \frac{1}{u_e}
    \sum_{i\in \cR_e} 
    r^e_i$. 
By the ENRT design, treatment assignment is independent across egos. The Neyman-type variance estimator is therefore \citep{ding2024first},
\begin{equation*}
    \widehat{V}_{Neyman}\left(\widehat{DE}_{adj}\right) = 
    \frac{1}{u_e^2}
    \sum_{i \in \cR_e}
    \left(r^e_i - \widehat{DE}_{adj}\right)^2.
\end{equation*}
This estimator assumes that the terms $r^e_i$ are independent. However, in the presence of contamination, this independence assumption is violated. 
If two egos $i$ and $j$ share a neighbor $k \in \cR_e$ in the population network $\bA$, then the terms $r^e_i$ and $r^e_j$ are not independent. This dependence induces pairwise covariances ($\text{Cov}_{\bZ}(r^e_i, r^e_j)$) that are omitted from  $\widehat{V}_{Neyman}\left(\widehat{DE}_{adj}\right)$.
To ensure valid inference, we propose a conservative estimator $\widehat{V}_{Conta.}\left(\widehat{DE}_{adj}\right)$ for the variance in $\widehat{DE}_{adj}$ attributed to contamination between ego-networks,
derived using the edge probabilities under the sensitivity model (presented in Section \ref{subsec:rho_param}) to account for these latent correlations.
The final corrected variance estimator for $\widehat{DE}_{adj}$ is given by
\begin{equation}
    \label{eq:de_adjust_var}
    \widehat{V}_{\bZ}\left(\widehat{DE}_{adj}\right) = 
    \widehat{V}_{Neyman}\left(\widehat{DE}_{adj}\right)
    + 
    \widehat{V}_{Conta.}\left(\widehat{DE}_{adj}\right).
\end{equation}
In Appendix~\ref{apdx_subsec:var_estimation} we provide the derivation of the variance estimator $\widehat{V}_{\bZ}\left(\widehat{DE}_{adj}\right)$ and, analogously to Section~\ref{subsec:ie}, show the asymptotic normality of $\widehat{DE}_{adj}$ and that Wald-type confidence intervals using the variance estimator $\widehat{V}_{\bZ}\left(\widehat{DE}_{adj}\right)$ achieve a valid, albeit conservative, asymptotic coverage rate.

\subsection{Outcome-model augmented estimators}
\label{subsec:augmented_cf}
% Augmenting the bias-corrected design-based estimators $\widehat{IE}_{adj}$ and $\widehat{DE}_{adj}$ with outcome models can improve efficiency and estimation accuracy.

The bias-corrected estimators $\widehat{IE}_{adj}$ and $\widehat{DE}_{adj}$ can be improved by incorporating outcome regression models that leverage information from the covariates $\bX_i$. 
Crucially, the incorporation of these models does not constitute a departure from our design-based paradigm. The regression models serve strictly as working models used to improve asymptotic precision, and the theoretical validity of the resulting augmented estimators relies solely on the randomization of the treatment assignment \citep{lin2013agnostic, abadie2020sampling, gao2025causal}.
As we demonstrate in our simulation study (Section~\ref{sec:sim}), this augmentation yields valid variance estimates, overcoming the conservative variance estimation of the non-augmented estimators, while maintaining negligible bias. 

To fully preserve this design-based validity, standard regression adjustments require careful implementation. While sample splitting (i.e., cross-fitting) is the standard solution to avoid finite-sample bias, standard splitting schemes can violate the design-based independence assumptions \citep{lu2025conditional}.
To address this, we employ a conditional two-fold cross-fitting procedure tailored to the ENRT design.
Specifically, we adapt Algorithm~1
of \citet{lu2025conditional}
to our setting. We describe here the algorithm for completeness, but refer readers to \citet{lu2025conditional} for additional details. We focus on the augmented bias-corrected estimators, but the naive estimators $\widehat{IE}$ and $\widehat{DE}$ can also be augmented similarly, albeit without the bias-correction weighting.

Let $\widehat{\mu}^a_i(f)$ denote the predicted value for alter $i$ under exposure $\widetilde{F}_i=f$, from an outcome model estimated using covariates $\bX_i$. 
For example, $\widehat{\mu}^a_i(f)$ can be the predicted value from a linear model of the form $Y_i \sim \widetilde{F}_i + \bX_i$.
Similarly, let $\widehat{\mu}^e_i(z)$ be the predicted value for ego $i$ under treatment $Z_i=z$, from an outcome model estimated using covariates $\bX_i$. The two-fold cross-fitting algorithm is as follows.
\begin{enumerate}
    \item Randomly split the observed ego-networks into two disjoint parts $S_0, S_1$ such that $\lvert S_0 \cup S_1 \rvert = n$ and $S_0 \cap S_1 = \emptyset$. Specifically, we assign each ego-network (ego and alters) into $S_1$ with probability $0.5$.
    \item For $q=0,1$: 
    \begin{enumerate}
        \item Estimate the outcome models for alters and egos using data from the complementary split $S_{1-q}$.
        \item Predict $\widehat{\mu}^a_i(f)$ and $\widehat{\mu}^e_i(z)$ on the units from $S_q$.
        \item Estimate $IE$ on units from $S_q$:
        \begin{equation*}
            \widehat{IE}^{aug}_{adj[q]} 
            = \frac{1}{n_{a[q]}}
            \sum_{i \in \cR_a \cap S_q}
            \frac{1-p_z}{1-\pi^a_i}
            \left[D^a_i +
            \widehat{\mu}^a_i(1) - 
            \widehat{\mu}^a_i(0) \right],
        \end{equation*}
        where $n_{a[q]} =\sum_{i \in \cR_a} \mathbb{I}\{i \in S_q\}$ is the number of alters in split $S_q$, and where
        \begin{equation*}
            D^a_i = \frac{
    \mathbb{I}\{
    \widetilde{F}_i=1\}\left(Y_i - \widehat{\mu}^a_i(1) \right)}{p_z} 
    -
    \frac{
    \mathbb{I}\{\widetilde{F}_i=0\}\left(Y_i - \widehat{\mu}^a_i(0) \right)}{1 - p_z}. 
        \end{equation*}
    \item Estimate $DE$ on units from $S_q$:
    \begin{equation*}
        \widehat{DE}^{aug}_{adj[q]} =
        \frac{1}{u_{e[q]}} 
    \sum_{i \in \cR_e \cap S_q}
    \left[
    D^e_i
    + \widehat{\mu}^e_i(1) - \widehat{\mu}^e_i(0) 
    \right],
    \end{equation*}
    where $u_{e[q]} =n_{e[q]}[1 + \overline{\pi}^{e[q]
    }(\kappa-1)]$ similarly to $u_e$ in Section~\ref{subsec:de}, with $n_{e[q]} = \sum_{i \in \cR_e} \mathbb{I}\{i \in S_q\}$ and 
    $\overline{\pi}^{e[q]} = \frac{1}{n_{e[q]}}\sum_{i \in \cR_e \cap S_q}  \pi_i^e$ 
    are the number of egos and the average exposure probability, respectively, in split $S_q$, and 
    \begin{equation*}
        D^e_i = \frac{
    \mathbb{I}\{
    Z_i=1\}\left(Y_i - \widehat{\mu}^e_i(1) \right)}{p_z} 
    -
    \frac{
    \mathbb{I}\{Z_i=0\}\left(Y_i - \widehat{\mu}^e_i(0) \right)}{1 - p_z}. 
    \end{equation*}
    \end{enumerate}
    \item Combine the estimates across splits
    \begin{equation}
    \label{eq:aug_esti}
        \begin{aligned}
            \widehat{IE}^{aug}_{adj}
            &= \sum_{q =0,1}\frac{n_{a[q]}}{n_a} \widehat{IE}^{aug}_{adj[q]}, 
            \\ 
            \widehat{DE}^{aug}_{adj}
            &= \sum_{q =0,1}\frac{u_{e[q]}}{u_e} \widehat{DE}^{aug}_{adj[q]}.
        \end{aligned}
    \end{equation}
\end{enumerate}
Variance estimation is a combination across splits of variance estimators \eqref{eq:ie_adj_var_esti} for the IE estimator and \eqref{eq:de_adjust_var} for the DE estimator \citep{lu2025conditional}.
In Appendix~\ref{apdx_subsec:aug_estimation}, we show that the augmented estimators $\widehat{IE}^{aug}_{adj}$ and $\widehat{DE}^{aug}_{adj}$ remain unbiased estimators of $IE$ and $DE$, respectively, derive their variance estimator, and establish their asymptotic normality, ensuring that Wald-type confidence intervals based on their associated variance estimators achieve at least the nominal asymptotic coverage rate.

\subsection{Specifying the sensitivity parameters}
\label{subsec:rho_param}
Propositions~\ref{prop:ie_sa} and \ref{prop:de_sa} establish that the bias-corrected estimators
of the indirect and direct effects depend on the unknown exposure probabilities
$\pi_i^a$ and $\pi_i^e$ of alters and egos, respectively, in the population network $\bA$. 
For the direct effect, the bias-corrected estimator also depends on the sensitivity parameter $\kappa$. 
We now illustrate how the computation of $\pi_i^a$ and $\pi_i^e$
reduces to specification of edge-level probabilities or the expected number
  of missing alter--ego and ego--ego edges. 
  These specifications form the basis of our sensitivity parameters.
Importantly, while we maintain the design-based framework, implying that the population network $\bA$ is fixed, its full structure is unobserved. Consequently, we treat the presence of missing edges as uncertain for the purpose of the sensitivity analysis. We do not posit a generative superpopulation model for $\bA$. Rather, we define sensitivity parameters to model our uncertainty about the missing alter--ego and ego--ego edges. Therefore, all subsequent probability statements regarding missing edges should be interpreted strictly as quantifications of this uncertainty.

By the exposure mapping specification (Assumption~\ref{ass:expos_map}) and 
the experimental design (Assumption~\ref{ass:bernoulli_design}),
the relevant edges in the population network $\bA$ that determine the probabilities $\pi_i^a$ and $\pi_i^e$ are 
only the alter--ego or ego--ego edges.
In the observed network $\widetilde{\bA}$, each alter $i \in \cR_a$ is connected only to its ego $e(i) \in \cR_e$.
Let $\cR_e^{j} = \cR_e \setminus \{j\}$ be the set of all egos other than $j \in \cR_e$.
The latent alter--ego edges $\left\{A_{ij} : i \in \cR_a, j \in \cR_e^{e(i)} \right\}$,
and ego--ego edges $\left\{A_{ij}: i,j  \in \cR_e, i \neq j \right\}$  are the edges missing from $\widetilde{\bA}$, relevant to the exposure probabilities.
% Define the edge-level sensitivity probabilities by
Let $\rho^e_{ij}$ and $\rho^a_{ij}$ represent the postulated probabilities that a missing edge exists, i.e.,
\begin{equation}
    \label{eq:sbm}
    \begin{aligned}
    \rho^e_{ij} &= \Pr\left(A_{ij}=1 \mid i,j \in\cR_e, i \neq j\right),
    \\ 
    \rho^a_{ij} &= \Pr\big(A_{ij}=1 \mid i \in \cR_a, j \in \cR_e^{e(i)} \big).
    \end{aligned}
\end{equation}
Using \eqref{eq:sbm}, we can derive expressions for $\pi^e_i$ and $\pi^a_i$.
By the law of total probability and Assumption~\ref{ass:bernoulli_design},
 $\Pr\left(Z_jA_{ij} = 1 \mid i\neq j \in \cR_e\right) =
    p_z\rho^e_{ij}$.
Assuming edge independence yields
\begin{equation}
    \label{eq:pi_ego}
    \pi^e_i = 1 - \prod_{j \in  \cR_e^i}
    \big(1 - p_z \rho^e_{ij}\big).
\end{equation}
Turning to $\pi^a_i$, alter $i\in\cR_a$ is exposed ($F_i=1$) only if its ego $e(i)$ or at least one of the other recruited egos connected to $i$ in the population network $\bA$ is treated.
Therefore, we obtain (Appendix~\ref{apdx_subsec:expos_prob_alters})
\begin{equation}
    \label{eq:pi_a}
    \pi^a_i = p_z + 
    (1-p_z)
    \Bigg[1- \prod_{j \in  \cR_e^{e(i)}}
    \big(1 - p_z \rho^a_{ij}\big)\Bigg].
    % = 
    % 1 - (1-p_z)\prod_{j \in  \cR_e^{e(i)}}
    % \big(1 - p_z \rho^a_{ij}\big),
\end{equation}
Eqs.~\eqref{eq:pi_ego} and \eqref{eq:pi_a} illustrate how the exposure probabilities 
can be computed given specifications of the edge-level sensitivity parameters $\rho^e_{ij}$ and $\rho^a_{ij}$.

We now discuss some prominent examples of how researchers can specify these parameters in practice. The examples include both homogeneous and heterogeneous edge probabilities,
as well as specifications based on the expected number of missing edges.  

\begin{exmp}[Homogeneous probabilities]
    \label{exmp:pi_homo}
    If researchers are willing to specify the same probabilities \eqref{eq:sbm} for all units, i.e., 
    $\rho^e_{ij}=\rho^e$ and $\rho^a_{ij}=\rho^a$, then
    the exposure probabilities \eqref{eq:pi_ego} and \eqref{eq:pi_a} simplify to
    \begin{equation*}
        \begin{aligned}
            \pi_i^e &= 1 - (1-p_z\rho^e)^{n_e-1}, \\
            \pi_i^a &= p_z + (1-p_z)\left[1- (1-p_z\rho^a)^{n_e-1}\right].
        \end{aligned}
    \end{equation*}
    For instance, if researchers suspect that about $1\%$ of the ego--ego edges are possible, they can set $\rho^e=0.01$. 
\end{exmp}

\begin{exmp}[Homogeneous number of edges]
    \label{exmp:pi_number}
    A special case of Example~\ref{exmp:pi_homo} occurs
     by specifying the approximate \emph{expected number} of latent edges, separately for the ego--ego and alter--ego edges. 
    Let $m^e \in \left\{1,\ldots, \binom{n_e}{2}\right\}$ be the total number of ego--ego edges that are believed to be missing. Setting 
    \begin{equation*}
        \rho^e_{ij} = \rho^e(m^e) = \frac{m^e}{\binom{n_e}{2}},
    \end{equation*}
    weights equally each ego--ego edge probability and yields $m^e$ expected ego--ego edges.
    \\
    There are $n_a(n_e-1)$ possible missing alter--ego edges. 
     If approximately $m^a$ such edges are expected to be missing, set the probability of each edge to
    \begin{equation*}
        \rho^a_{ij} = \rho^a(m^a) = \frac{m^a}{n_a(n_e-1)}.
    \end{equation*} 
    For instance, if researchers suspect that each alter 
    is connected to approximately $c^a \geq 0$ additional egos, 
    they can select $m^a = c^a \times n_a$.     
    Similarly, if each ego is believed to be connected to approximately $c^e\geq 0$ additional egos,
    then $m^e = \frac{c^e \times n_e}{2}$.
\end{exmp}

\begin{exmp}[Heterogeneous probabilities]
    \label{exmp:pi_hetero}
 Example~\ref{exmp:pi_homo} can be extended by allowing edge probabilities to vary based on unit-level covariates $\bX_i$. 
 This captures the well-known phenomenon in network science of homophily \citep{mcpherson2001homophily}, where units with similar characteristics are more likely to interact.
 
 We begin with a baseline probability parameter, $\rho^e_{base}$, analogous to the homogeneous parameter $\rho^e$ in Example~\ref{exmp:pi_homo}.
 We then adjust this baseline using a pairwise dissimilarity measure $d_{ij} = d(\bX_i, \bX_j)$. Common choices include the $L^p$ norm, $d_{ij}=\lVert\bX_i - \bX_j\rVert_p$, or the Cosine distance, $d_{ij}=1 - \frac{\bX_i^T\bX_j}{\lVert \bX_i \rVert \lVert \bX_j \rVert}$.
 To incorporate these distances while preserving symmetry ($\rho^e_{ij} = \rho^e_{ji}$), we represent the edge probabilities using the logistic link function:
 \begin{equation*}
     \text{logit}(\rho^e_{ij}) = \text{logit}(\rho^e_{base}) - \gamma^e \left(d_{ij} - \overline{d}^e\right),
    \end{equation*}
    where $\overline{d}^e = \frac{2}{n_e(n_e-1)}\sum_{i<j \in \cR_e} d_{ij}$ is the average distance between all ego pairs.
    Here, $\gamma^e \geq 0$ is a sensitivity parameter controlling the \emph{strength of homophily}.
    By centering the distances (subtracting $\overline{d}^e$), we decouple the network density from the homophily structure such that
    %\begin{itemize}
        %\item
        $\rho^e_{base}$ controls the overall network density and
        %\item 
        $\gamma^e$ controls how strongly edges concentrate among similar units. Setting $\gamma^e = 0$ eliminates the distance term, recovering the homogeneous probabilities of Example~\ref{exmp:pi_homo} ($\rho^e_{ij} = \rho^e_{base}$). Larger values of $\gamma^e$ decrease the probability of edges between dissimilar units (where $d_{ij} > \overline{d}^e$) and increase it for similar units.
 %   \end{itemize}
  
    Analogous definitions apply for alter--ego edges using a baseline $\rho^a_{base}$, a sensitivity parameter $\gamma^a$, and centered distances between alters and potential egos. 
\end{exmp}

\begin{exmp}[Heterogeneous number of missing edges]
\label{exmp:pi_hetero_number}
    Examples~\ref{exmp:pi_number} and \ref{exmp:pi_hetero} can be combined to allow 
    pair-specific heterogeneous probabilities while fixing the expected number of missing edges. Let $w_{ij}^e = \exp(-\gamma^e d_{ij}^e)$ and $w_{ij}^a=\exp(-\gamma^a d_{ij}^a)$ be the exponential of the negative measures defined in Example~\ref{exmp:pi_hetero}, for fixed values of $\gamma^e$ and $\gamma^a$.
    Larger values of $w_{ij}^e$ and $w_{ij}^a$ indicate higher similarity between the covariates of $i$ and $j$.
    Define the total sum of weights among all distinct ego--ego and alter--ego pairs with missing edges:
    \begin{equation*}
     W^e=\sum_{i,j\in \cR_e, i<j}w_{ij}^e, \qquad W^a = \sum_{i\in \cR_a}\sum_{j \in \cR_e^{e(i)}} w_{ij}^a.   
    \end{equation*}
    Similar to Example~\ref{exmp:pi_number}, researchers can specify $m^e$ and $m^a$, the approximate number of missing ego--ego and alter--ego edges, respectively, and set:
    \begin{equation*}
        \rho^e_{ij}=\frac{m^e}{W^e}\,w^e_{ij},
        \qquad
        \rho^a_{ij}=\frac{m^a}{W^a}\,w^a_{ij}.
    \end{equation*}
    This specification reflects that the expected number of missing edges is approximately $m^e$ and $m^a$, while allowing for heterogeneous edge probabilities.
    Setting all weights ($w_{ij}^e$ and $w_{ij}^a$) to one recovers Example~\ref{exmp:pi_number}.
\end{exmp}

\subsection{Running the sensitivity analysis in practice}
\label{subsec:practice}
To assess the impact of contamination on the causal estimates, 
researchers can utilize the bias-corrected estimators for the indirect effect
and for the direct effect.
The bias-corrected estimators depend on the sensitivity parameters
that define the edge-level probabilities \eqref{eq:sbm} and the parameter $\kappa$ for the direct effect.
Examples~\ref{exmp:pi_homo}-\ref{exmp:pi_hetero_number} provide practical guidelines on how to specify these parameters in practice.
When the true values of such parameters are unknown and cannot be estimated from the data, researchers can resort to sensitivity analysis by exploring a range of plausible values. We consider here two approaches: grid sensitivity analysis (GSA) and probabilistic bias analysis (PBA).
 
 In GSA, researchers specify a grid of values for the sensitivity parameters 
 and compute the bias-corrected estimates for each combination
  of parameters \citep{rosenbaum1983assessing,greenland1996basic}.
 That results in a grid of bias-corrected estimates.
 Such estimates can be coupled with the associated confidence interval given each
 value of the sensitivity parameters in the specified grid. Researchers can then summarize the sensitivity analysis results by reporting the answers for questions such as ``what is the largest number of missing alter-ego edges such that the confidence interval for $IE$ excludes the value zero?'' or ``are there combinations of $\rho^e_{ij}$ and $\kappa$ such that $\widehat{DE}_{adj}$ flips its sign compared to the naive estimator?''.

 In PBA, researchers specify a distribution for the
 sensitivity parameters and compute the distribution of bias-corrected estimates 
 over this distribution \citep{Greenland2005,Lash2014,Fox2021}. 
 In practice, this distribution is approximated using
  Monte Carlo by sampling from the specified distribution, and summarized
    through averages and intervals of the resulting bias-corrected estimates.
 This distribution of bias-corrected estimates accounts for uncertainty
  in the sensitivity parameters (bias uncertainty). Furthermore, as described below, it is possible to compute a distribution of estimates that accounts for both statistical uncertainty and bias uncertainty. 
  In our design-based framework, statistical uncertainty arises solely from the treatment assignment distribution.
  A review of PBA is given in Appendix~\ref{apdx.sec:pba}.
  
We now describe the GSA and PBA approaches in more details.

\paragraph{Grid sensitivity analysis.}
Given a grid of sensitivity parameters, the full procedure is as follows:
\begin{enumerate}
    \item \label{sa:step1} (Optional.)
    Get the predicted values for the outcome models
    $\widehat{\mu}^a_i(f)$ and $\widehat{\mu}^e_i(z)$ using the cross-fitting procedure described in Section~\ref{subsec:augmented_cf}.
     These predicted values are calculated only once and used for all sensitivity parameter values.
    \item \label{sa:step2} For each parameter value in the grid of sensitivity parameters:
        \begin{enumerate}
            \item Compute bias-corrected estimates $\widehat{IE}_{adj}$ or $\widehat{IE}^{aug}_{adj}$
                and $\widehat{DE}_{adj}$ or $\widehat{DE}^{aug}_{adj}$.
                \item Estimate the variance of each bias-corrected estimator.
                \item Compute confidence intervals for each estimator.
        \end{enumerate}    
\end{enumerate}
The results are then summarized by presenting the estimators and associated confidence intervals as a function of the sensitivity parameter(s), as we demonstrate in the data analysis (Section \ref{sec:data}).
Details on estimating variance and confidence intervals are provided in Sections~\ref{subsec:ie}-\ref{subsec:augmented_cf}.

% Details regarding the variance estimators and confidence intervals are provided in Appendix~\ref{apdx_sec:proof}.

\paragraph{Probabilistic bias analysis.}
Instead of specifying a grid of sensitivity parameters,
researchers assign a distribution for each parameter.
For instance, under the setting in Example~\ref{exmp:pi_homo}, if researchers believe that approximately 1\% of the possible ego--ego edges are missing from $\widetilde{\bA}$,
they can set $\rho^e \sim \text{Beta}(a,b)$ and choose the values of $a$ and $b$ such that the mean is $\frac{a}{a+b}=0.01$, and the variance reflects the level of uncertainty about $\rho^e$.
Let $p(\cdot)$ be the specified distribution for the sensitivity parameters.
In practice, researchers can specify a separate distribution for each parameter.

Let $B > 0$ be the number of Monte Carlo samples.
The PBA procedure that accounts for both bias and design uncertainty proceeds as follows.
\begin{enumerate}
    \item \label{pba:step1} For each $b=1,\ldots,B$:
    \begin{enumerate}
        \item Sample sensitivity parameters from the distribution $p(\cdot)$.
        \item Compute the bias-corrected estimates $\widehat{IE}_{adj}^{(b)}$ and $\widehat{DE}_{adj}^{(b)}$ under the sampled values.
        \item Draw from the (approximate) distribution of the estimator
            \begin{equation*}
                \widehat{IE}_{adj}^{(b,rand)} \sim
                 N\left(\widehat{IE}_{adj}^{(b)},
                  \widehat{V}_{\bZ}\left(\widehat{IE}_{adj}\right)\right),
            \end{equation*}             
            and similarly for the other estimators.
    \end{enumerate}
    \item Summarize the distribution of bias-corrected estimates. For example,
    by computing the average $\frac{1}{B}\sum_{b=1}^B \widehat{IE}_{adj}^{(b,rand)}$
    and empirical percentiles for intervals.
\end{enumerate}
If researchers use the augmented estimators  $\widehat{IE}_{adj}^{aug}$ and $\widehat{DE}_{adj}^{aug}$ (Section~\ref{subsec:augmented_cf}), the outcome models can be pre-computed as in GSA step \ref{sa:step1}.
Step \ref{pba:step1}c propagates the statistical uncertainty (w.r.t. $\bZ$) into the estimators by simulating a draw from the estimators' asymptotic normal distribution. 
Therefore, the distribution of estimates $\widehat{IE}_{adj}^{(b,rand)}$ 
accounts for both bias uncertainty about the sensitivity parameters and statistical uncertainty resulting from the treatments assignment \citep{Greenland2005, Fox2021}.
Further technical details and motivation of the PBA procedure are provided in Appendix~\ref{apdx.sec:pba}. 

\subsection{Estimation of sensitivity parameters with internal validation data}
\label{subsec:int_validation_data}

Until this section, the exposure probabilities $\pi_i^a$ and $\pi_i^e$ (or the basic components of latent edge probabilities $\rho_{ij}^a$ and $\rho_{ij}^e$) and $\kappa$ have been treated as unknown sensitivity parameters. However, these parameters can be directly estimated with different types of internal validation data, thus alleviating the reliance on pure sensitivity analysis. We now describe the specific data types required for estimating these parameters, separately for alters (network members) and egos.

\paragraph{Internal validation data for alters.} 
Recall that by ENRT design and the exposure mapping specification (Assumption~\ref{ass:expos_map}), alters' exposure misclassification is one-sided. That is, alters with observed exposure $\widetilde{F}_i=1$ will have a correctly classified exposure, while those with $\widetilde{F}_i=0$ can have a true exposure of $F_i=0$ or $F_i=1$. Therefore, in the study follow-up period, researchers can use a validation subset of alters from the non-treated ego-networks and ask them to recall information about the intervention.
For example, in the HPTN 037 study,  the treated egos underwent training sessions that included a number of specific phrases and terms designed to assess the diffusion of intervention messages. 
Thus, alters connected to a non-treated ego should have no recollection of these terms. In the follow-up period, alters within a validation subset were asked to recall terms associated with the intervention to assess possible contamination \citep{Simmons2015, aroke2023, Chao2023}.
This information can be used to estimate the conditional probability $\Pr(F_i=1 \mid \widetilde{F}_i=0, i\in\cR_a)$ and subsequently the exposure probabilities $\pi_i^a$ or, in the case of homogeneous probabilities (Examples~\ref{exmp:pi_homo}-\ref{exmp:pi_number}), the edge probabilities $\rho^a$. We describe this in more detail in Appendix~\ref{apdx_subsec:int_valid_data}.
In Section~\ref{sec:data}, we compare the bias-adjusted indirect effect estimates resulting from the calibration using the internal validation data of the study to the sensitivity analysis described in the preceding subsections. 

\paragraph{Internal validation data for egos.}
Similar recall data described for the alters can be used to assess the exposure probabilities of egos, in the case of homogeneous exposure probabilities. Non-treated egos should have no recollection of the terms described in the intervention.
Therefore, non-treated egos that recall intervention-related information in the follow-up period ostensibly obtained this information due to contamination from other egos.
Under homogeneous exposure probabilities, both $\pi^e$ and $\rho^e$ can be estimated using the proportion of non-treated egos in the follow-up period that recalled intervention-related information (Appendix~\ref{apdx_subsec:int_valid_data}).

Alternatively, researchers can ask the egos more direct and informed questions. For example, instead of recalling \emph{any} terms associated with the intervention, researchers can ask the egos if they learned about the intervention from other friends or sources. That will provide information regarding the true exposure status of egos in the validation subsample. 
These internal validation data can be used to estimate both $\pi_i^e$ and $\kappa$ (Appendix~\ref{apdx_subsec:int_valid_data}).
This type of information was not collected in the HPTN 037 study and can be of valuable importance for future studies.

\section{Simulation study}
\label{sec:sim}

We conducted a simulation study to evaluate the performance of the bias-corrected 
estimators $\widehat{IE}_{adj},\widehat{IE}_{adj}^{aug}, \widehat{DE}_{adj},$ and $\widehat{DE}_{adj}^{aug}$. 
We assessed the bias, variance estimation, and empirical coverage of the confidence intervals under the true values of the sensitivity parameters. 
Furthermore, we evaluated how misspecification of $\kappa$ values or violation of Assumption~\ref{ass:cov_pi_de_egos} affects the estimation of the direct effect.

The simulation study is based on a data-generating process
 that partly mimics the settings of the HPTN 037 study \citep{Latkin2009}.
We generated $n_e=200$ disjoint ego-networks with two alters each,
which was the average number of alters per ego-network in HPTN 037,
resulting in $n_a=400$.
Contamination between the ego-networks was generated
 by specifying the expected number of missing
alter--ego and ego--ego edges, as in Examples~\ref{exmp:pi_number} and \ref{exmp:pi_hetero_number},
and drawing the latent edges independently. 
Specifically, we set the expected number of missing alter--ego edges to $m^a \in \{100,200\}$,
and ego--ego edges to $m^e \in \{150,250\}$.
Under each setup, a single network was generated and was fixed throughout the simulation replications.
The potential outcomes were generated once according to the models
    \begin{equation*}
        \begin{aligned}
            Y_i(z,f) &= -0.5 + 2 z + 0.5 f + zf + \bX_i^T \bbeta_e + \varepsilon_i,   & \text{for} \; i \in \cR_e,
            \\ 
            Y_i(0,f) &= -0.5 + 2 f + \bX_i^T \bbeta_a +
            \varepsilon_i, \;  &   \text{for} \; i \in \cR_a,        
        \end{aligned}
    \end{equation*}
where $\varepsilon_i \sim N(0,1)$ are independent noise terms.
This data-generating process yields true effects of $DE =IE=2$, and ratio of direct effects among exposed versus non-exposed egos of $\kappa = 1.5$ as defined in \eqref{eq:kappa}.
The generation process of the three covariates $\bX_i$ and the specification of $\bbeta_e$ and $\bbeta_a$
are provided in Appendix~\ref{apdx.sec:sim}.
Treatments were assigned according to a Bernoulli design with $p_z=0.5$.
We performed $5 \times 10^3$ simulation replications by random draws of treatment assignments.

We report here the results when contamination was generated using  
heterogeneous edge probabilities as in Example~\ref{exmp:pi_hetero_number}
with Euclidean distance and $\gamma^e=\gamma^a=1$.
We compare the bias-corrected estimators under the specification that uses the correct edge probabilities (``Heterogeneous''), under the specification that uses a misspecified model with the same edge probabilities as in Example~\ref{exmp:pi_number} (``Homogeneous'') and the uncorrected estimators $\widehat{IE}$ and $\widehat{DE}$ (``Naive''). We considered both the HT estimators and the augmented estimators using an outcome model estimated via linear regression with two-fold cross-fitting (as described in Section~\ref{subsec:augmented_cf}).

Table~\ref{tab:combined_results_main} summarizes the results for the indirect and direct effect estimation. As the level of contamination ($m^a$ and $m^e$) increases, the naive estimators exhibit substantial bias and poor coverage.
In contrast, all bias-corrected estimators
exhibit negligible bias across all scenarios,
illustrating that even a misspecified model for the edge probabilities
(homogeneous specification) can effectively correct for the bias.
For the non-augmented estimators, the variance estimates were conservative, particularly for the direct effect in high contamination settings (SD/SE $= 0.574$). This is comparable to the conservativeness of the variance estimators in other papers in the network interference literature \citep[e.g.,][]{aronow_estimating_2017, Weinstein2026}.
The augmented indirect effect estimators, on the other hand, yielded valid variance estimates and achieved nominal coverage rates. 
The augmented direct effect estimators tend to slightly overestimate the variance (SD/SE $< 1$), which is attributed to 
their conservative variance estimator (Section~\ref{subsec:de}).
Results for homogeneous contamination settings are provided in Appendix~\ref{apdx.sec:sim} and exhibit similar patterns.

We also evaluated the impacts of misspecifying the value of $\kappa$ or violating Assumption~\ref{ass:cov_pi_de_egos}.
We found that violations of Assumption~\ref{ass:cov_pi_de_egos}, induced by heterogeneous treatment-exposure interactions, resulted in negligible bias for the corrected direct effect estimators, albeit with slight variance underestimation.
The impact of $\kappa$ misspecification depended on the direction of the misspecification compared to the true value.
If $\kappa$ is misspecified in the wrong direction (e.g., specifying $\kappa < 1$ when $\kappa > 1$), the bias-corrected estimator can exhibit severe bias and lower coverage than the naive estimator. Conversely, when $\kappa$ is misspecified in magnitude but in the correct direction, the bias-corrected estimator outperforms the naive estimator.
Moreover, the performance of the proposed bias-corrected estimators under binary potential outcomes was also assessed, yielding results consistent with those under the continuous potential outcome model.
See Appendix~\ref{apdx.sec:sim} for additional details and results.

{\renewcommand{\arraystretch}{0.4} % Shrink row height to 60%
\begin{table}[!hbtp]
    \footnotesize
\caption{Simulation results for indirect and direct effects under heterogeneous contamination.
Bias is the empirical bias, Coverage is the empirical coverage of the 95\% confidence intervals,
and SD/SE is the ratio of the standard deviation of the estimates to the mean standard error estimates.
Specification ``Heterogeneous" uses the correct edge probabilities,
``Homogeneous" uses the same probability for all edges,
and ``Naive" is the uncorrected estimator.
Augmented indicates whether the estimators augmented with an outcome model are used.
True effects are $IE=2$ and $DE=2$.
}
\centering
\begin{tabular}[!hbtp]{clccccc}
\toprule
Estimand & Scenario & Specification & Augmented & Bias & Coverage & SD/SE\\
\midrule
% IE Block
\multirow{12}{*}[-8ex]{$IE$} & & & FALSE & -0.006 & 0.963 & 0.948\\
\cmidrule{4-7}
 & & \multirow{-2}{*}{\centering\arraybackslash Heterogeneous} & TRUE & -0.007 & 0.948 & 0.988\\
\cmidrule{3-7}
 & & & FALSE & -0.004 & 0.963 & 0.949\\
\cmidrule{4-7}
 & & \multirow{-2}{*}{\centering\arraybackslash Homogeneous} & TRUE & -0.005 & 0.949 & 0.987\\
\cmidrule{3-7}
 & & & FALSE & -0.239 & 0.507 & 0.949\\
\cmidrule{4-7}
 & \multirow{-6}{*}[1\dimexpr\aboverulesep+\belowrulesep+\cmidrulewidth]{\raggedright\arraybackslash $m^a=100$} & \multirow{-2}{*}{\centering\arraybackslash Naive} & TRUE & -0.240 & 0.403 & 0.987\\
\cmidrule{2-7}
 & & & FALSE & 0.002 & 0.970 & 0.902\\
\cmidrule{4-7}
 & & \multirow{-2}{*}{\centering\arraybackslash Heterogeneous} & TRUE & 0.000 & 0.955 & 0.988\\
\cmidrule{3-7}
 & & & FALSE & 0.011 & 0.971 & 0.904\\
\cmidrule{4-7}
 & & \multirow{-2}{*}{\centering\arraybackslash Homogeneous} & TRUE & 0.009 & 0.954 & 0.987\\
\cmidrule{3-7}
 & & & FALSE & -0.434 & 0.074 & 0.904\\
\cmidrule{4-7}
 & \multirow{-6}{*}[1\dimexpr\aboverulesep+\belowrulesep+\cmidrulewidth]{\raggedright\arraybackslash $m^a=200$} & \multirow{-2}{*}{\centering\arraybackslash Naive} & TRUE & -0.435 & 0.026 & 0.987\\
\midrule
% DE Block
\multirow{12}{*}[-8ex]{$DE$} & & & FALSE &  0.000 & 0.986 & 0.780\\
\cmidrule{4-7}
 & & \multirow{-2}{*}{\centering\arraybackslash Heterogeneous} & TRUE & -0.005 & 0.960 & 0.952\\
\cmidrule{3-7}
 & & & FALSE & -0.011 & 0.984 & 0.802\\
\cmidrule{4-7}
 & & \multirow{-2}{*}{\centering\arraybackslash Homogeneous} & TRUE & -0.016 & 0.956 & 0.960\\
\cmidrule{3-7}
 & & & FALSE & 0.514 & 0.279 & 1.068\\
\cmidrule{4-7}
 & \multirow{-6}{*}[1\dimexpr\aboverulesep+\belowrulesep+\cmidrulewidth]{\raggedright\arraybackslash $m^e=150, \kappa=1.5$} & \multirow{-2}{*}{\centering\arraybackslash Naive} & TRUE & 0.508 & 0.175 & 1.045\\
\cmidrule{2-7}
 & & & FALSE & 0.000 & 0.999 & 0.574\\
\cmidrule{4-7}
 & & \multirow{-2}{*}{\centering\arraybackslash Heterogeneous} & TRUE & -0.005 & 0.981 & 0.829\\
\cmidrule{3-7}
 & & & FALSE & -0.019 & 0.998 & 0.596\\
\cmidrule{4-7}
 & & \multirow{-2}{*}{\centering\arraybackslash Homogeneous} & TRUE & -0.024 & 0.976 & 0.840\\
\cmidrule{3-7}
 & & & FALSE & 0.689 & 0.102 & 1.064\\
\cmidrule{4-7}
 & \multirow{-6}{*}[1\dimexpr\aboverulesep+\belowrulesep+\cmidrulewidth]{\raggedright\arraybackslash $m^e=250, \kappa = 1.5$} & \multirow{-2}{*}{\centering\arraybackslash Naive} & TRUE & 0.682 & 0.023 & 1.029\\
\bottomrule
\label{tab:combined_results_main}
\end{tabular}
\end{table}
}

\section{Data analysis}
\label{sec:data}

HPTN 037 was an ENRT that evaluated the efficacy of a peer education intervention
in reducing HIV risk behaviors among people who inject drugs and their drug and sexual networks \citep{Latkin2009}.
Egos were randomized ($p_z =0.5$) to receive a peer education training
and were encouraged to share the HIV risk reduction information with their drug and sexual network members (alters).
Egos in the control arm received standard HIV prevention counseling.
The study was conducted between 2002 and 2006 in Philadelphia, Pennsylvania, US and Chiang Mai, Thailand.
The study recruited $414$ egos (index participants) who 
were asked to recruit at least one drug or sex network member (alter) into the study.
Researchers enforced that alters were not also egos, and that each alter was connected to only one ego in the observed network
\citep{Latkin2009}.
Following previous analyses \citep{Buchanan2018,Buchanan_2024,Chao2023}, we limit our analysis to the Pennsylvania site 
and to a 12-month follow-up,
yielding a sample of $n_e=150$ egos and $n_a=263$ alters,
 with an average of $2$ alters per ego-network.
The primary outcome is a binary indicator of any injection-related risk behavior in the past month,
 including sharing injection equipment, front and back loading, and injecting with people not well known or
 in a public space.

Contamination across ego-networks is a concern in ENRT.
That is, alters might inject drugs or have sexual relations with egos from other ego-networks,
 or egos might be connected to each other. 
 Using recall data collected after the trial,
  \citet{Simmons2015} found evidence of contamination.
  \citet{Chao2023} developed a bias-correction method to adjust for (homogeneous) exposure misclassification,
  only for the indirect effect,
 building on the recall data and the assumption of non-overlapping ego-networks.
 That is, as in most ENRT studies, the analysis in \citet{Chao2023} assumed that, in the population network, alters can only be connected to one ego,
 and egos can not be connected to each other.

 We relax this assumption and apply our proposed sensitivity analysis methods to assess the potential impact of contamination on both indirect and direct effect estimates.
  For both alter--ego and ego--ego edges, we specify the expected number of missing edges as in Examples~\ref{exmp:pi_number} and \ref{exmp:pi_hetero_number}.
   Covariates used in the analysis include risk behaviors at baseline, age, gender, and race.
   In the outcome models, we also included the average neighbors' covariates for each unit, as was also done by \citet{Buchanan_2024}.
   Quantitative covariates were scaled to have a mean of zero and a variance of one.
 For the specification of heterogeneous edge probabilities using the approximate expected number of missing edges (Example~\ref{exmp:pi_hetero_number}),
 we use the Euclidean distance and set $\gamma^e=\gamma^a=1$. 
 Results with $\gamma^e=\gamma^a=2$ are given in Appendix~\ref{apdx.sec:data} and were consistent with the results shown here, albeit with higher variance estimates.
Note that here $\gamma^e$ and $\gamma^a$ are hyperparameters and not the sensitivity parameters, which are $\kappa$, $m^a$, and $m^e$.
 For the direct effect, we specify values larger than $1$ to the sensitivity parameter $\kappa$, addressing the concern that the direct effect is larger, on average, among egos exposed to at least one other treated ego.
We apply both the GSA and PBA described in Section~\ref{subsec:practice}. Variance estimates and 95\% confidence intervals were calculated as described in Sections~\ref{subsec:ie}-\ref{subsec:augmented_cf}.

Furthermore, we compare the results of the sensitivity analysis with those obtained using point estimates of parameters from internal validation data, as described in Section~\ref{subsec:int_validation_data}. 
During the 6-month follow-up of the study, participants were asked to recall terms introduced in the intervention training program and unlikely to be known outside of it. Following previous analyses \citep{aroke2023, Chao2023}, we assume that a participant was indirectly exposed to the intervention if she recalled at least one of the five terms associated with the intervention. 
For greater accuracy, we limit the validation subset only to participants who recalled the positive term and none of the negative terms \citep{aroke2023, Chao2023}.
As described in Section~\ref{subsec:int_validation_data}, we estimate $\pi^a$ from non-exposed alters ($\widetilde{F}_i=0, i \in \cR_a$) and $\pi^e$ from non-treated egos ($Z_i=0, i \in \cR_e$), both under homogeneous exposure probabilities. 
Among $22$ alters and $15$ egos in the validation subset, only $4$ alters and $3$ egos recalled the terms, mapping to an estimated exposure probabilities of $\widehat{\pi}^a =0.59$ and $\widehat{\pi}^e=0.2$. From Example~\ref{exmp:pi_number} this is equivalent to approximately $\widehat{m}^a=105.5$ 
missing ego-alter edges and $\widehat{m}^e=33.4$ missing ego-ego edges.

\subsection{Grid sensitivity analysis}
\label{subsec:hptn_sa}
For the indirect effect, we consider a grid of the approximate expected number of missing edges with
$m^a \in [10, 500]$ and $m^e \in [10, 150]$, representing the concern that, in the homogeneous case, 
each alter and ego are approximately connected to at most two additional egos (Example~\ref{exmp:pi_number}).
In addition, we consider $\kappa \in [1, 2]$ reflecting the suspicion that the direct effect
among egos connected to at least one treated ego is up to two times larger, on average, than the 
effect among egos that are not connected to other treated egos.
We report here the results for the augmented estimators $\widehat{IE}_{adj}^{aug}$ and $\widehat{DE}_{adj}^{aug}$ using two-fold cross-fitting (Section~\ref{subsec:augmented_cf}) and logistic regression for the outcome models.
We consider both homogeneous and heterogeneous edge probability specifications.
Results for the non-augmented estimators $\widehat{IE}_{adj}$ and $\widehat{DE}_{adj}$ are provided in Appendix~\ref{apdx.sec:data}, and were consistent with the results shown here.

\begin{figure}[!hbtp]
    \centering
    \includegraphics[width=0.55\linewidth]{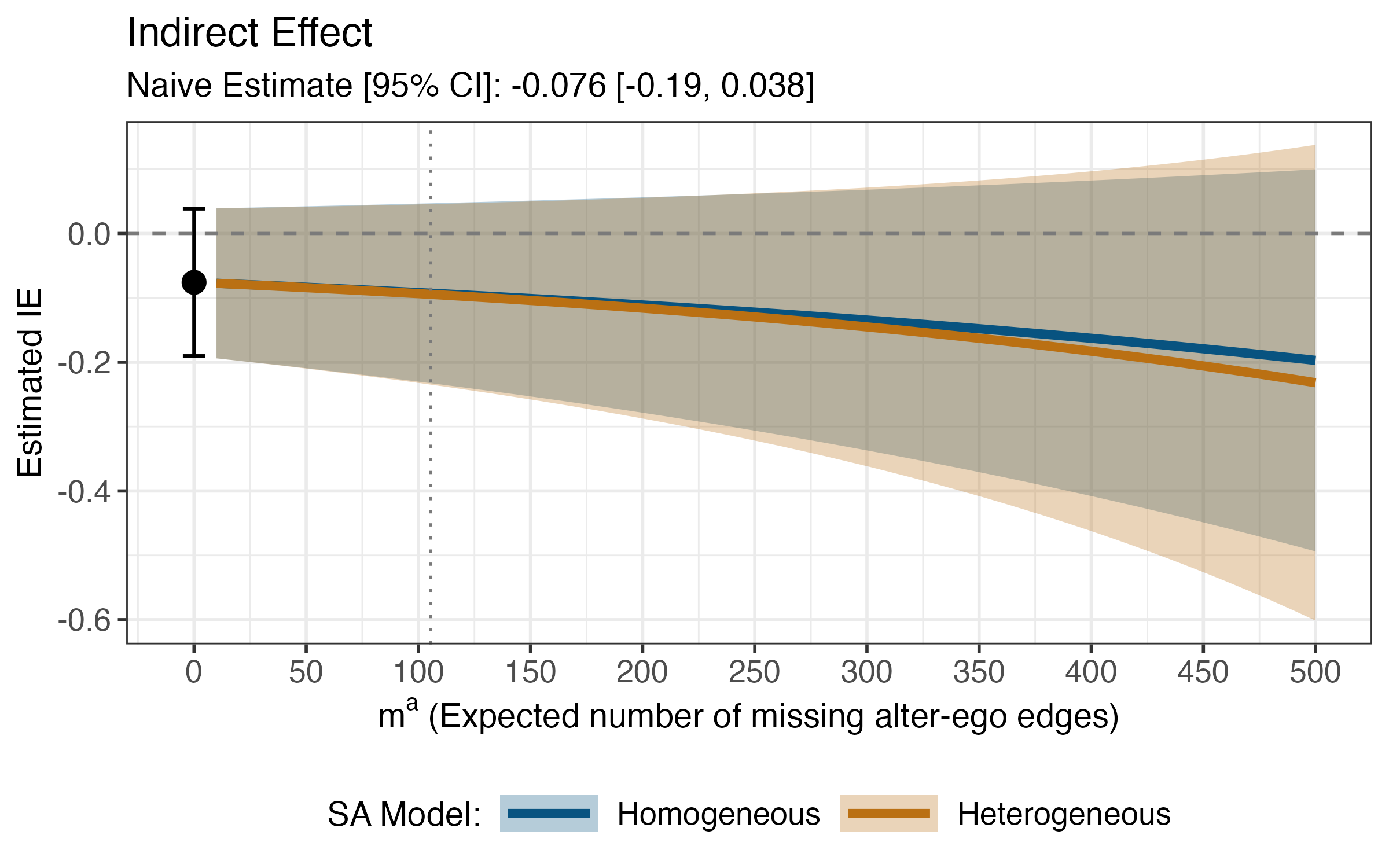}
    \caption{Grid sensitivity analysis results for indirect effect
    using the augmented estimator $\widehat{IE}_{adj}^{aug}$.
    The x-axis represents the approximate total number of missing alter--ego edges $m^a$,
    while the y-axis represents the estimates.
    The dashed vertical line at $105.5$ is the estimate of $m^a$ from the internal validation data.
    Results are shown for both homogeneous (blue) and heterogeneous (orange)
     edge probabilities specifications.}
    \label{fig:hptn_sa_ie_augmented}
\end{figure}

\begin{figure}[!hbtp]
    \centering
    \includegraphics[width=0.7\linewidth]{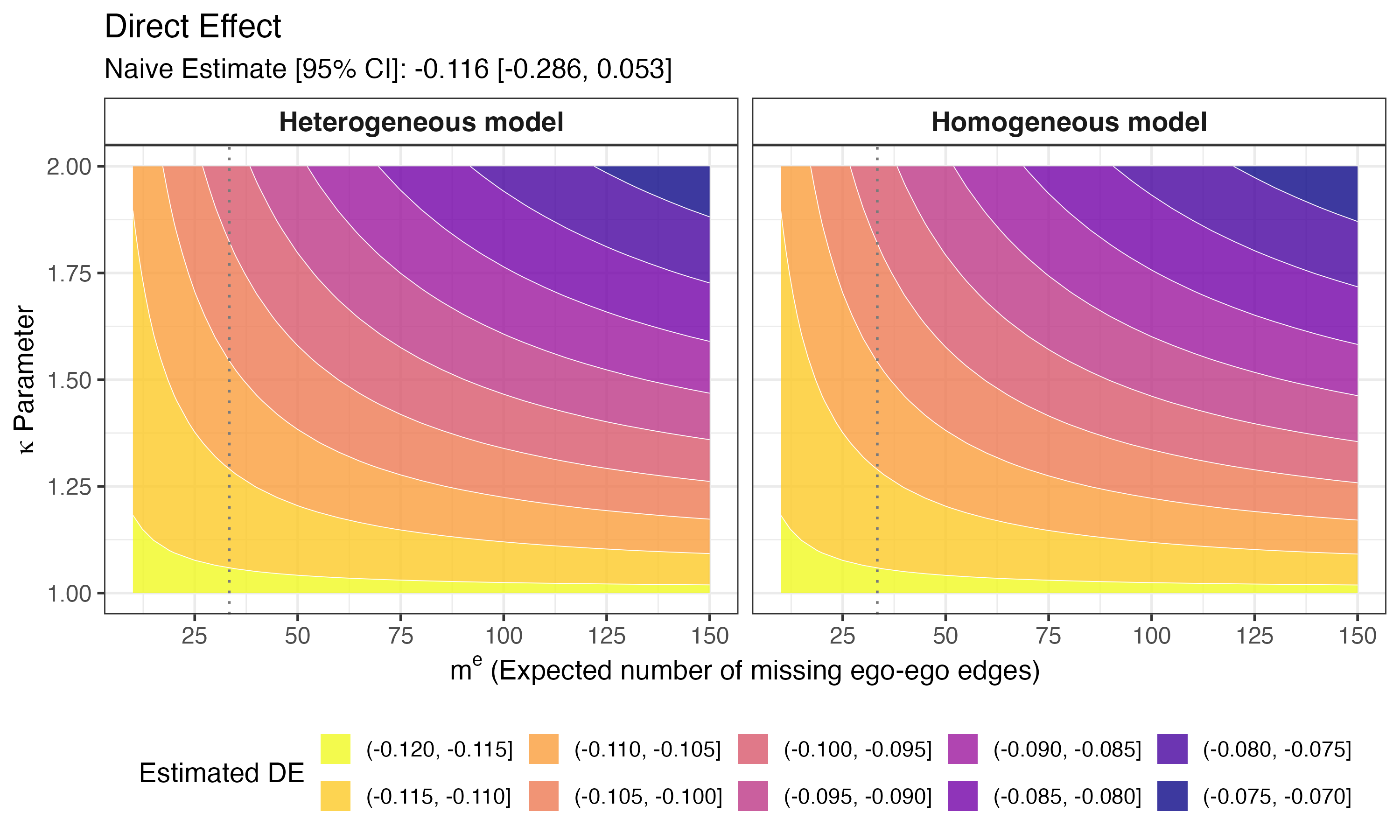}
    \caption{Grid sensitivity analysis results for direct effect
    using the augmented estimator $\widehat{DE}_{adj}^{aug}$.
    The x-axis represents the approximate total number of missing ego--ego edges $m^e$,
    while the y-axis represents $\kappa$, the interaction sensitivity parameter.
    The dashed vertical line at $33.4$ is the estimate of $m^e$ from the internal validation data.
    Contour lines and color shading represent the estimated values.
    Results are shown for both heterogeneous (left) and homogeneous (right) edge probability specifications.}
    \label{fig:hptn_sa_de_augmented}
\end{figure}

Figures~\ref{fig:hptn_sa_ie_augmented} and \ref{fig:hptn_sa_de_augmented}
 summarize the GSA results for indirect and direct effects, respectively.
 For the indirect effect, both homogeneous and heterogeneous specifications
  yield similar results.
The results suggest that, for instance, if each alter is connected to up to one additional ego on average ($m^a=263$) in the homogeneous case, the bias-corrected estimate is around $-0.194$ compared to the naive estimate of $-0.076$, and a bias-corrected estimate of around $-0.093$ using the estimate $\widehat{m}^a=105.5$ from the internal validation data. Taking the estimate $\widehat{m}^a=105.5$ as a lower bound for the true $m^a$, implies that the naive estimator underestimate $IE$ by an order of at least $18\%$, indicating a substantially larger indirect effect than previously reported.

For the direct effect, the bias-corrected estimates under the homogeneous and heterogeneous specifications yielded similar estimates that are closer to zero than the naive estimate of $-0.116$ across the entire grid of sensitivity parameters.
For example, if $\kappa=1.5$ and each ego is connected to up to one additional ego on average ($m^e=75$), the bias-corrected estimate is around $-0.097$ in the homogeneous case, compared to the bias-corrected estimate of around $-0.105$ using the estimate $\widehat{m}^e=33.4$ from the internal validation data.
These results imply that the naive estimates of $DE$ might be overestimated by over $10.5\%$, indicating a smaller direct effect than previously found.
% These results suggest that assuming no contamination may lead to overestimation of the magnitude of the direct effect.
 % However, both indirect and direct bias-corrected estimates remain statistically insignificant at the $5\%$ level across the entire grid of sensitivity parameters for both homogeneous and heterogeneous specifications. 

\subsection{Probabilistic bias analysis}
\label{subsec:hptn_pba}
For the PBA, we specify distributions for the sensitivity parameters,
representing different degrees of uncertainty about their values.
For $m^a$ and $m^e$, we compare three uncertainty distributions. 
The first is a discrete uniform distribution over the ranges used in the GSA:
$m^a \sim \text{Uniform}\{1,500\}$ and $m^e \sim \text{Uniform}\{1,150\}$,
representing maximum entropy, but with a capped right tail.
The second is a Poisson distribution with a mean equal to half of the maximum value in the GSA grid:
$m^a \sim \text{Poisson}(250)$ and $m^e \sim \text{Poisson}(75)$,
representing moderate uncertainty with a thin but unbounded tail.
The third is a Negative Binomial (NB) distribution with a mean equal to half the maximum value in the GSA grid
 and a size parameter equal to $10$, 
% $m^a \sim \text{NB}(10,1/21)$ and $m^e \sim \text{NB}(10,1/16)$,
representing a heavier tail than the Poisson distribution.
For the sensitivity parameter $\kappa$, we consider two uncertainty distributions. The first is a
continuous uniform distribution,
$\kappa \sim \text{Uniform}[1,2]$, and the second is 
 a log-normal distribution with expected value $1.5$ and standard deviation equal to $0.2$ on the log scale,
resulting in a heavier right tail than the uniform distribution.
We run the PBA procedure with $B=10^4$ Monte Carlo samples and account for both bias and statistical uncertainty, as described in Section~\ref{subsec:practice}.

\begin{figure}[!hbtp]
    \centering
    \begin{minipage}{0.48\linewidth}
        \centering
        \includegraphics[width=\linewidth]{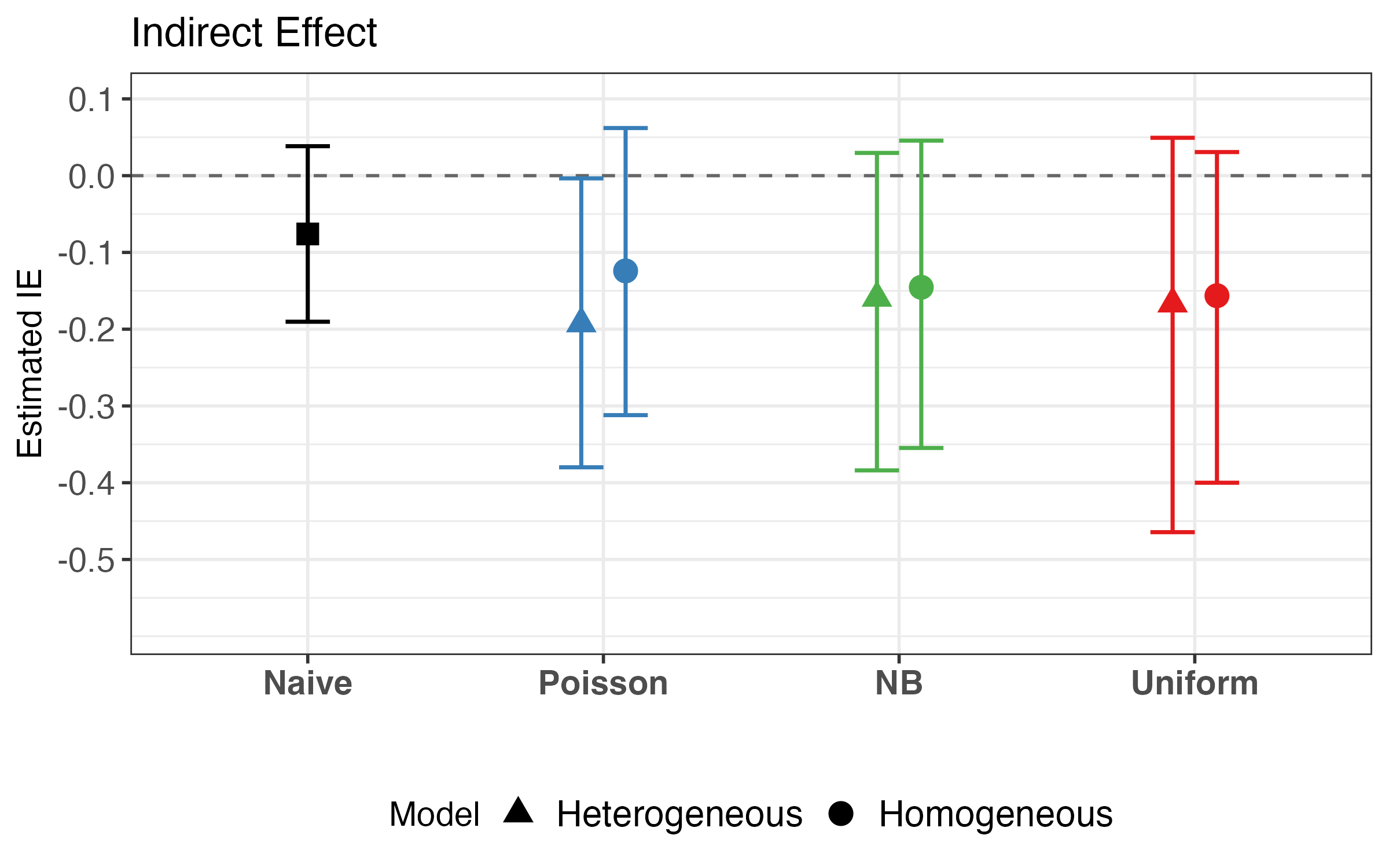}
    
    \end{minipage}
    % \hfill
    \begin{minipage}{0.48\linewidth}
        \centering
        \includegraphics[width=\linewidth]{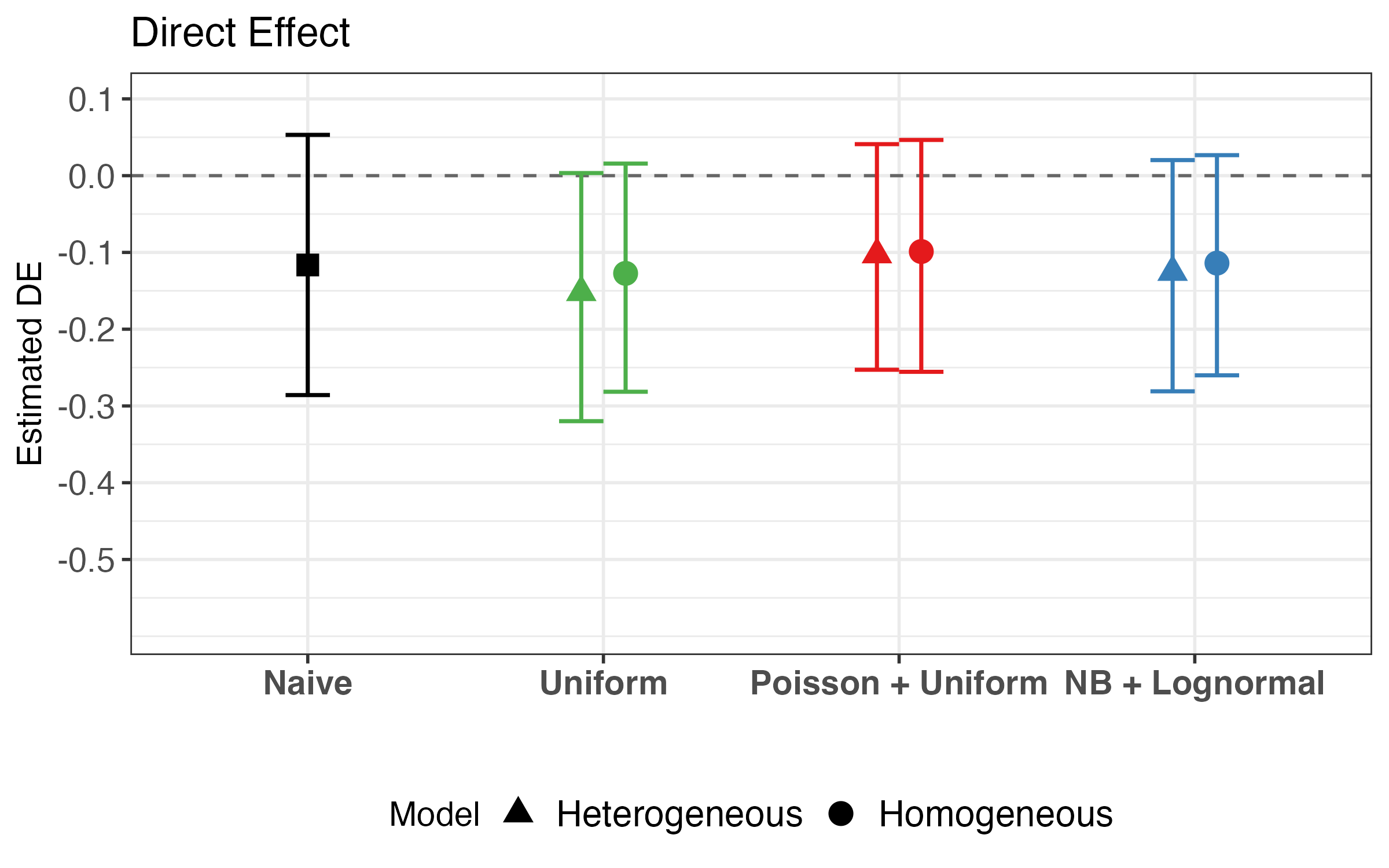}
    \end{minipage}
    \caption{Probabilistic bias analysis results for indirect effect
    (left) using the augmented estimator $\widehat{IE}_{adj}^{aug}$,
    and direct effect (right) using the augmented estimator $\widehat{DE}_{adj}^{aug}$.
    Results are shown as mean and $95\%$ intervals for both homogeneous and heterogeneous edge probability specifications, accounting for both bias and statistical uncertainty.
    Colors represent different specified distributions for the total number of missing edges and the sensitivity parameter $\kappa$.}
    \label{fig:hptn_pba_augmented}
\end{figure}

% Figures~\ref{fig:hptn_pba_ie_not_augmented} and \ref{fig:hptn_pba_de_not_augmented}
Figure~\ref{fig:hptn_pba_augmented}
 summarizes the PBA results for the indirect and direct effects
 with the augmented estimators.
 For the indirect effect, the three distributions for $m^a$ yield similar average estimates, with a slightly larger magnitude for the heterogeneous specification.
 The uniform distribution results in the widest intervals, while the Poisson distribution yields the narrowest intervals.
All bias-corrected estimates are lower (average estimates around $-0.158$) than the naive estimate of $-0.076$, suggesting that contamination leads to severe underestimation of the magnitude of the indirect effect.
For the direct effect, all specified distributions for $m^e$ and $\kappa$ yield similar average estimates, with the Poisson and uniform combination resulting in estimates closer to zero than the rest. The uniform distribution yields the widest intervals. The averages of the bias-corrected estimates are closer to zero (average estimates around $-0.12$ in the heterogeneous specification) than the naive estimate of $-0.116$, suggesting that contamination leads to slight
overestimation of the direct effect.
Nevertheless, apart from the heterogeneous specification with the Poisson distribution for indirect effect, the indirect and direct $95\%$ PBA intervals include zero in all distribution and edge probabilities specifications.

% \begin{figure}[H]
%     \centering
%     \includegraphics[width=0.5\linewidth]{figures/hptn/pba_ie_not_augmented.png}
%     \caption{Probabilistic bias analysis results for indirect effects
%     using the estimator \eqref{eq:IE_corrected}.
%     Results are shown for both homogeneous and heterogeneous edge probabilities specifications (shapes).
%     Different colors represent different prior distributions for the expected number of missing alter--ego edges $m^a$.
%     }
%     \label{fig:hptn_pba_ie_not_augmented}
% \end{figure}

% \begin{figure}[H]
%     \centering
%     \includegraphics[width=0.5\linewidth]{figures/hptn/pba_de_not_augmented.png}
%     \caption{
% Probabilistic bias analysis results for direct effects
%     using the estimator \eqref{eq:DE_corrected}.
%     Results are shown for both homogeneous and heterogeneous edge probabilities specifications (shapes).
%     Different colors represent different prior distributions for the
%      expected number of missing ego--ego edges $m^e$ and the interaction sensitivity parameter $\kappa$.
%     }
%     \label{fig:hptn_pba_de_not_augmented}
% \end{figure}

\section{Discussion}
\label{sec:discussion}
In this paper, we developed sensitivity analysis methods to assess the impact of contamination between ego-networks on the causal estimates in egocentric network randomized trials. Such trials are increasingly used to evaluate interventions in hard-to-reach populations, where full network information is often unavailable or difficult to obtain. Our methods build on a general formulation of contamination as exposure misclassification due to missing network edges, and provide bias-corrected estimators for both indirect and direct effects.
Our bias-corrected estimators rely on two types of sensitivity parameters. The first 
characterizes the magnitude of contamination.
The second, specific to the direct effect, represents the ratio of the average direct effect among egos exposed to at least one treated neighbor to those not exposed.
We provided practical guidelines for specifying these sensitivity parameters
and described two sensitivity analysis approaches: GSA and PBA.

Our re-analysis of the HPTN 037 trial underscores the consequences of unmeasured contamination in ENRTs.
Our findings demonstrate that standard estimators can severely underestimate the indirect effect while slightly overestimating the direct effect. For policymakers, this highlights that peer-driven interventions may be substantially more effective than naive estimates suggest, altering the cost-benefit calculus for deploying network-based public health strategies.

A limitation of our design-based framework is the focus on sample-level causal estimands \eqref{eq:indirect_effect} and \eqref{eq:directed_effect}, which restrict our inference to the finite sample of recruited egos and alters rather than the broader population. Extensions to population-level estimands require specifying the exact mechanism of the egocentric network sampling, such as the inclusion probabilities of the egos. However, many ENRTs, including the HPTN 037 study, involve hard-to-reach populations where egos are typically recruited via convenience sampling rather than a strict probability sample. While recent methodological work has considered population-level estimands in ENRTs, doing so generally relies on the assumption of simple random sampling of units from the population \citep{Fang2023}. Consequently, our focus on sample-level estimands reflects a methodological trade-off: we prioritize robust internal validity and assumption-lean causal inference within the sample, without imposing strong unverifiable conditions on the recruitment process.

A natural alternative to our sensitivity framework might be to explicitly infer the missing alter-ego and ego-ego edges by assuming a stochastic generative model for the population network $\bA$ and estimating its parameters from the observed network $\widetilde{\bA}$. However, this approach presents significant challenges. First, it requires both positing a valid population-level network model and specifying the egocentric sampling mechanism.
Second, even if these mechanisms could be specified, a fundamental challenge remains.  Fitting a model to a network of $n$ nodes does not necessarily recover the parameters governing the population network of $N$ nodes. Assuming that edge formation probabilities scale directly from a sample to the population requires the network model to satisfy properties such as consistency under sampling \citep{Shalizi2013, Crane2021}. 
Our design-based framework coupled with sensitivity analysis offers a reliable alternative. By treating the missing edges as fixed but uncertain, we can quantify the impact of contamination while bypassing the need to posit assumptions on the data-generating and sampling processes.

We assumed a two-level exposure mapping specification, indicating 
whether a unit is connected to at least one treated neighbor (Assumption~\ref{ass:expos_map}).
This specification is sensible when the treatment effect is saturated at one treated neighbor (regardless of who it is).
That is, when exposure to additional treated neighbors does not further affect the outcome.
In Appendix~\ref{apdx.sec:three_level_em}, we extend our methods to a three-level exposure mapping specification, distinguishing between no treated neighbors, one treated neighbor, and two-or-more treated neighbors.
In this case, for example, an alter with an observed exposure level of one treated neighbor
may have a true exposure level of one or two-or-more treated neighbors.
We limit our analysis for the indirect effect $IE$,
and derive bias expressions for the naive estimator $\widehat{IE}$.
We also provide bias-corrected estimators that depend on an additional sensitivity parameter that captures the marginal unit-level exposure effect. 
We show that the two-level exposure mapping model is a special case of the three-level model with saturation of effect after the first exposure. 

The exposure mapping specification we took in this paper relies on a first-order neighborhood interference assumption, restricting the diffusion of treatment effects to direct neighbors. 
A valuable direction for future research is relaxing this specification to accommodate higher-order interference. For example, under second-order interference, an alter could be indirectly affected by the treatment through a connection to another alter in a treated ego-network. In such settings, an ego might also have indirect exposure if its recruited alters are connected to other treated egos. Adapting our proposed sensitivity analysis framework to these scenarios would require specifying sensitivity parameters for latent alter-alter edges across different ego-networks. 
While conceptually natural, this extension would introduce notable combinatorial complexity to the computation of the exposure probabilities.

ENRT is an attractive design 
that allows for the estimation of spillovers and the indirect effect 
without the requirement for sociocentric network measurements. 
It enables researchers to estimate causal effects under interference 
with limited network data.
However, extending these designs to observational studies 
where units are sampled from a larger network (e.g., through egocentric sampling),
but treatments are not randomly assigned by the researchers, is a more challenging problem.
The primary concern is that non-recruited units can be treated as well, leading to treatment interference between
recruited and non-recruited units. Since information on the edges connecting 
such units in the population is rarely available, addressing interference in observational studies with egocentric sampling remains a significant challenge. 

% More generally, network sampling introduces fundamental challenges under interference, complicating the interpretation and identification of population-level estimands and inducing complex missing data structures. Further rigorous examination of the implications of network sampling is a desirable frontier for causal inference under interference. 

\section*{Funding}
This work was supported by the Israel Science Foundation (ISF grant No. 2300/25); BW is supported by the Data Science Fellowship granted by the Israeli Council for Higher Education. 

\section*{Acknowledgements}
We thank Dr. Ashley Buchanan for sharing code and assisting in the analysis of the HPTN 037 study. We also thank two anonymous reviewers for helpful comments and suggestions that improved the paper.

\section*{Conflict of interest}
None declared.

\section*{Data Availability}
Analysis code applied to a simulated dataset similar to the design of HPTN 037 can be found at \url{https://github.com/barwein/ENRT_SA}.
The HPTN 037 study datasets are publicly available and can be requested from the Statistical Center for HIV/AIDS Research and Prevention through \url{https://atlas.scharp.org/project/home/begin.view}.

\bibliographystyle{apalike}
\bibliography{all_biblo}

\newpage

\appendix

\renewcommand{\theequation}{\thesection.\arabic{equation}}
\numberwithin{equation}{section}

\renewcommand{\theassumption}{\thesection.\arabic{assumption}}
\numberwithin{assumption}{section}

% \renewcommand{\theproposition}{\thesection.\arabic{proposition}}
% \numberwithin{proposition}{section}

\renewcommand\thefigure{\thesection.\arabic{figure}}    
\counterwithin{figure}{section}

\renewcommand\thetable{\thesection.\arabic{table}}    
\counterwithin{table}{section}

% 1. Start recording contents for a group named "appendix"
\startcontents[appendix]

% 2. Print the contents
% Syntax: \printcontents[name]{prefix}{start-level}{code-before}
\printcontents[appendix]{l}{1}{\section*{Appendix -- Table of Contents}}

\section{Proofs and technical details}
\label{apdx_sec:proof}

\subsection{Proof of Proposition~1}
\label{apdx_subsec:proof_prop1}

Recall that the HT estimator of the indirect effect based on the observed data \eqref{eq:IE_naive_HT} is
\begin{equation*}
    \widehat{IE} = 
    \frac{1}{n_a}
    \sum_{i \in \cR_a}
    \left[
    \frac{
    \mathbb{I}\{\widetilde{F}_i=1\}Y_i}{p_z} 
    -
    \frac{
    \mathbb{I}\{\widetilde{F}_i=0\}Y_i}{1 - p_z} 
    \right]
\end{equation*}
By the assumed experimental design (Assumption~1), the probability that an alter is assigned to treatment is zero,
 that is, $\Pr(Z_i =1 \mid i\in \cR_a)=0$. In addition, by design $Z_i$ and $\widetilde{F}_i = F(\bZ_{-i},\widetilde{\bA})$ are independent. Thus,  for an alter $i \in \cR_a$, we have $\mathbb{I}\{\widetilde{F}_i=f\}=\mathbb{I}\{Z_i=0,\widetilde{F}_i=f\}$ with probability one.
Starting with the first inner term, for $i \in \cR_a$,
\begin{equation}
    \label{apdx.eq:alter_0_1}
    \begin{aligned}        
    \EX_{\bZ}&\left[\frac{
    \mathbb{I}\{\widetilde{F}_i=1\}Y_i}{p_z} \right]\\[1ex]
    &\overset{(i)}{=}
     \frac{1}{p_z}
     \EX_{\bZ}
     \left[
    \mathbb{I}\{F_i=1,\widetilde{F}_i=1\}Y_i(0,1) 
    +
    \mathbb{I}\{F_i=0,\widetilde{F}_i=1\}Y_i(0,0) 
    \right]
    \\[1ex] &\overset{(ii)}{=}
    \frac{\Pr(F_i=1, \widetilde{F}_i=1 \mid i \in \cR_a)}{p_z}Y_i(0,1)
     + 
    \frac{\Pr(F_i=0, \widetilde{F}_i=1 \mid i \in \cR_a)}{p_z}Y_i(0,0)
    \\[1ex] &\overset{(iii)}{=}
    \Pr(F_i = 1 \mid \widetilde{F}_i=1, i\in \cR_a) Y_i(0,1) + 
    \Pr(F_i = 0 \mid \widetilde{F}_i=1, i\in \cR_a) Y_i(0,0)
    \\[1ex] &\overset{(iv)}{=}
    Y_i(0,1),
    \end{aligned}
\end{equation}
where $(i)$ follows since $Z_i=0$ with probability one for alters 
 and consistency (Assumption~3) that connects the observed outcomes $Y_i$ to the potential outcomes $Y_i(z,f)$; $(ii)$ from taking expectation with respect to the design $\bZ$ (recalling that the distribution of $\bZ$ in this paper is conditional on the sample); $(iii)$ simply uses that $p_z$ is equal to $\Pr(\widetilde{F}_i=1\mid i \in \cR_a)$ and $\frac{\Pr(F_i=f, \widetilde{F}_i=1 \mid i \in \cR_a)}{\Pr(\widetilde{F}_i=1\mid i \in \cR_a)} = \Pr(F_i = f \mid \widetilde{F}_i=1, i\in \cR_a)$; and $(iv)$ follows since $\Pr(F_i=1 \mid \widetilde{F}_i=1, i \in \cR_a) = 1$. To see this, note that $\sum_{j\neq i}Z_j \widetilde{A}_{ij} \leq \sum_{j\neq i}Z_j A_{ij} $ with probability one, which implies that $\widetilde{F}_i \leq F_i$ with probability one as well. Since $F_i$ is binary, if $\widetilde{F}_i=1$, it must also be $F_i=1$.
\\
\\ 
Moving to the second inner term, a similar derivation yields
\begin{align*}
     \EX_{\bZ}\left[\frac{
    \mathbb{I}\{\widetilde{F}_i=0\}Y_i}{1-p_z} \right]
    &= 
    \Pr(F_i = 1 \mid \widetilde{F}_i=0, i\in \cR_a) Y_i(0,1)
    \\ & +
    \Pr(F_i = 0 \mid \widetilde{F}_i=0, i\in \cR_a) Y_i(0,0).
\end{align*}
By Bayes' rule, 
\begin{equation}
   \label{eq.apdx:pr_f0_tildef0_alt}
    \begin{aligned}
            \Pr(F_i = 0 \mid \widetilde{F}_i=0, i \in \cR_a)
    &= 
    \frac{\Pr(\widetilde{F}_i=0 \mid F_i=0, i \in \cR_a) \Pr(F_i=0 \mid i \in \cR_a)}
    {\Pr(\widetilde{F}_i=0 \mid i \in \cR_a)}
    \\ &= 
    \frac{\Pr(F_i=0 \mid i \in \cR_a)}
    {\Pr(\widetilde{F}_i=0 \mid i \in \cR_a)}
    \\ &=
    \frac{1-\pi^a_i}{1-p_z},
    \end{aligned}
\end{equation}
as $\Pr(\widetilde{F}_i=0 \mid F_i=0, i \in \cR_a) = 1$ since $\widetilde{F}_i \leq F_i$ with probability one.
Setting $\Pr(\widetilde{F}_i=1 \mid F_i=0, i \in \cR_a) = 1 - \Pr(\widetilde{F}_i=0 \mid F_i=0, i \in \cR_a)$ and rearranging terms yields
\begin{equation}
    \label{apdx.eq:alter_0_0}
     \EX_{\bZ}\left[\frac{
    \mathbb{I}\{\widetilde{F}_i=0\}Y_i}{1-p_z} \right]
    =
    \frac{\left(\pi_i^a -p_z\right)Y_i(0,1) + \left(1 - \pi_i^a\right)Y_i(0,0)}{1 - p_z}.
\end{equation}
Consequently, combining \eqref{apdx.eq:alter_0_1} and \eqref{apdx.eq:alter_0_0},
\begin{equation}
    \label{apdx.eq:alter_bias_ht}
    \begin{aligned}
        \EX_{\bZ}\left[\widehat{IE}\right]
        &= 
         \frac{1}{n_a}
    \sum_{i \in \cR_a}
    \left[
    Y_i(0,1)
    -
   \frac{\left(\pi_i^a -p_z\right)Y_i(0,1) + \left(1 - \pi_i^a\right)Y_i(0,0)}{1 - p_z}
    \right]
    \\ &=
    \frac{1}{n_a}
    \sum_{i \in \cR_a}
    \frac{1-\pi_i^a}{1-p_z}
    \left[
    Y_i(0,1) - Y_i(0,0)
    \right]
    \\ &=
    IE + \frac{1}{n_a}
    \sum_{i \in \cR_a}
    \frac{p_z-\pi_i^a}{1-p_z}
    \left[
    Y_i(0,1) - Y_i(0,0)
    \right],
    \end{aligned}
\end{equation}
which completes the proof of the first part of Proposition~1.
\\
\\
Moving to the direct effect estimator. Recall that the HT estimator based on the observed data \eqref{eq:DE_naive_HT} is
\begin{equation*}
    \widehat{DE} = 
    \frac{1}{n_e}
    \sum_{i \in \cR_e}
    \left[
    \frac{
    \mathbb{I}\{Z_i=1\}Y_i}{p_z} 
    -
    \frac{
    \mathbb{I}\{Z_i=0\}Y_i}{1 - p_z} 
    \right].
\end{equation*}
By the assumed experimental design (Assumption~1), the observed exposure of all egos is zero $\Pr(\widetilde{F}_i =0 \mid i \in \cR_e) =1$, and  among the egos $i \in \cR_e$,  we have that $\mathbb{I}\{Z_i=z\}=\mathbb{I}\{Z_i=z, \widetilde{F}_i =0\}$  with probability one.
The expectation of the first inner term of $\widehat{DE}$, for $i \in \cR_e$, is as follows.
\begin{equation*}
    % \label{apdx.eq:egos_z_0}
    \begin{aligned}        
    \EX_{\bZ}\left[\frac{
    \mathbb{I}\{Z_i=1\}Y_i}{p_z} \right]
    &\overset{(i)}{=}
    \EX_{\bZ}\left[\frac{
    \mathbb{I}\{Z_i=1, \widetilde{F}_i =0\}Y_i}{p_z} \right]
    \\ &\overset{(ii)}{=}
     \frac{1}{p_z}
     \EX_{\bZ}
     \left[
    \mathbb{I}\{Z_i=1,F_i=1,\widetilde{F}_i=0\}Y_i(1,1) 
    \right]
    \\ &\quad +
    \frac{1}{p_z}
     \EX_{\bZ}
     \left[
    \mathbb{I}\{Z_i=1,F_i=0,\widetilde{F}_i=0\}Y_i(1,0) 
    \right]
    \\ &\overset{(iii)}{=}
    \Pr(F_i=1, \widetilde{F}_i=0 \mid i \in \cR_e)Y_i(1,1)
    \\ &\quad + 
    \Pr(F_i=0, \widetilde{F}_i=0 \mid i \in \cR_e)Y_i(1,0)
    \\ &\overset{(iv)}{=}
    \Pr(F_i=0 \mid i \in \cR_e)
    Y_i(1,1)
    \\ &\quad +
    \Pr(F_i=1 \mid i \in \cR_e)
    Y_i(1,0),
    \end{aligned}
\end{equation*}
where $(i)$ follows from the explanation described above; $(ii)$ from consistency (Assumption~3);
$(iii)$ since $Z_i$ is independent of both $F_i$ and $\widetilde{F}_i$ by Assumption~1, and $\Pr(Z_i=1\mid i \in \cR_e)=p_z$; and $(iv)$ since $\widetilde{F}_i$ is equal to $0$ with probability one.
Recalling that $\pi^e_i = \Pr(F_i =1 \mid i \in \cR_e)$ and 
plugging it into the above expectation yields 
\begin{equation}
    \label{apdx.eq:ego_1_0}
    \EX_{\bZ}\left[\frac{
    \mathbb{I}\{Z_i=1\}Y_i}{p_z} \right]
    = \pi_i^e Y_i(1,1) + (1-\pi_i^e)Y_i(1,0).
    \end{equation}
Replacing $Z_i=1$ with $Z_i=0$ and $p_z$ with $1-p_z$ above, yields similar results. Specifically, for $z=0$ we have
\begin{equation}
    \label{apdx.eq:ego_0_0}
    \EX_{\bZ}\left[\frac{
    \mathbb{I}\{Z_i=0\}Y_i}{1-p_z} \right]
    = \pi_i^e Y_i(0,1) + (1-\pi_i^e)Y_i(0,0).
    \end{equation}
Plugging \eqref{apdx.eq:ego_1_0} and \eqref{apdx.eq:ego_0_0} into $\EX_{\bZ}\left[\widehat{DE}\right]$ yields
\begin{equation}
    \label{apdx.eq:ego_ht_bias}
    \begin{aligned}
    \EX_{\bZ}\left[\widehat{DE}\right]
    &=
     \frac{1}{n_e}
    \sum_{i \in \cR_e}
    \left[
    \pi_i^e\left\{Y_i(1,1) - Y_i(0,1)
    \right\}
    +
    (1-\pi_i^e)\left\{Y_i(1,0) - Y_i(0,0)
    \right\}
    \right]
    \\ &=
    DE + 
    \frac{1}{n_e} 
    \sum_{i\in\cR_e}
    \pi_i^e
    \left[
    \left\{ Y_i(1, 1) - Y_i(0,1)\right\} 
    - \left\{ Y_i(1,0) - Y_i(0,0)\right\}
    \right],
    \end{aligned}
\end{equation}
which completes the second part of Proposition~1.
\qed

\subsection{Exposure probabilities of alters}
\label{apdx_subsec:expos_prob_alters}
Alter $i\in\cR_a$ is exposed ($F_i=1$) if its ego $e(i)$ or at least one of the other recruited egos connected to $i$ in the population network $\bA$ is treated. We have
\begin{align}
\begin{split}
\label{Eq:pi_help}
    \pi^a_i &= \Pr(F_i=1 \mid i\in \cR_a) 
    \\ &=
    \Pr(F_i=1 \mid Z_{e(i)}=1, i\in \cR_a)\Pr(Z_{e(i)}=1\mid i \in \cR_a) 
    \\ &+ \Pr(F_i=1 \mid Z_{e(i)}=0, i\in \cR_a)\Pr(Z_{e(i)}=0\mid i \in \cR_a),
\end{split}
\end{align}
but $\Pr(F_i=1 \mid Z_{e(i)}=1, i\in \cR_a)=1$ and $\Pr(Z_{e(i)}=1\mid i \in \cR_a)=p_z \mathbb{I}\{e(i) \in \cR_e\}=p_z$ by Assumptions~1 and 2. Since only recruited egos have a non-zero probability of being assigned to treatment, then for alter $i\in \cR_a$, $F_i = \sum_{j\neq i}Z_jA_{ij}=\sum_{j \in \cR_e}Z_jA_{ij}$ with probability one. Therefore,
\begin{align*}
    \Pr(F_i=0 \mid Z_{e(i)}=0, i\in \cR_a) &=
    \Pr
    \Big(
    \sum_{j \in \cR_e}Z_jA_{ij}=0 \mid Z_{e(i)}=0, i \in \cR_a
    \Big) 
    \\ &=
    \Pr
    \Big(
    \sum_{j \in \cR_e \setminus e(i)}Z_jA_{ij}=0 \mid i \in \cR_a
    \Big) 
    \\ &=
    \prod_{j \in  \cR_e \setminus \{e(i)\}}
    \big(1 - p_z \rho^a_{ij}\big).
\end{align*}
Substituting $\Pr(F_i=1 \mid Z_{e(i)}=0, i\in \cR_a)= 1- \Pr(F_i=0 \mid Z_{e(i)}=0, i\in \cR_a)$ in \eqref{Eq:pi_help} yields \eqref{eq:pi_a}.

\subsection{Proof of Proposition~2}
\label{apdx.subsec:proof_prop2}
The bias-corrected estimator is
\begin{equation*}
    \widehat{IE}_{adj} = 
    \frac{1}{n_a}
    \sum_{i \in \cR_a}
    \frac{1-p_z}{1-\pi_i^a}
    \left[
    \frac{
    \mathbb{I}\{\widetilde{F}_i=1\}Y_i}{p_z} 
    -
    \frac{
    \mathbb{I}\{\widetilde{F}_i=0\}Y_i}{1 - p_z} 
    \right].
\end{equation*}
As the weights $\frac{1-p_z}{1-\pi_i^a}$ are fixed and therefore independent of $\bZ$, taking the expectation $\EX_{\bZ}\left[\widehat{IE}_{adj}\right]$
will result in the second row of \eqref{apdx.eq:alter_bias_ht} where the terms $\frac{1-\pi_i^a}{1-p_z}$ and $\frac{1-p_z}{1-\pi_i^a}$ are multiplied and therefore canceled. That implies $\EX_{\bZ}\left[\widehat{IE}_{adj}\right]= IE$ and completes the proof of Proposition~2.
\qed

\subsection{Proof of Proposition~3}
\label{apdx_subsec:proof_prop3}
Recall that \eqref{eq:kappa} states that
\begin{equation*}
    \overline{\Delta}(1) = \kappa \overline{\Delta}(0),
\end{equation*}
where $\overline{\Delta}(f) = \frac{1}{n_e}\sum_{i\in\cR_e} \Delta_i(f)$ and $\Delta_i(f) = Y_i(1,f) - Y_i(0,f)$.
The bias-adjusted direct effect estimator is
\begin{equation*}
   \widehat{DE}_{adj} =
    \frac{1}{1 + \overline{\pi}^e(\kappa-1)}
    \widehat{DE} = 
    \frac{1}{u_e}
    \sum_{i \in \cR_e}
    \left[
    \frac{
    \mathbb{I}\{Z_i=1\}Y_i}{p_z} 
    -
    \frac{
    \mathbb{I}\{Z_i=0\}Y_i}{1 - p_z} 
    \right]
    ,
\end{equation*}
where $u_e = n_e \left[1 + \overline{\pi}^e(\kappa-1) 
    \right]$ and $\overline{\pi}^e = n_e^{-1}\sum_{i \in \cR_e} \pi_i^e$ is the average probability of exposure among the egos.

The inner product of two vectors $\boldsymbol{a}, \boldsymbol{b} \in \mathbb{R}^m$ can be written as $\frac{1}{m}\sum_i a_ib_i = \overline{a}\overline{b} + S_{\boldsymbol{a}, \boldsymbol{b}}$ where $S_{\boldsymbol{a}, \boldsymbol{b}} = \frac{1}{m}\sum_i (a_i - \overline{a})(b_i - \overline{b})$.
Since both $\kappa$ and $\pi_i^e$ are fixed and therefore independent of $\bZ$, we can use \eqref{apdx.eq:ego_ht_bias} and write the expectation of the naive estimator as
\begin{equation}
\label{apdx.eq:expec_de_naive}
\begin{aligned}
    \EX_{\bZ}\left[\widehat{DE}\right]
    &=
    DE + 
    \frac{1}{n_e} 
    \sum_{i\in\cR_e}
    \pi_i^e
    \left[
    \left\{ Y_i(1, 1) - Y_i(0,1)\right\} 
    - \left\{ Y_i(1,0) - Y_i(0,0)\right\}
    \right]
    \\ &=
    DE + 
    \frac{1}{n_e} 
    \sum_{i\in\cR_e}
    \pi_i^e
    \left[
    \Delta_i(1) 
    - \Delta_i(0)
    \right]    
    \\&=
    DE + 
    \overline{\pi}^e \left(\overline{\Delta}(1) - \overline{\Delta}(0)\right) 
    + S_{\boldsymbol{\pi}^e, \boldsymbol{\Delta(1)}} 
    - S_{\boldsymbol{\pi}^e, \boldsymbol{\Delta(0)}}
    \\ &=
    \left[1 + \overline{\pi}^e(\kappa-1)\right]DE 
    +
    S_{\boldsymbol{\pi}^e, \boldsymbol{\Delta(1)}} 
    - S_{\boldsymbol{\pi}^e, \boldsymbol{\Delta(0)}},
\end{aligned}
\end{equation}
where we used $\overline{\Delta}(1) = \kappa \overline{\Delta}(0)$ and the fact that $\overline{\Delta}(0)=DE$.
Thus, the expected value of the bias-corrected estimator is
\begin{equation*}
    \EX_{\bZ}\left[\widehat{DE}_{adj}\right] = DE 
    + 
    \frac{S_{\boldsymbol{\pi}^e, \boldsymbol{\Delta(1)}} 
    - S_{\boldsymbol{\pi}^e, \boldsymbol{\Delta(0)}}}{
    1 + \overline{\pi}^e(\kappa-1)
    }.
\end{equation*}
Finally, under Assumption~4, $S_{\boldsymbol{\pi}^e, \boldsymbol{\Delta(1)}}  = S_{\boldsymbol{\pi}^e, \boldsymbol{\Delta(0)}}$, and we obtain $\EX_{\bZ}\left[\widehat{DE}_{adj}\right] = DE$.
That completes the proof of Proposition~3. 
\qed

In the case of homogeneous exposure probabilities $\pi_i^e=\pi^e$ for all $i \in \cR_e$ for some constant $\pi^e \in [0,1]$, we have $S_{\boldsymbol{\pi}^e, \boldsymbol{\Delta(1)}} = S_{\boldsymbol{\pi}^e, \boldsymbol{\Delta(0)}}=0$, therefore $\widehat{DE}_{adj}$ is an unbiased estimator of $DE$.
\\
\\
\paragraph{Conditions under which the naive HT estimator is unbiased.} If $\pi_i^e=0$ for all $i \in \cR_e$ or $\kappa=1$, the bias-corrected estimator $\widehat{DE}_{adj}$ reduces to the naive HT estimator $\widehat{DE}$.
If $\pi_i^e=0$, then from Proposition~1 the naive HT estimator will be an unbiased estimator of $DE$. 
If $\kappa=1$, we have two options
\begin{enumerate}
    \item If the exposure probabilities are homogeneous ($\pi_i^e=\pi^e$), then $S_{\boldsymbol{\pi}^e, \boldsymbol{\Delta(1)}} = S_{\boldsymbol{\pi}^e, \boldsymbol{\Delta(0)}}=0$ and $\widehat{DE}$ is an unbiased estimator of $DE$. That will be true even if there is some contamination between the egos, i.e., if $\pi^e > 0$.
    \item As described in the main text, under the saturated potential outcome model
    \begin{equation*}
        Y_i(z,f) = \beta_{0i} + \beta_{1i}z + \beta_{2i}f + \beta_{3i}zf,
    \end{equation*}
    Assumption~4 reduces to
    \begin{equation*}
        S_{\boldsymbol{\pi}^e, \boldsymbol{\beta}_{3}} = 
        \frac{1}{n_e}\sum_{i \in \cR_e}
        (\pi_i^e - \overline{\pi}^e)(\beta_{3i} - \overline{\beta_{3}}) = 0.
    \end{equation*}
    Setting $\kappa=1$ implies that $\sum_{i\in \cR_e} \beta_{3i} =0$. 
    Thus, $\widehat{DE}$ will be an unbiased estimator of $DE$ if $\sum_{i \in \cR_e}
        \pi_i^e\beta_{3i}=0$.
\end{enumerate}

\subsection{Direction of bias relative to the null}
\label{apdx_subsec:direction_bias_to_null}
We now show the direction of bias of the naive estimators $\widehat{IE}$ and $\widehat{DE}$ relative to the null of no effects.

\paragraph*{Indirect effect estimator}
Starting with $\widehat{IE}$.
If $Y_i(0,1) \geq Y_i(0,0)$ for all $i \in \cR_a$ such that $IE\geq0$, 
then since $p_z - \pi_i^a \leq 0$, from \eqref{apdx.eq:alter_bias_ht}, 
we have $0\leq \EX_{\bZ}\left[\widehat{IE}\right] \leq IE$. 
Vice versa, if $Y_i(0,1) \leq Y_i(0,0)$ for all $i \in \cR_a$,
 then $IE \leq 0$ and $0 \geq \EX_{\bZ}\left[\widehat{IE}\right] \geq IE$. 
 That is, the estimator $\widehat{IE}$ is, in expectation, biased towards the null of no indirect effect.
 That will hold in both the homogeneous and heterogeneous scenarios of $\rho_{ij}^a$, as $p_z - \pi_i^a \leq 0$ for all $i \in \cR_a$.

 \paragraph*{Direct effect estimator}
We show the direction of bias relative to the null of the naive direct effect estimator $\widehat{DE}$ using the representation $\kappa$ as defined in \eqref{eq:kappa}. 
From \eqref{apdx.eq:expec_de_naive}
\begin{equation*}
    \EX_{\bZ}\left[\widehat{DE}\right]
    =
    \left[1 + \overline{\pi}^e(\kappa-1)\right]DE 
    +
    S_{\boldsymbol{\pi}^e, \boldsymbol{\Delta(1)}} 
    - S_{\boldsymbol{\pi}^e, \boldsymbol{\Delta(0)}}.
\end{equation*}
Under Assumption~4, the covariance terms are equal, thus 
\begin{equation*}
    \EX_{\bZ}\left[\widehat{DE}\right]
    =
    \left[1 + \overline{\pi}^e(\kappa-1)\right]DE 
\end{equation*}
Since $\overline{\pi}^e \geq 0$, we have
\begin{itemize}
    \item If $DE \geq 0$ 
        \begin{align*}
    \begin{cases}
        \EX_{\bZ}\big[\widehat{DE}\big] \geq DE \geq 0, & \kappa > 1, \\
        DE \geq \EX_{\bZ}\big[\widehat{DE}\big]  \geq 0, & 
            \kappa \in \left( 1 - \frac{1}{\overline{\pi}^e},
             1 \right), \\
        \EX_{\bZ}\big[\widehat{DE}\big] \leq 0, & \kappa < 1 - \frac{1}{\overline{\pi}^e}.
    \end{cases}
\end{align*}
    \item If $DE \leq 0$
        \begin{align*}
    \begin{cases}
        \EX_{\bZ}\big[\widehat{DE}\big] \leq DE \leq 0, & \kappa > 1, \\
         DE \leq \EX_{\bZ}\big[\widehat{DE}\big] \leq 0, & \kappa \in
             \left( 1 - \frac{1}{\overline{\pi}^e},
             1 \right), \\
        \EX_{\bZ}\big[\widehat{DE}\big] \geq 0, & \kappa < 1 - \frac{1}{\overline{\pi}^e}.
    \end{cases}
    \end{align*}
\end{itemize}
Therefore, 
the estimator $\widehat{DE}$ is biased \emph{away} from the null of no direct effect
if $\kappa > 1$, biased \emph{towards} the null if $\kappa \in \left( 1 - \frac{1}{\overline{\pi}^e}, 1 \right)$, and changes sign if $\kappa < 1 - \frac{1}{\overline{\pi}^e}$.
Since $\pi_i^e$ is typically small, then $1 - \frac{1}{\overline{\pi}^e}$ is a large negative number, and it is unlikely that $\kappa$ is smaller than that value in practice.
Thus, in general, we expect that the estimator $\widehat{DE}$ is biased away from the null if $\kappa > 1$ and towards the null if $\kappa < 1$
for many realistic scenarios.

\subsection{Direction and magnitude of bias-corrected estimators}
\label{apdx_subsec:sign_pres}
We now derive the direction and effect magnitude of the bias-corrected estimators $\widehat{IE}_{adj}$ and $\widehat{DE}_{adj}$ compared to their respective naive estimators $\widehat{IE}$ and $\widehat{DE}$.

\paragraph*{Indirect effect estimators.}
Eq.~(14) implies that $\pi^a_i \geq p_z$, therefore $\frac{1-p_z}{1-\pi^a_i} \geq 1$.
If $\pi^a_i = \pi^a$ for all $i\in\cR_a$, that is, the same for all alters, then $\widehat{IE}_{adj}=\frac{1-p_z}{1-\pi^a}\widehat{IE}$. Thus, $\text{sign}(\widehat{IE}_{adj})=\text{sign}(\widehat{IE})$ and 
\begin{align*}
    \begin{cases}
        \widehat{IE}_{adj} \geq \widehat{IE}, & \widehat{IE} > 0, \\
        \widehat{IE}_{adj} \leq \widehat{IE}, & \widehat{IE} < 0.
    \end{cases}
\end{align*}

\paragraph*{Direct effect estimators.}
Recall that $\widehat{DE}_{adj} = \frac{1}{1+\overline{\pi}^e(\kappa-1)}\widehat{DE}$. 
Thus, if $1 + \overline{\pi}^e(\kappa-1) > 1 \Rightarrow \kappa > 1 - \frac{1}{\overline{\pi}^e}$, 
    then $\text{sign}(\widehat{DE}_{adj}) = \text{sign}(\widehat{DE})$.
Otherwise, if $\kappa < 1 - \frac{1}{\overline{\pi}^e}$, then $\text{sign}(\widehat{DE}_{adj}) = -\text{sign}(\widehat{DE})$.
However, for example, if $\overline{\pi}^e \approx 0.01$, $1 - \frac{1}{\overline{\pi}^e} \approx -99$,
 and it is unlikely that $\kappa$ is smaller than that value in practice.
 So in general, we expect $\text{sign}(\widehat{DE}_{adj}) = \text{sign}(\widehat{DE})$.
Moreover, if $\kappa > 1$ then
\begin{align*}
    \begin{cases}
        \widehat{DE}_{adj} < \widehat{DE}, & \widehat{DE} > 0, \\
        \widehat{DE}_{adj} > \widehat{DE}, & \widehat{DE} < 0,
    \end{cases}
\end{align*}
so the adjusted estimator is, in magnitude, closer to zero than the unadjusted estimator.
Vice versa, if $\kappa \in \left(1-\frac{1}{\overline{\pi}^e}, 1\right)$ then
\begin{align*}
    \begin{cases}
        \widehat{DE}_{adj} > \widehat{DE}, & \widehat{DE} > 0, \\
        \widehat{DE}_{adj} < \widehat{DE}, & \widehat{DE} < 0.
    \end{cases}
\end{align*}
Thus, the adjusted estimator is, in magnitude, further away from zero than the unadjusted estimator.

\subsection{Variance estimation and asymptotic distribution (no cross-fitting)} 
\label{apdx_subsec:var_estimation}
We derive variance estimators for the non-augmented (but bias-adjusted) estimators. 
\paragraph*{Indirect effect estimator.} 
Starting with $\widehat{IE}_{adj}$. Let $r^a_i = \frac{1-p_z}{1-\pi_i^a}\left[\frac{
    \mathbb{I}\{\widetilde{F}_i=1\}Y_i}{p_z} 
    -
    \frac{
    \mathbb{I}\{\widetilde{F}_i=0\}Y_i}{1 - p_z}\right]$ such that $\widehat{IE}_{adj}=\frac{1}{n_a}\sum_{i\in \cR_a} r^a_i$. 
By the ENRT design, alters in the same ego-network all have the same treatment assignments and exposure status. Thus, we can calculate the variance by using the aggregated outcomes in each ego-network. To this end, we can write
\begin{align*}
    \widehat{IE}_{adj} &= 
    \frac{1}{n_a}
    \sum_{i \in \cR_e}
    \sum_{j \in \cR_a : e(j)=i} 
    r^a_j 
    \\ &=
    \frac{1}{n_a}
    \sum_{i \in \cR_e} T_i 
    \\ &=
    \frac{n_e}{n_a} \overline{T},
\end{align*}
where $T_i =\sum_{j \in \cR_a : e(j)=i} 
    r^a_j $    and $\overline{T} = \frac{1}{n_e}\sum_{i \in \cR_e} T_i$ is the average over ego-networks. 
    Therefore, the variance is 
    $V_{\bZ}\left(\widehat{IE}_{adj}\right) = 
    \frac{n_e^2}{n_a^2}V_{\bZ}\left(\overline{T}\right)$, which can be conservatively estimated with 
    the design-based variance estimator
    \citep{ding2024first}
\begin{equation*}
\widehat{V}_{\bZ}\left(\widehat{IE}_{adj}\right) = 
\frac{1}{n_a^2}\sum_{i \in \cR_e}
\left(T_i - \overline{T}\right)^2. 
\end{equation*}
Under standard regularity conditions, the estimator $\widehat{IE}_{adj}$ is asymptotically normally distributed with mean $IE$ \citep{ding2024first}.
This motivates a Wald-type confidence interval for $IE$
based on the variance estimator $\widehat{V}_{\bZ}\left(\widehat{IE}_{adj}\right)$
\begin{equation*}
    \widehat{IE}_{adj} \pm z_{1-\alpha/2}
    \sqrt{\widehat{V}_{\bZ}\left(\widehat{IE}_{adj}\right)},
\end{equation*}
where $z_{1-\alpha/2}$ is the $1- \alpha/2$ upper quantile of a standard normal distribution.
However, since the estimator $\widehat{V}_{\bZ}\left(\widehat{IE}_{adj}\right)$ is a conservative estimator of $V_{\bZ}\left(\widehat{IE}_{adj}\right)$,
constructing confidence intervals based on it results in a conservative confidence interval with coverage rate of at least $1-\alpha$.

\paragraph{Direct effect estimator.} 
Moving to $\widehat{DE}_{adj}$. Let 
$r^e_i = \frac{
    \mathbb{I}\{Z_i=1\}Y_i}{p_z} 
    -
    \frac{
    \mathbb{I}\{Z_i=0\}Y_i}{1 - p_z}$ 
    such that $\widehat{DE}_{adj}=
    \frac{1}{u_e}
    \sum_{i\in \cR_e} 
    r^e_i$. 
By the ENRT design, treatment assignment is independent across egos.
We can write the variance of $\widehat{DE}_{adj}$ as 
\begin{equation}
    \label{apdx.eq:var_de_adj}
    V_{\bZ}\left(\widehat{DE}_{adj}\right) 
    = 
    \frac{1}{u_e^2}\sum_{i \in \cR_e} V_{\bZ}(r^e_i) + 
    \frac{1}{u_e^2}\sum_{i,j \in \cR_e:
    i\neq j} \text{Cov}_{\bZ}(r^e_i, r^e_j),
\end{equation}
where $V_{\bZ}(r^e_i)$ is the variance of the terms $r^e_i$ w.r.t. the design $\bZ$.
The Neyman-type variance estimator is \citep{ding2024first}
\begin{equation*}
    \widehat{V}_{Neyman}\left(\widehat{DE}_{adj}\right) = 
    \frac{1}{u_e^2}
    \sum_{i \in \cR_e}
    \left(r^e_i - \widehat{DE}_{adj}\right)^2.
\end{equation*}
This estimator assumes that the terms $r^e_i$ are independent, and is therefore a conservative estimator of the first term in $V_{\bZ}\left(\widehat{DE}_{adj}\right)$. However, in the presence of contamination, this independence assumption is violated. 
From the exposure mapping specification (Assumption~2) and consistency (Assumption~3), we can write the terms $r^e_i$ as
\begin{equation*}
% \label{apdx.eq:r_i_ego}
    r^e_i = 
    \frac{Z_i}{p_z}
    \left[F_i Y_i(1,1)
    + (1-F_i)Y_i(1,0)
    \right]
    - 
    \frac{1-Z_i}{1-p_z}
    \left[F_i Y_i(0,1)
    + (1-F_i)Y_i(0,0)
    \right],
\end{equation*}
where $F_i = F(\bZ_{-i},\bA)$ are the true exposures of ego $i$.
Therefore, if egos $i$ and $j$ share an ego neighbor $k$ in the population network $\bA$, the terms $r^e_i$ and $r^e_j$ will be dependent, as $Z_k$ will influence both $F_i$ and $F_j$. These dependencies induce pairwise covariances $\text{Cov}_{\bZ}(r^e_i, r^e_j)$ that are omitted by the Neyman-type estimator $\widehat{V}_{Neyman}\left(\widehat{DE}_{adj}\right)$.
Thus, $\widehat{V}_{Neyman}\left(\widehat{DE}_{adj}\right)$ captures only the sum of marginal variances, and ignoring the covariance terms that arise under contamination might lead to underestimation of the overall variance.

To ensure valid inference, we propose a conservative variance correction that accounts for these latent correlations. 
The terms $r^e_i$ and $r^e_j$ for two egos $i,j \in \cR_e$ are correlated if the egos share a common (latent) neighbor $k \in \cR_e$.
Let $K^{\bA}_{ij}(k)=\mathbb{I}\{A_{ik}=1,A_{jk}=1\}$ be the indicator that in the population network $\bA$, the ego $k \in \cR_e$ is a common neighbor of both $i$ and $j$ while $k \neq i,j$. Define the number of shared neighbors between ego $i$ and $j$ by in the population network $\bA$ by
\begin{equation*}
    K_{ij}^{\bA} = \sum_{k \in \cR_e, k\neq i,j} K_{ij}^{\bA}(k).
\end{equation*}
We construct a dependency graph $\bG^{\bA}_{dep} =\left(\cR_e, \bE^{\bA}_{dep}\right)$ where an edge exists between two egos $i,j \in \cR_e$ if the terms $r^e_i$ and $r^e_j$ are dependent. 
That is, $\bE^{\bA}_{dep}(i,j)=1$ occurs if $A_{ik}=A_{jk}=1$ for some other ego $k \in \cR_e$. 
Thus, we can write
$\mathbb{I}\{\bE^{\bA}_{dep}(i,j)=1\} =
\mathbb{I}\{K^{\bA}_{ij}>0\}$.
However, the ego-ego edges in $\bA$ are missing from the observed network $\widetilde{\bA}$. But we can compute the probability of the dependence under the sensitivity model (Section~3.4).
Each of the missing ego-ego edges $A_{ik}$ is postulated to exist independently with probability $\rho^e_{ik}$. Therefore,
taking expectation with regard to the epistemic uncertainty from the sensitivity model yields,
\begin{equation*}
    \begin{aligned}
        \Pr(\bE^{\bA}_{dep}(i,j)=1) 
        &=
        \Pr(K^{\bA}_{ij}>0) 
        \\ &=
        1 - \Pr(K^{\bA}_{ij} =0 ) 
        \\ &=
        1 - \prod_{k \in \cR_e : k\neq i,j}
        (1-\rho^e_{ik}\rho^e_{jk})
        \\ &= 
        \xi_{ij}.
    \end{aligned}
\end{equation*}
As the network $\bA$ is assumed to be fixed, but partly unknown, the expressions $\text{Cov}_{\bZ}(r^e_i, r^e_j)$ in \eqref{apdx.eq:var_de_adj} are essentially conditioned on it. That is, written explicitly, $\text{Cov}_{\bZ}(r^e_i, r^e_j)$ should be viewed as $\text{Cov}_{\bZ}(r^e_i, r^e_j \mid \bA)$.
As the terms $r^e_i$ and $r^e_j$ are dependent only under the event $\mathbb{I}\{\bE^{\bA}_{dep}(i,j)=1\}$, for a fixed population network $\bA$, we can write 
$\text{Cov}_{\bZ}(r^e_i, r^e_j \mid \bA)$ as $\mathbb{I}\{\bE^{\bA}_{dep}(i,j)=1\}
    \text{Cov}_{\bZ}(r^e_i, r^e_j \mid \bA)$.
By the Cauchy-Schwarz inequality, we can bound the covariance term
\begin{equation*}
   \mathbb{I}\{\bE^{\bA}_{dep}(i,j)=1\}
    \lvert \text{Cov}_{\bZ}(r^e_i, r^e_j \mid \bA) \rvert 
    \leq
    \mathbb{I}\{\bE^{\bA}_{dep}(i,j)=1\}
    \sqrt{V_{\bZ}(r^e_i)V_{\bZ}(r^e_j)},
\end{equation*}
where $V_{\bZ}(r^e_i)$ is the variance of the terms $r^e_i$ w.r.t. the design $\bZ$.
Since $\mathbb{I}\{\bE^{\bA}_{dep}(i,j)=1\}$ is unknown due to the missing ego-ego edges in the observed network, we replace it with its expected value $\Pr(\bE^{\bA}_{dep}(i,j)=1) = \xi_{ij}$ under the sensitivity model.
Therefore, 
\begin{equation*}
    \xi_{ij}  \lvert \text{Cov}_{\bZ}(r^e_i, r^e_j \mid \bA) \rvert\leq 
    \xi_{ij}
    \sqrt{V_{\bZ}(r^e_i)V_{\bZ}(r^e_j)}
\end{equation*}
We use the squared residuals $\hat{v}_i = (r^e_i - \widehat{DE}_{adj})^2$ as conservative estimators for the unit-level variances $V_{\bZ}(r^e_i)$.
This yields a conservative estimator for the variance in $\widehat{DE}_{adj}$ due to contamination between ego-networks
\begin{equation}
    \label{apdx.eq:de_cov_conserv}
    \widehat{V}_{Conta.}\left(\widehat{DE}_{adj}\right) =\frac{1}{u_e^2} \sum_{i \neq j}
   \xi_{ij}
    \sqrt{\hat{v}_i \hat{v}_j}.
\end{equation}
The final corrected variance estimator for $\widehat{DE}_{adj}$ is given by
\begin{equation}
    \label{apdx.eq:de_adjust_var}
    \widehat{V}_{\bZ}\left(\widehat{DE}_{adj}\right) = 
    \widehat{V}_{Neyman}\left(\widehat{DE}_{adj}\right)
    + 
    \widehat{V}_{Conta.}\left(\widehat{DE}_{adj}\right).
\end{equation}
% Note that the contamination variance correction $\widehat{V}_{Conta.}\left(\widehat{DE}_{adj}\right)$ in the variance estimator $\widehat{V}_{\bZ}\left(\widehat{DE}_{adj}\right)$ 
% relies on the expected topology of the network defined by the sensitivity parameters, as no data on the missing ego--ego edges is available.

We establish the asymptotic normality of $\widehat{DE}_{adj}$ by
adapting standard asymptotic normality results for design-based estimators \citep{ding2024first} to the dependent data setting induced by our sensitivity model.
We present a sketch of a proof based on the Central Limit Theorem for dependency graphs established by \citet{baldi1989normal}.
Under the design-based framework, statistical dependence between the summands
$r_i^e$ in $\widehat{DE}_{adj}$ arises solely from the random treatment assignments of shared neighbors in the latent population network. Specifically, the dependence arises from the true exposure values $F_i = F_i(\bZ_{-i},\bA_{i})$ due to possible contamination 
under the sensitivity model.

Consider the dependency graph $\bG^{\bA}_{dep} =\left(\cR_e, \bE^{\bA}_{dep}\right)$ presented above. 
Define the degree of each ego $i \in \cR_e$ in this graph by $\text{Deg}_{n_e, i}$, and let $\text{Deg}_{n_e, max} =\max_{i \in \cR_e}\text{Deg}_{n_e, i}$ denote the maximum degree.
Assume that the terms $r_i^e$ are bounded, and that the variance of the sum converges linearly $n_e$, i.e., $V_{\bZ}\left(\sum_{i \in \cR_e} r_i^e \right) = \Theta(n_e)$, indicating that the variance of the sum is bounded both above and below by $n_e$ asymptotically.
Therefore, Corollary 2 of
\citet{baldi1989normal} 
implies that the error in the normal approximation is bounded by a term of order $\text{Deg}_{n_e, max} / n_e^{1/4}$.
Consequently, the validity of the normal approximation requires that the size of the dependency neighborhoods grows sufficiently slowly relative to the sample size, specifically $\text{Deg}_{n_e, max} = o(n_e^{1/4})$.

In the context of our sensitivity analysis, 
as the true population network $\bA$ is only partially known, this imposes a constraint on the postulated latent network in the specified sensitivity model. If the specified probabilities $\rho^e_{ij}$ are sufficiently large such that the latent network become dense (e.g., $\rho^e_{ij} \approx 1$ for all egos $i,j \in \cR_e$), this causes the maximal degree of the dependency graph to scale linearly with the sample size ($\text{Deg}_{n_e, max} \approx n_e$), violating the condition $\text{Deg}_{n_e, max} = o(n_e^{1/4})$.
Therefore, the reported confidence intervals are valid only for values of the sensitivity parameters consistent with a sufficiently sparse latent network topology.

Following a similar argument to $\widehat{IE}_{adj}$, the Wald-type confidence interval for $DE$ based on the conservative variance estimator $\widehat{V}_{\bZ}\left(\widehat{DE}_{adj}\right)$
\begin{equation*}
    \widehat{DE}_{adj} \pm z_{1-\alpha/2}
    \sqrt{\widehat{V}_{\bZ}\left(\widehat{DE}_{adj}\right)},
\end{equation*}
results in a conservative confidence interval for $DE$ with coverage of at least $1-\alpha$.

\subsection{Augmented estimation via cross-fitting}
\label{apdx_subsec:aug_estimation}
The augmented estimators combine the outcome $Y_i$ and predicted outcome models for estimating direct and indirect effects. 
Estimating the outcome models requires cross-fitting. 
We utilize a two-fold cross-fitting algorithm tailored for design-based estimators \citep{lu2025conditional}.
We also provide variance estimators for the augmented estimators obtained from this cross-fitting procedure.
Finally, we describe the asymptotic distribution of the estimators.

We adapt \citet[][Algorithm~1]{lu2025conditional} to our setting.
Specifically, from Assumption~1 of Bernoulli design for each ego-network, \citet[][Proposition~1]{lu2025conditional} motivates the following two-fold cross-fitting algorithm. 
\begin{enumerate}
    \item Randomly split ego-networks into two disjoint parts $S_0, S_1$ such that $\lvert S_0 \cup S_1 \rvert = n$ and $S_0 \cap S_1 = \emptyset$. Specifically, we assign each ego-network (ego and alters) into $S_1$ with probability $0.5$.
    \item For $q=0,1$ do
    \begin{enumerate}
        \item Estimate the outcome functions for alters $\widehat{\mu}^a_i(f)$ and egos $\widehat{\mu}^e_i(z)$ using data from the complementary split $S_{1-q}$.
        \item Predict $\widehat{\mu}^a_i(f)$ and $\widehat{\mu}^e_i(z)$ on the units from $S_q$.
        \item Estimate $IE$ on units from $S_q$:
        \begin{equation*}
            \widehat{IE}^{aug}_{adj[q]} 
            = \frac{1}{n_{a[q]}}
            \sum_{i \in \cR_a \cap S_q}
            \frac{1-p_z}{1-\pi^a_i}
            \left[D^a_i +
            \widehat{\mu}^a_i(1) - 
            \widehat{\mu}^a_i(0) \right],
        \end{equation*}
        where $n_{a[q]} =\sum_{i \in \cR_a} \mathbb{I}\{i \in S_q\}$ is the number of alters in split $S_q$, and where
        \begin{equation*}
            D^a_i = \frac{
    \mathbb{I}\{
    \widetilde{F}_i=1\}\left(Y_i - \widehat{\mu}^a_i(1) \right)}{p_z} 
    -
    \frac{
    \mathbb{I}\{\widetilde{F}_i=0\}\left(Y_i - \widehat{\mu}^a_i(0) \right)}{1 - p_z}. 
        \end{equation*}
     \item Estimate $DE$ on units from $S_q$:
    \begin{equation*}
        \widehat{DE}^{aug}_{adj[q]} =
        \frac{1}{u_{e[q]}} 
    \sum_{i \in \cR_e \cap S_q}
    \left[
    D^e_i
    + \widehat{\mu}^e_i(1) - \widehat{\mu}^e_i(0) 
    \right],
    \end{equation*}
    where $u_{e[q]} =n_{e[q]}\left[1 + \overline{\pi}^{e[q]
    }(\kappa-1)\right]$ similarly to $u_e$ in Section~3.2, with $n_{e[q]} = \sum_{i \in \cR_e} \mathbb{I}\{i \in S_q\}$ and 
    $\overline{\pi}^{e[q]} = \frac{1}{n_{e[q]}}\sum_{i \in \cR_e \cap S_q}  \pi_i^e$ 
    are the number of egos and the average exposure probability, respectively, in split $S_q$, and 
    \begin{equation*}
        D^e_i = \frac{
    \mathbb{I}\{
    Z_i=1\}\left(Y_i - \widehat{\mu}^e_i(1) \right)}{p_z} 
    -
    \frac{
    \mathbb{I}\{Z_i=0\}\left(Y_i - \widehat{\mu}^e_i(0) \right)}{1 - p_z}. 
    \end{equation*}
    \end{enumerate}
    \item Combine the estimates across splits
    \begin{equation*}
    % \label{eq:aug_esti}
        \begin{aligned}
            \widehat{IE}^{aug}_{adj}
            &= \sum_{q =0,1}\frac{n_{a[q]}}{n_a} \widehat{IE}^{aug}_{adj[q]},
            \\ 
            \widehat{DE}^{aug}_{adj}
            &= \sum_{q =0,1}\frac{u_{e[q]}}{u_e} \widehat{DE}^{aug}_{adj[q]}.
        \end{aligned}
    \end{equation*}
\end{enumerate}
Note that the weighting in the combination of the direct effect estimators uses $u_e$ (as defined in Section~3.2) since $u_e=u_{e[0]} + u_{e[1]}$.
As the units in the splits $S_0$ and $S_1$ are independent (w.r.t. $\bZ$), variance estimation will also be a combination of variance estimates in each split \citep{lu2025conditional}.
For the indirect effect estimates, we can write the variance estimate in split $q$ by
\begin{equation*}
    \widehat{V}_{\bZ}\left(\widehat{IE}^{aug}_{adj[q]}\right) = 
    \frac{1}{n_{a[q]}^2} 
    \sum_{i \in \cR_e \cap S_q} 
    \left(
    T^{aug}_i - \overline{T}^{aug}_{[q]}
    \right)^2,
\end{equation*}
where $T^{aug}_i = \sum_{j \in \cR_a\cap S_q : e(j)=i} \frac{1-p_z}{1-\pi^a_j} D^a_j$ \
is the average of weighted residuals $D^a_j$ of alter $j$ in ego-network of ego $i \in \cR_e$ in split $q$, and $\overline{T}^{aug}_{[q]} = \frac{1}{n_{e[q]}} \sum_{i \in \cR_e\cap S_q}T^{aug}_i$ is the average over ego-networks in split $q$.
The outcome model terms $\widehat{\mu}^e_i(1) - \widehat{\mu}^e_i(0)$ are not included in the variance estimator as they are fixed w.r.t. $\bZ$ and since the splitting rule ensures independence between the two folds.
The combined variance estimator of $\widehat{IE}^{aug}_{adj}$ is
\begin{equation}
    \label{apdx.eq:ie_var_esti_cf}
    \widehat{V}_{\bZ}\left(\widehat{IE}^{aug}_{adj}\right) = 
    \sum_{q =0,1}\frac{n_{a[q]}^2}{n_a^2}
    \widehat{V}_{\bZ}\left(\widehat{IE}^{aug}_{adj[q]}\right).
\end{equation}
For direct effects, the Neyman-type variance estimator in split $q$ is
\begin{equation*}
    \widehat{V}_{Neyman}\left(\widehat{DE}^{aug}_{adj[q]} \right) =
    \frac{1}{u_{e[q]}^2}
    \sum_{i \in \cR_e \cap S_q}
    \left( 
    D^e_i -
    \overline{D^e}_{[q]}
    \right)^2,
\end{equation*}
where 
$\overline{D^e}_{[q]} = 
\frac{1}{u_{e[q]}} 
\sum_{i \in \cR_e \cap S_q}
D^e_i
$.
Therefore, similarly to \eqref{apdx.eq:ie_var_esti_cf}, the Neyman variance estimator of $\widehat{DE}^{aug}_{adj}$ is 
\begin{equation}
    \label{apdx.eq:de_var_esti_cf}
    \widehat{V}_{Neyman}\left(\widehat{DE}^{aug}_{adj}\right) = 
    \sum_{q =0,1}\frac{u_{e[q]}^2}{u_e^2}
    \widehat{V}_{\bZ}\left(\widehat{DE}^{aug}_{adj[q]}\right).
\end{equation}
We apply the same conservative contamination-attributed variance correction \eqref{apdx.eq:de_adjust_var} derived for the adjusted estimator, adapting it to the augmented residuals. Specifically, we compute the conservative contamination-attributed variance estimator \eqref{apdx.eq:de_cov_conserv} for each fold, and combine it similarly to \eqref{apdx.eq:ie_var_esti_cf}.

From \citet[][Theorem~4]{lu2025conditional},
 under common regularity conditions, both estimators $\widehat{IE}^{aug}_{adj}$ and $\widehat{DE}^{aug}_{adj}$
are approximately normal and normal-based confidence intervals using the variance estimators achieve a valid (asymptotic) coverage rate.

In addition, the random splitting minimizes the dependence between the two folds. While strict independence (w.r.t. $\bZ$) implies no interference between the splits, the contamination between ego-networks is assumed to be sparse. Consequently, the dependence of the estimated outcome models on the treatment assignments of units in the complementary fold is expected to be negligible.
Under this approximation, the estimated outcome models at each split can be treated as fixed w.r.t. the design $\bZ$ \citep{lin2013agnostic, lu2025conditional}. 
Therefore, the derivations shown in Sections~\ref{apdx.subsec:proof_prop2}-\ref{apdx_subsec:proof_prop3} imply that 
 $\widehat{IE}_{adj}^{aug}$ and $\widehat{DE}_{adj}^{aug}$ remain approximately unbiased estimators of $IE$ and $DE$, respectively, provided that the number of edges connecting units in the two folds (ego--ego and alter--ego) is small relative to the sample size.

\subsection{Relative risk estimand and estimators}
\label{apdx_subsec:rr_estimand}
In the case of binary outcomes, we can also consider the relative risk (RR) estimand
\begin{equation}
    \label{apdx.eq:IE_ratio}
    IE^{RR} = \frac{
    \sum_{i \in \cR_a}Y_i(0,1)}{
    \sum_{i \in \cR_a}Y_i(0,0)
    }.
\end{equation}
Analogously to \eqref{eq:IE_naive_HT}, the estimator based on the observed data is therefore
\begin{equation}
    \label{apdx.eq:ie_ratio_ht}
    \widehat{IE}^{RR} =
    \frac{
    \sum_{i \in \cR_a}\frac{
    \mathbb{I}\{\widetilde{F}_i=1\}Y_i}{p_z}}{
    \sum_{i \in \cR_a}\frac{
    \mathbb{I}\{\widetilde{F}_i=0\}Y_i}{1 - p_z}
    }.
\end{equation}
We saw in \eqref{apdx.eq:alter_0_1} that the numerator in \eqref{apdx.eq:ie_ratio_ht} is an unbiased estimator of the numerator in \eqref{apdx.eq:IE_ratio}.
However, from \eqref{apdx.eq:alter_0_0} the denominator is biased. Rearranging \eqref{apdx.eq:alter_0_0}, 
\begin{equation*}
      \frac{1-p_z}{1-\pi_i^a}\EX_{\bZ}\left[\frac{
    \mathbb{I}\{\widetilde{F}_i=0\}Y_i}{1-p_z} \right]
    =
    Y_i(0,0) +
    \frac{\pi_i^a - p_z}{1-\pi_i^a}Y_i(0,1).
\end{equation*}
Thus, using \eqref{apdx.eq:alter_0_1}, we can unbiasedly estimate $Y_i(0,0)$ through
\begin{equation*}
     \frac{1-p_z}{1-\pi_i^a}
     \frac{
    \mathbb{I}\{\widetilde{F}_i=0\}Y_i}{1-p_z} -
     \frac{\pi_i^a - p_z}{1-\pi_i^a} \frac{
    \mathbb{I}\{\widetilde{F}_i=1\}Y_i}{p_z}.
\end{equation*}
That motivates the adjusted estimator
\begin{equation}
    \label{apdx.eq:IE_ratio_adjusted_esti}
    \widehat{IE}^{RR}_{adj} =
    \frac{
    \sum_{i \in \cR_a}\frac{
    \mathbb{I}\{\widetilde{F}_i=1\}Y_i}{p_z}}{
    \sum_{i \in \cR_a}\frac{1-p_z}{1-\pi_i^a}
     \frac{
    \mathbb{I}\{\widetilde{F}_i=0\}Y_i}{1-p_z} -
     \frac{\pi_i^a - p_z}{1-\pi_i^a} \frac{
    \mathbb{I}\{\widetilde{F}_i=1\}Y_i}{p_z}
    }.
\end{equation}
Being a ratio estimator, \eqref{apdx.eq:IE_ratio_adjusted_esti} is biased, but the bias can be easily bounded \citep{Saerndal2003}.

\subsection{Relaxing Assumption~4}
\label{apdx_subsec:relaxing_ass4}
We provide two approaches to relax Assumption~4, which states that 
the empirical covariance between exposure probabilities $\pi_i^e$ and the direct effect among exposed egos $\Delta_i(1)$ is equal to the empirical covariance between $\pi_i^e$ and the direct effect among non-exposed egos $\Delta_i(0)$, that is,
\begin{equation*}
    S_{\boldsymbol{\pi}^e, \boldsymbol{\Delta}(1)}=S_{\boldsymbol{\pi}^e, \boldsymbol{\Delta}(0)}.
\end{equation*}
In the first approach, we derive sharp bounds for $DE$ as a function of the sensitivity parameters $\pi_i^e$ and $\kappa$ and the observed data. That enables researchers to obtain bounds, instead of point estimates, for $DE$, or obtain bounds for $DE$ given estimates of $\pi_i^e$ and $\kappa$ from validation data, as described in Section~3.6.

In the second approach, we modify the sample average proportionality expressed in \eqref{eq:kappa}, to a stronger assumption of unit-level proportionality. That produces a modified bias-corrected estimator for $DE$ that is an unbiased estimator without requiring Assumption~4 to hold.

\paragraph{Bounds for $DE$ without Assumption~4.}
We can relax Assumption~~4 by using bounds instead of an unbiased estimator given values of $\pi_i^e$ and $\kappa$. 
Note that we can write
\begin{equation*}
    S_{\boldsymbol{\pi}^e, \boldsymbol{\Delta}(1)} 
    - S_{\boldsymbol{\pi}^e, \boldsymbol{\Delta}(0)} = S_{\boldsymbol{\pi}^e, \boldsymbol{\Delta}(1)-\boldsymbol{\Delta}(0)}.
\end{equation*}
Therefore, Assumption~4 is equivalent to assuming $S_{\boldsymbol{\pi}^e, \boldsymbol{\Delta}(1)-\boldsymbol{\Delta}(0)} = 0$.
Note that from \eqref{apdx.eq:expec_de_naive} the expectation of the naive HT estimator is
\begin{equation*}
\begin{aligned}
    \EX_{\bZ}\left[\widehat{DE}\right] 
    &= 
    \left[1 + \overline{\pi}^e(\kappa-1)\right]DE 
    +
    S_{\boldsymbol{\pi}^e, \boldsymbol{\Delta(1)}} 
    - S_{\boldsymbol{\pi}^e, \boldsymbol{\Delta(0)}}
    \\ &=
    \left[1 + \overline{\pi}^e(\kappa-1)\right]DE 
    + S_{\boldsymbol{\pi}^e, \boldsymbol{\Delta}(1)-\boldsymbol{\Delta}(0)}
\end{aligned}
\end{equation*}
By the Cauchy-Schwarz inequality, we have
\begin{equation*}
    \lvert S_{\boldsymbol{\pi}^e, \boldsymbol{\Delta}(1)-\boldsymbol{\Delta}(0)} \rvert \leq 
    S_{\boldsymbol{\pi}^e}
    S_{\boldsymbol{\Delta}(1)-\boldsymbol{\Delta}(0)},
\end{equation*}
where the first component $S_{\boldsymbol{\pi}^e}$ is the empirical standard deviation of $\pi_i^e$ and can be computed given a specific value of the sensitivity parameters.
The second standard deviation can be bounded as follows.
Assume that there exist two constants $b_0\leq b_1$ such that $b_0 \leq Y_i(z,f) \leq b_1$ for all $i,z,f$.
Then $\lvert \Delta_i(f) \lvert \leq b_1 - b_0$ and $\lvert \Delta_i(1)-\Delta_i(0) \lvert \leq 2(b_1 - b_0)$. Therefore, by the Bhatia–Davis inequality, a sharp bound for the variance $S^
2_{\boldsymbol{\Delta}(1)-\boldsymbol{\Delta}(0)}$ is
\begin{equation*}
\begin{aligned}
    S^2_{\boldsymbol{\Delta}(1)-\boldsymbol{\Delta}(0)} 
    &\leq
    \left[2(b_1-b_0) -(\overline{\Delta}(1)-\overline{\Delta}(0))\right]
    \left[
    \overline{\Delta}(1)-\overline{\Delta}(0) 
    + 2(b_1-b_0)
    \right]
    \\ &= 
    4(b_1-b_0)^2 - \left(\overline{\Delta}(1)-\overline{\Delta}(0)\right)^2.
\end{aligned}
\end{equation*}
But from \eqref{eq:kappa}, we can write $\overline{\Delta}(1)-\overline{\Delta}(0) = (\kappa-1)\overline{\Delta}(0) = (\kappa-1)DE$. Thus, the variance bound is
\begin{equation*}
    S^2_{\boldsymbol{\Delta}(1)-\boldsymbol{\Delta}(0)}  \leq 4(b_1-b_0)^2 - \left[(\kappa-1)DE\right]^2. 
\end{equation*}
That yields a bound for the standard deviation 
\begin{equation*}
    S_{\boldsymbol{\Delta}(1)-\boldsymbol{\Delta}(0)} \leq
    \sqrt{4(b_1-b_0)^2 - \left[(\kappa-1)DE\right]^2} 
    \leq 2(b_1-b_0),
\end{equation*}
as $\left[(\kappa-1)DE\right]^2 \geq 0$.
Therefore, we obtain that the empirical covariance $S_{\boldsymbol{\pi}^e, \boldsymbol{\Delta}(1)-\boldsymbol{\Delta}(0)}$ is bounded by 
\begin{equation*}
    \lvert S_{\boldsymbol{\pi}^e, \boldsymbol{\Delta}(1)-\boldsymbol{\Delta}(0)} \rvert \leq 2(b_1-b_0)     S_{\boldsymbol{\pi}^e}.
\end{equation*}
Thus, we can bound $DE$ as follows,
\begin{equation*}
    \frac{\EX_{\bZ}\left[\widehat{DE}\right] - 2(b_1-b_0) S_{\boldsymbol{\pi}^e}}{1 + \overline{\pi}^e(\kappa-1)}
    \leq DE 
    \leq 
    \frac{\EX_{\bZ}\left[\widehat{DE}\right] + 2(b_1-b_0) S_{\boldsymbol{\pi}^e}}{1 + \overline{\pi}^e(\kappa-1)},
\end{equation*}
or, since $\EX_{\bZ}\left[\widehat{DE}_{adj}\right] = 
\frac{\EX_{\bZ}\left[\widehat{DE}\right]}{1 + \overline{\pi}^e(\kappa-1)}$, we can write the bounds in terms of the bias-corrected estimator
\begin{equation*}
    \EX_{\bZ}\left[\widehat{DE}_{adj}\right] -\frac{2(b_1-b_0) S_{\boldsymbol{\pi}^e}}{1 + \overline{\pi}^e(\kappa-1)}
    \leq DE 
    \leq 
    \EX_{\bZ}\left[\widehat{DE}_{adj}\right] + \frac{2(b_1-b_0) S_{\boldsymbol{\pi}^e}}{1 + \overline{\pi}^e(\kappa-1)}.
\end{equation*}
For example, for binary outcomes, $b_0=0, b_1=1$, and we obtain
\begin{equation*}
    \frac{\EX_{\bZ}\left[\widehat{DE}\right] - 2 S_{\boldsymbol{\pi}^e}}{1 + \overline{\pi}^e(\kappa-1)}
    \leq DE 
    \leq 
    \frac{\EX_{\bZ}\left[\widehat{DE}\right] + 2 S_{\boldsymbol{\pi}^e}}{1 + \overline{\pi}^e(\kappa-1)}.
\end{equation*}
Where in practice it is possible to use the naive estimator $\widehat{DE}$ instead of its expectation in the bounds and construct confidence intervals for the partially identified parameter $DE$ \citep[e.g., as proposed by][]{imbens2004}.

\paragraph{Unit-level proportionality assumption.}
Recall that in the main text we defined $\kappa$ as the proportion between the sample average direct effects among exposed egos $\overline{\Delta}(1)$ to the effect among non-exposed egos $\overline{\Delta}(0)$. By the design-based inference framework, this proportionality is only a notation and holds by design. The bias-corrected estimator $\widehat{DE}_{adj}$, as defined in \eqref{eq:DE_corrected}, is unbiased estimator of $DE$ only given additional conditions, as formally expressed in Proposition~3. Here, replacing Assumption~4 with a different assumption produce a modified bias-corrected estimator.

Specifically, assume that instead of constant $\kappa$ that link $\Delta_i(1)$ and $\Delta_i(0)$ on average over the egos ($\overline{\Delta}(1) = \kappa \overline{\Delta}(0)$), there is a constant $\kappa^\ast$ that control the unit-level proportionality
\begin{assumption}[Unit-level proportionaliity]
    \label{ass.apdx:kappa_unit}
    There exist a constant $\kappa^\ast$ such that 
    $\Delta_i(1) = \kappa^\ast \Delta_i(0)$ 
    for all $i \in \cR_e$.
\end{assumption}
That is, Assumption~\ref{ass.apdx:kappa_unit} 
 states that $Y_i(1,1)-Y_i(0,1) = \kappa^\ast \left\{Y_i(1,0) - Y_i(0,0) \right\}$ for all egos. 
 The interpretation of the additional sensitivity parameter $\kappa$ in
 Assumption~\ref{ass.apdx:kappa_unit} is related to the interaction between treatment and exposure in the potential outcomes of the egos.
  Under the saturated potential outcomes model
\begin{equation*}
    Y_i(z,f) = \beta_{0i} + \beta_{1i}z + \beta_{2i}f + \beta_{3i}zf,
\end{equation*}
Assumption~\ref{ass.apdx:kappa_unit} is equivalent to $\beta_{3i} = (\kappa^\ast-1)\beta_{1i}$. 
That is, the interaction term $\beta_{3i}$ is proportional to the direct effect term $\beta_{1i}$ for all egos.
Thus, while $\kappa$ in \eqref{eq:kappa} expressed an \emph{average-level} relation, $\kappa^\ast$ in Assumption~\ref{ass.apdx:kappa_unit} posit \emph{unit-level} proportionality.
However, for binary potential outcomes, Assumption~\ref{ass.apdx:kappa_unit} implies only a specific set of possible values for $\kappa^\ast$, and not any constant value, thereby restricting the valid values to specify for $\kappa^\ast$ when running the sensitivity analysis described below.
 
We can use a modified bias-corrected direct effect estimator 
\begin{equation*}
       \widehat{DE}_{adj}^\ast =
    \frac{1}{n_e}
    \sum_{i \in \cR_e}
    \frac{1}{1 + \pi_i^e(\kappa^\ast-1)}
    \left[
    \frac{
    \mathbb{I}\{Z_i=1\}Y_i}{p_z} 
    -
    \frac{
    \mathbb{I}\{Z_i=0\}Y_i}{1 - p_z} 
    \right].
\end{equation*}
Since both $\kappa^\ast$ and $\pi_i^e$ are fixed and therefore independent of $\bZ$, we can use \eqref{apdx.eq:ego_1_0}, \eqref{apdx.eq:ego_0_0}, and \eqref{apdx.eq:ego_ht_bias} to obtain
\begin{equation*}
    \begin{aligned}
        \EX \left[\widehat{DE}_{adj}^\ast \right]
        &=
        \frac{1}{n_e}
    \sum_{i \in \cR_e}
    \frac{1}{1 + \pi_i^e(\kappa^\ast-1)}
    \left[
    \pi_i^e\left\{Y_i(1,1) - Y_i(0,1)
    \right\}
    +
    (1-\pi_i^e)\left\{Y_i(1,0) - Y_i(0,0)
    \right\}
    \right]
    \\ &\overset{(i)}{=}
      \frac{1}{n_e}
    \sum_{i \in \cR_e}
    \frac{1}{1 + \pi_i^e(\kappa^\ast-1)}
    \left[
    \pi_i^e\kappa^\ast\left\{Y_i(1,0) - Y_i(0,0)
    \right\}
    +
    (1-\pi_i^e)\left\{Y_i(1,0) - Y_i(0,0)
    \right\}
    \right]
    \\ &=
    \frac{1}{n_e}
    \sum_{i \in \cR_e}
    [Y_i(1,0) - Y_i(0,0)]
    \\ &=
    DE,
    \end{aligned}
\end{equation*}
where in $(i)$ we used Assumption~\ref{ass.apdx:kappa_unit}.
Therefore, $\widehat{DE}_{adj}^\ast$ is an unbiased estimator of $DE$ under Assumptions~1-3 and  Assumption~\ref{ass.apdx:kappa_unit} instead of assuming homogeneous exposure probabilities or Assumption~4 in Proposition~3.

The estimation of variance for $\widehat{DE}_{adj}^\ast$ can be performed similarly to the estimator described in Section~\ref{apdx_subsec:var_estimation} by replacing $r_i^e$ with $r^{e\ast}_i = \frac{1}{1 + \pi_i^e(\kappa^\ast-1)} \left[\frac{
    \mathbb{I}\{Z_i=1\}Y_i}{p_z} 
    -
    \frac{
    \mathbb{I}\{Z_i=0\}Y_i}{1 - p_z}\right]$ 
    such that $\widehat{DE}_{adj}^\ast=
    \frac{1}{n_e}
    \sum_{i\in \cR_e} 
    r^{e\ast}_i$. 
In addition, outcome-augmented estimation for $\widehat{DE}_{adj}^\ast$ is also similar to the methods described in Section~\ref{apdx_subsec:aug_estimation} for $\widehat{DE}_{adj}$. 

Running the sensitivity analysis (GSA or PBA) with this modified estimator follows the same logic described in Section~3.5, by replacing $\kappa$ with $\kappa^\ast$, albeit with a different interpretation, of course. 

In addition, Assumption~\ref{ass.apdx:kappa_unit} can be relaxed by allowing conditional proportionality given covariates $\bX_i$. 
That is, replacing $\kappa^\ast$ with $\kappa^\ast_i = \kappa^\ast(\bX_i)$ in Assumption~\ref{ass.apdx:kappa_unit} enables researchers to specify some structure on the proportionality instead of assuming the same level of homogeneity for all units.
For example, for a categorical covariate $X_{i1} \in \{1,2,3\}$, researchers can specify $\kappa^\ast_1$ for $X_{i1}=1$ and set $\kappa^\ast_k= \tau_k \kappa^\ast_1$ for constant $\tau_k$ for $k=2,3$. This relaxes the homogeneous proportionality by allowing each subgroup (determined by $X_{i1}$) to have different values. For instance, $X_{i1}$ can represent some proxy of risky behavior or susceptibility to be influenced by peer advice, such as age group. 
In that case, researchers can treat $\kappa^\ast_1$ as the sensitivity parameters and either assume fixed values for $\tau_k$ or instead specify multiple values for $\tau_k$ as well. That can be easily done in PBA, where researchers only need to specify a distribution for these parameter values.

\subsection{Estimating the sensitivity parameters with internal validation data}
\label{apdx_subsec:int_valid_data}

\paragraph{Alters.} Assume we have an internal validation subsample of alters $\mathcal{V}_a \subset \cR_a$ that are asked to recall information about the intervention. Assume that alters who recall such information have a true exposure $F_i=1$ as they were exposed to at least one treated ego. 
This essentially assumes that the recall data in the validation subset perfectly equate the true exposure status. This might not hold if recall is not perfect, but can be strengthened through stringent selection criteria for the validation subsample \citep{Chao2023}.
The conditional probability of true exposure given observed non-exposure for alters $\Pr(F_i=1 \mid \widetilde{F_i}=0, i \in \cR_a)$ is linked to the marginal exposure probability $\pi_i^a$ through \eqref{eq.apdx:pr_f0_tildef0_alt}:
\begin{equation*}
    \pi_i^a = p_z + (1-p_z)\Pr(F_i=1 \mid \widetilde{F_i}=0, i \in \cR_a).
\end{equation*}
Assuming homogeneous exposure probabilities $\pi_i^a = \pi^a$, we can estimate this conditional probability in validation subsample through
\begin{equation*}
    \widehat{q}^a = \frac{\sum_{i \in \mathcal{V}_a} \mathbb{I} \left\{F_i=1, \widetilde{F}_i=0 \right\}}{\sum_{i \in \mathcal{V}_a} \mathbb{I} \left\{ \widetilde{F}_i=0 \right\}},
\end{equation*}
and use $\widehat{q}^a$ to estimate $\pi^a$
\begin{equation*}
    \widehat{\pi}^a = p_z + (1-p_z)\widehat{q}^a.
\end{equation*}
Given $\widehat{\pi}^a$, we can estimate the latent ego-alter edges probabilities $\rho^a$ as well. By Example~1, $\pi^a = p_z + (1-p_z)\left[1- (1-p_z\rho^a)^{n_e-1}\right]$. Substituting $\pi^a$ with $\widehat{\pi}^a$ and solving for $\rho^a$ yields
\begin{equation*}
    \widehat{\rho}^a = \frac{1 - \left(1 - \widehat{q}^a\right)^{\frac{1}{n_e - 1}}}{p_z}.
\end{equation*}
A similar derivation can yield an estimate of $m^a$, the expected number of missing ego-alter edges (Example~2).

Under heterogeneous exposure probabilities, we can regress $F_i$ on $\widetilde{F_i}=0$ and $\bX_i$ among the validation subsample to learn $\Pr(F_i=1 \mid \widetilde{F}_i=0, \bX_i, i \in \mathcal{V}_a)$ and use the learned model to predict the exposure probabilities for the other alters in the main study. In this case, it is impossible to infer the edge probabilities $\rho_{ij}^a$ as $\pi_i^a$ is a convolution of these non-identical values.  However, given the estimated $\pi_i^a$ values, researchers can still use the bias-adjusted indirect effect estimator \eqref{eq:IE_corrected}. Accounting for uncertainty in estimating $\pi_i^a$ in the final indirect effect estimates can be achieved using standard methods from the measurement error literature \citep[e.g., as described by][]{Chao2023}.

\paragraph{Egos.}
For egos, we separate between two types of internal validation data gathered on a validation subsample $\mathcal{V}_e \subset \cR_e$ of egos
\begin{enumerate}
    \item Asking egos to recall \emph{any} information related to the intervention.
    \item Asking egos if they heard or learned about the intervention from other friends or sources.
\end{enumerate}
The first type is similar to the validation data described for the alters. The second is more directly related to indirect exposure to treatment through neighbors.

Treated egos ($Z_i=1$) in the validation subset $\mathcal{V}_e$ of the first type should recall any information related to the intervention, as they were randomized to participate in it. However, non-treated egos ($Z_i=0$) would have learned about the intervention only through their relations to other treated egos. Thus, any non-treated ego that recalled information about the intervention in the validation subsample can be viewed as having been truly exposed to at least one treated ego, that is, $F_i=1$.
By the experimental design (Assumption~1) and the exposure mapping specification (Assumption~2), the treatment status $Z_i$ is independent of the exposure $F_i$. Therefore, $\pi_i^e = \Pr(F_i = 1 \mid i \in \cR_e) = \Pr(F_i = 1 \mid Z_i=0, i \in \cR_e)$. Under homogeneous exposure probabilities $\pi_i^e = \pi^e$ we can estimate $\pi_i^e$ from the validation subsample through
\begin{equation*}
    \widehat{q}^e =
    \widehat{\Pr}(F_i = 1 \mid Z_i=0, i \in \mathcal{V}_e)
    =
    \frac{\sum_{i \in \mathcal{V}_e} \mathbb{I} \left\{F_i=1, Z_i=0 \right\}}{\sum_{i \in \mathcal{V}_e} \mathbb{I} \left\{ Z_i=0 \right\}}.
\end{equation*}
Similarly for the derivation given to the alters, we can use $\widehat{q}^e$ to estimate $\rho^e$ (in Example~1) or $m^e$ (in Example~2). We can also estimate heterogeneous exposure probabilities by learning $\Pr(F_i = 1 \mid Z_i=0, \bX_i, i \in \mathcal{V}_e)$.

In the second validation data type, we have access to the true exposure status $F_i$ of all egos in the validation subsample. Thus, we can learn $\pi^e$ or $\pi_i^e$ from the entire validation subsample, instead of only non-treated egos, as described above.
Moreover, $\kappa$ can also be learned on this validation subsample, given that researchers are willing to assume that the validation subsample is representative of the full sample and does not suffer from selection challenges. Recall that $\kappa = \frac{\overline{\Delta}(1)}{\overline{\Delta}(0)}$. Therefore, an estimator of $\kappa$ from the validation subsample is
\begin{equation*}
    \widehat{\kappa} = \frac{
    \overline{Y}^{\mathcal{V}_e}(1,1) 
    - 
    \overline{Y}^{\mathcal{V}_e}(0,1)
    }{
    \overline{Y}^{\mathcal{V}_e}(1,0)
    - 
    \overline{Y}^{\mathcal{V}_e}(0,0)
    }, 
\end{equation*}
where
$\overline{Y}^{\mathcal{V}_e}(z,f) = \frac{
    \sum_{i \in \mathcal{V}_e}
    \mathbb{I}\left\{Z_i=z, F_i=f\right\} Y_i 
    }
    {\sum_{i \in \mathcal{V}_e}
    \mathbb{I}\left\{Z_i=z, F_i=f\right\}}
$,
is the average outcome in the validation subsample for egos with treatment $z \in \{0,1\}$ and exposure $f \in \{0,1\}$.
Given the estimators of $\pi_i^e$ and $\kappa$ from the validation subset, we can either obtain point estimates of $DE$ using $\widehat{DE}^{adj}$ or have bounds for $DE$, as described in Appendix~\ref{apdx_subsec:relaxing_ass4}.

\section{Extension to three-level exposure mapping}
\label{apdx.sec:three_level_em}
\subsection{Preliminaries}

We extend the framework to a three-level exposure mapping for alters, $F_i \in \{0, 1, 2+\}$,
 indicating exposure to zero, exactly one, or two-or-more treated neighbors, respectively. 
 We consider only the indirect effect estimand and estimators.
 We derive bias expressions for the naive HT estimator
 and propose an adjusted estimator to correct for the bias.
 The adjusted estimators rely on the same sensitivity parameter of alter--ego edges probabilities $\rho^a_{ij}$ \eqref{eq:sbm}
  defined in Section~3.4.
 However, we require an additional sensitivity parameter, similar to the unit-level interaction 
 parameter $\kappa^\ast$ for the bias-corrected direct effect estimator presented in Appendix~\ref{apdx_subsec:relaxing_ass4}.

Let $S_i = \sum_{j\neq i} Z_jA_{ij}$ be the true number of 
treated neighbors of unit $i$ in the population network $\bA$. 
For alters $i \in \cR_a$, we can write $S_i = \sum_{j \in \cR_e} Z_jA_{ij}$.
We first modify the exposure mapping (Assumption~2).
\begin{assumption}[Three-Level Exposure Mapping]
\label{apdx.ass:expos_map_three_level}
    Let
    \begin{equation*}
        F(\bZ_{-i},\bA) =
        \begin{cases}
            0, & S_i = 0 \\
            1, & S_i = 1 \\
            2+, & S_i \ge 2
        \end{cases}
    \end{equation*}
    For any $\bz, \bz' \in \mathcal{Z}_{\cR}$, 
    if $z_i = z'_i$ and $F(\bz_{-i},\bA) = F(\bz'_{-i},\bA)$, then $Y_i(\bz) = Y_i(\bz')$.
\end{assumption}
By Assumption~\ref{apdx.ass:expos_map_three_level}, we can write $Y_i(\bZ) = Y_i(Z_i, F_i)$, where $Z_i \in  \{0,1\}$ and $F_i \in \{0, 1, 2+\}$.
Thus, each unit now has six potential outcomes, instead of four.
The consistency assumption (Assumption~3) is also modified.
\begin{assumption}[Consistency (Three-Level)]
\label{apdx.ass:consis_three_level} The observed outcomes are given by
\begin{equation*}
    Y_i = \sum_{z \in \{0,1\}}\sum_{f \in \{0, 1, 2+\}}Y_i(z,f) \mathbb{I}\left\{Z_i=z, F_i=f\right\}.
\end{equation*}
\end{assumption}
We focus on the estimand  of the sample average indirect effect 
of exposure to a single versus no treated neighbors \eqref{eq:indirect_effect}
\begin{equation*}
    IE = \frac{1}{n_a} \sum_{i\in\cR_a} Y_i(0, 1) - Y_i(0,0).
\end{equation*}

\subsection{Bias of the naive estimator}
The naive HT estimator $\widehat{IE}$ remains as defined in \eqref{eq:IE_naive_HT}, as the observed exposures $\widetilde{F}_i$ are zero or one.  Its expectation is $\EX_{\bZ}\left[\widehat{IE}\right] = \frac{1}{n_a} \sum_{i \in \cR_a} \left( E_{i,1} - E_{i,0} \right)$, where
\begin{align*}
    E_{i,1} &= \EX_{\bZ}\left[\frac{\mathbb{I}\{\widetilde{F}_i=1\}Y_i}{p_z} \right] \\
    E_{i,0} &= \EX_{\bZ}\left[\frac{\mathbb{I}\{\widetilde{F}_i=0\}Y_i}{1-p_z} \right].
\end{align*}
The issue is that observed exposures $\widetilde{F}_i=1$ 
can correspond to true exposures of either $F_i=1$ or $F_i=2+$,
and observed exposures $\widetilde{F}_i=0$ can correspond to true exposures of $F_i=0$, $F_i=1$, or $F_i=2+$.
Therefore, the observed outcomes $Y_i$ under observed exposures 
    $\widetilde{F}_i$ are mixtures of potential outcomes under different true exposures $F_i$.

Define $S_{i, -e(i)} = \sum_{j \in \cR_e^{e(i)}} Z_j A_{ij}$ to be the 
 number of treated neighbors of alter $i$ \emph{other than} its own ego $e(i)$.
  This represents the possible exposure from contamination between the ego-networks.

To derive the bias, we first write $E_{i,0}$ and $E_{i,1}$ as functions of the potential outcomes and the exposure probabilities, which in turn are used to obtain an expression for $\EX_{\bZ}\left[\widehat{IE}\right]$.

\paragraph*{Derivation of $E_{i,1}$:}
If $\widetilde{F}_i=1$, then $Z_{e(i)}=1$. The true number of treated neighbors is $S_i = 1 + S_{i, -e(i)}$.
Given that $Z_{e(i)}=1$,
\begin{itemize}
    \item The true exposure is $F_i=1$ if $S_i = 1$, which implies $S_{i, -e(i)} = 0$.
    \item The true exposure is $F_i=2+$ if $S_i \ge 2$, which implies $S_{i, -e(i)} \ge 1$.
\end{itemize}
By Assumptions~\ref{apdx.ass:expos_map_three_level}-\ref{apdx.ass:consis_three_level},
\begin{align*}
    E_{i,1} &= \frac{1}{p_z} \EX_{\bZ}\left[ 
        \mathbb{I}\{\widetilde{F}_i=1, F_i=1\} Y_i(0, 1)
         + \mathbb{I}\{\widetilde{F}_i=1, F_i=2+\}Y_i(0, 2+)
             \right] \\
        &=
        \frac{1}{p_z} \EX_{\bZ}\left[ 
        \mathbb{I}\{Z_{e(i)}=1, S_{i, -e(i)}=0\} Y_i(0, 1)
         + \mathbb{I}\{Z_{e(i)}=1, S_{i, -e(i)}\geq 1\}Y_i(0, 2+)
             \right] \\
    &= \Pr(S_{i, -e(i)}=0) Y_i(0, 1) + \Pr(S_{i, -e(i)} \geq 1) Y_i(0, 2+).
\end{align*}
The last step holds because $S_{i, -e(i)}$ is independent of $Z_{e(i)}$ under Assumption~1,
and $\Pr(Z_{e(i)}=1) = p_z$.

\paragraph*{Derivation of $E_{i,0}$:}
If $\widetilde{F}_i=0$, then $Z_{e(i)}=0$. The true number of treated neighbors is $S_i = S_{i, -e(i)}$.
Given that $Z_{e(i)}=0$,
\begin{itemize}
    \item The true exposure is $F_i=0$ if $S_i = 0$, which implies $S_{i, -e(i)} = 0$.
    \item The true exposure is $F_i=1$ if $S_i = 1$, which implies $S_{i, -e(i)} = 1$.
    \item The true exposure is $F_i=2+$ if $S_i \geq 2$, which implies $S_{i, -e(i)} \geq 2$.
\end{itemize}
Therefore,
\begin{align*}
    E_{i,0} &= 
    \frac{1}{1-p_z} \EX_{\bZ}\left[ 
       \sum_{f \in \{0,1,2+\}} \mathbb{I}\{\widetilde{F}_i=0, F_i=f\} Y_i(0, f)
        \right] 
    \\ &= 
    \Pr(S_{i, -e(i)}=0) Y_i(0, 0) + \Pr(S_{i, -e(i)}=1) Y_i(0, 1) + \Pr(S_{i, -e(i)} \ge 2) Y_i(0, 2+).
\end{align*}

\paragraph*{Exposures probabilities.}
As in Section~3.4, the sensitivity model assumes conditional edge independence, with $\rho^a_{ij}$ defined in \eqref{eq:sbm}.
The random variable $S_{i, -e(i)}$ follows a Poisson-Binomial distribution
 with $n_e-1$ probabilities $\{p_z \rho^a_{ij}\}_{j \in \cR_e^{e(i)}}$. 
 Let $\pi_{i,k}^* = \Pr(S_{i, -e(i)} = k)$. The required probabilities are
\begin{align*}
    \pi_{i,0}^* &= \Pr(S_{i, -e(i)} = 0) =
     \prod_{j \in \cR_e^{e(i)}} (1 - p_z \rho^a_{ij}) \\
    \pi_{i,1}^* &= \Pr(S_{i, -e(i)} = 1) = 
    \sum_{j \in \cR_e^{e(i)}} 
    \left[ p_z \rho^a_{ij}
    \prod_{l \in \cR_e^{e(i)} \setminus \{j\}} (1 - p_z \rho^a_{il})
     \right] \\
    \pi_{i,2+}^* &= \Pr(S_{i, -e(i)} \geq 2) = 1 - \pi_{i,0}^* - \pi_{i,1}^*.
\end{align*}
For large $n_e$, analytic computation of $\pi_{i,1}^*$ can be challenging,
but approximation methods exist \citep{hong2013}.

\paragraph*{Bias expression.}
First, note that $\Pr(S_{i, -e(i)}\geq 1) = 1 - \Pr(S_{i, -e(i)}=0) = 1 - \pi_{i,0}^*$.
Substituting these probabilities into the expressions we obtained above for $E_{i,0}$ and $E_{i,1}$, the expectations of the two components of the naive estimator are:
\begin{align*}
    E_{i,1} &= \pi_{i,0}^* Y_i(0, 1) + (1 - \pi_{i,0}^*) Y_i(0, 2+)  \\
    E_{i,0} &= \pi_{i,0}^* Y_i(0, 0) + \pi_{i,1}^* Y_i(0, 1) + 
    (1 - \pi_{i,0}^* - \pi_{i,1}^*)^* Y_i(0, 2+) 
\end{align*}
The expectation of the naive estimator is:
\begin{equation}
    \label{apdx.eq:bias_three_level}
    \begin{aligned}
        \EX_{\bZ}\left[\widehat{IE}\right] &=
        \frac{1}{n_a} \sum_{i \in \cR_a} \left( E_{i,1} - E_{i,0} \right)
        \\ &=
        \frac{1}{n_a} \sum_{i \in \cR_a} \left[ -\pi_{i,0}^* Y_i(0, 0) + 
        (\pi_{i,0}^* - \pi_{i,1}^*) Y_i(0, 1) +
         \pi_{i,1}^* Y_i(0, 2+) \right]
         \\ &=
         IE + \frac{1}{n_a} \sum_{i \in \cR_a} \left[
         \pi_{i,1}^*\left\{ Y_i(0, 2+) - Y_i(0, 1) \right\}
         - (1 - \pi_{i,0}^*) \left\{ Y_i(0, 1) - Y_i(0, 0) \right\}
         \right],
    \end{aligned}
\end{equation}
which implies that the naive estimator is also biased for this indirect effect estimand.
The bias depends on the average of the unit-level indirect effect of $2+$ exposures relative to $1$ exposures
and of $1$ exposures relative to $0$ exposures,
weighted by the exposure probabilities $\pi_{i,1}^*$ and $1 - \pi_{i,0}^*$, respectively.

\subsection{Bias-corrected estimator}
The resulting expected value \eqref{apdx.eq:bias_three_level} is a linear combination of the 
unknown potential outcomes $\{Y_i(0,0), Y_i(0,1), Y_i(0,2+)\}$.
The problem is that the observed data do not provide enough information to identify all three potential outcomes
for each alter $i \in \cR_a$.
To achieve identification, we introduce an additional sensitivity parameter $\delta$ via 
a unit-level structural assumption on the potential outcomes, analogous to Assumption~\ref{ass.apdx:kappa_unit} of unit-level proportionality for the egos.
\begin{assumption}
    \label{ass:alters_delta}
    There exists a constant $\delta$ such that
    $Y_i(0, 2+) - Y_i(0, 1) = \delta \left\{Y_i(0, 1) - Y_i(0, 0)\right\}$ for all alters $i \in \cR_a$.
\end{assumption}
The sensitivity parameter $\delta$ in Assumption~\ref{ass:alters_delta} provides an intuitive way to model the dose-response relationship of alter exposure.
 It is best interpreted through the \emph{marginal effect of additional exposure}.
  The effect of first exposure is $Y_i(0, 1) - Y_i(0, 0)$.
  The \emph{additional} effect of the second-and-higher exposure is $Y_i(0, 2+) - Y_i(0, 1)$.
  Assumption~\ref{ass:alters_delta} implies a direct relationship.
  Researchers can therefore specify plausible values for $\delta$ based on their domain knowledge.
\begin{itemize}
    \item $\delta = 0$. The effect \emph{saturates} after one exposure, and any further contamination is irrelevant. This recovers the binary exposure model.
    \item $\delta > 0$. There is a \emph{reinforcing} cumulative dose effect, where a second exposure adds benefit. 
        For example, $\delta=1$ implies the same marginal effect between first and second-and-higher exposure.
        However, $\delta \in (0,1)$ implies that the marginal effect between first and zero exposure
        is higher than the marginal effect between second-or-higher and first exposure.
    \item $\delta < 0$. Additional exposures \emph{diminish} the effect of the first.
       For example, exposure to more than one treated neighbor may create confusion or mistrust, reducing the benefit of exposure. 
\end{itemize}
This framework allows researchers to conduct sensitivity analysis by 
    specifying a plausible range for the marginal effect of exposures,
     rather than the absolute value of $Y_i(0, 2+)$ (e.g., by deriving sharp, yet informative, bounds).
 Rearranging the statement in Assumption~\ref{ass:alters_delta}, we have
\begin{equation}
    \label{apdx.eq:alter_y_0_1plus}
    Y_i(0, 2+) = (\delta+1) Y_i(0, 1) -\delta Y_i(0, 0).
\end{equation}
Using \eqref{apdx.eq:alter_y_0_1plus}, the expectation in \eqref{apdx.eq:bias_three_level} 
can be written as,
\begin{equation}
    \label{apdx.eq:ex_ie_under_delta}
    \EX_{\bZ}\left[\widehat{IE}\right] =
    \frac{1}{n_a}
    \sum_{i \in \cR_a}
    \left[
        \pi_{i,0}^* + \pi_{i,1}^*\delta
    \right]
    \left[
        Y_i(0,1) - Y_i(0,0)
    \right],
\end{equation}
which implies the bias-corrected estimator of indirect effect under 
the three-level exposures
\begin{equation}
    \label{apdx.eq:ie_adjust_three_level}
    \widehat{IE}_{adj} =
    \frac{1}{n_a}
    \sum_{i \in \cR_a}
    \frac{1}{
        \pi_{i,0}^* + \pi_{i,1}^*\delta
    }
    \left[
    \frac{\mathbb{I}\{\widetilde{F}_i=1\}Y_i}{p_z} 
    - 
    \frac{\mathbb{I}\{\widetilde{F}_i=0\}Y_i}{1-p_z} 
    \right].
\end{equation}
Following the same strategy as in the proof of Proposition 2, and utilizing the derivations above, it can be shown that the adjusted estimator $\widehat{IE}_{adj}$ is unbiased for the indirect effect estimand $IE$
\begin{equation*}
    \EX_{\bZ}\left[\widehat{IE}_{adj}\right] = IE.
\end{equation*}
Moreover, \eqref{apdx.eq:ex_ie_under_delta} implies that the naive estimator $\widehat{IE}$ is unbiased 
if 
\begin{itemize}
    \item $\pi_{i,0}^*=1$. This corresponds to no contamination between ego-networks
     (no additional treated neighbors beyond the ego).
     That is equivalent to $\rho^a_{ij} = 0$ for all alter--ego edges.
    \item For all $i \in \cR_a$, we have that $\delta = \frac{1-\pi_{i,0}^*}{\pi_{i,1}^*} = 1 + \frac{\pi_{i,2+}^*}{\pi_{i,1}^*}$.
         This implies that the marginal effect of additional exposure is exactly offset by the contamination between ego-networks.
\end{itemize}

\subsection{Relation to the two-level exposure model}
Note that if we set $\delta=0$ in \eqref{apdx.eq:alter_y_0_1plus},
then $Y_i(0,2+) = Y_i(0,1)$, and 
the additional exposure level $2+$ collapses into level $1$.
Recall that $\pi_i^a = \Pr(F_i=1 \mid i \in \cR_a)$ is the probability
that alter $i$ is exposed to \emph{at least} one treated neighbor under the two-level exposure mapping.
We can write 
\begin{equation*}
    \begin{aligned}
        1 - \pi_i^a 
        &= \Pr(F_i=0 \mid Z_{e(i)}=1 \in \cR_a)\Pr(Z_{e(i)}=1 \mid i \in \cR_a) 
        \\ & \quad +
        \Pr(F_i=0 \mid Z_{e(i)}=0 \in \cR_a)\Pr(Z_{e(i)}=0 \mid i \in \cR_a) 
        \\  &= 
         \Pr(S_{i, -e(i)}=0)(1-p_z)
         \\ &= 
          \pi_{i,0}^* (1-p_z).
    \end{aligned}
\end{equation*}
That is, $\pi_{i,0}^* = \frac{1 - \pi_i^a}{1-p_z}$.
Substituting this into \eqref{apdx.eq:ie_adjust_three_level} with $\delta=0$,
we recover the bias-corrected estimator for the two-level exposure mapping
given in \eqref{eq:IE_corrected}.
Therefore, the two-level exposure model can be 
viewed as a special case of the three-level exposure model
with saturation of effect after the first exposure.

\subsection{Direction of bias, sign preservation, and effect magnitude}
If $Y_i(0,1) - Y_i(0,0) \geq 0$ for all alters $i \in \cR_a$, then
\begin{align*}
    \begin{cases}
        \EX_{\bZ}\left[\widehat{IE}\right] \geq IE \geq 0, & 
            \delta \geq \max_{i \in \cR_a}\left\{1 + \frac{\pi_{i,2+}^*}{\pi_{i,1}^*}\right\}, \\
        0 \leq \EX_{\bZ}\left[\widehat{IE}\right] \leq IE, & 
            \delta \in \left( \max_{i \in \cR_a}\left\{ 
              1 - \frac{1 - \pi_{i,2+}^*}{\pi_{i,1}^*} \right\}, 
                \min_{i \in \cR_a}\left\{1 + \frac{\pi_{i,2+}^*}{\pi_{i,1}^*}\right\} \right), \\
        \EX_{\bZ}\left[\widehat{IE}\right] \leq 0 \leq IE, & 
            \delta \leq \min_{i \in \cR_a}\left\{ 1 - \frac{1 - \pi_{i,2+}^*}{\pi_{i,1}^*} \right\}. 
    \end{cases}
\end{align*}
Similarly, if $Y_i(0,1) - Y_i(0,0) \leq 0$ for all alters $i \in \cR_a$, we obtain the opposite inequalities.
Therefore, given monotonicity (in either direction) of the potential outcomes $Y_i(0,1)$ and $Y_i(0,0)$, 
the estimator $\widehat{IE}$ is biased away, towards, or changes sign relative to the true indirect effect $IE$,
 depending on the value of the sensitivity parameter $\delta$.

Similarly to the augmented bias-corrected estimator (Section~3.3), 
we can define an augmented version of the three-level exposure bias-corrected estimator
by adding outcome models
$\widehat{\mu}_i^a(f)$ for $f \in \{0,1\}$.
Variance estimation can be achieved using
the same approaches described in Section~\ref{apdx_subsec:var_estimation} for the bias-corrected 
indirect effect estimators under two-level exposures.

\subsection{Sensitivity analysis}
The bias-corrected estimator \eqref{apdx.eq:ie_adjust_three_level}
can be used in the same way as the bias-corrected estimators 
under two-level exposure mapping described in Section~3.5.
That is,  based on their domain knowledge, researchers can specify plausible ranges for the sensitivity parameters $\delta$ and the basic components that result in
$\rho^a_{ij}$ and conduct GSA or PBA.
The main difference is the specification of the additional sensitivity parameter $\delta$, and a slightly more complicated computation of the exposure probabilities $\pi_{i,0}^*$ and $\pi_{i,1}^*$.
The latter can be easily computed in \texttt{R} using the approximations 
developed by \citet{hong2013}, for example, as implemented in the \texttt{poisbinom} package. 

\section{Probabilistic bias analysis}
\label{apdx.sec:pba}
We provide a brief overview of the technical motivation behind PBA.
For a more detailed treatment, we refer the reader to \citet{Greenland2005} and \citet{Fox2021}.

Assume researchers observe data $\bD$ and are interested in estimating a parameter $\theta$ (e.g., a causal effect).
They construct an estimator $\widehat{\theta}(\bD)$ based on the observed data, but know that it is biased.
Namely, 
\begin{equation*}
    \EX\left[\widehat{\theta}(\bD)\right] \neq \theta.
\end{equation*}
Suppose there exists a vector of sensitivity or bias parameters $\rho$ such that the researchers can construct an adjusted estimator
$\widehat{\theta}_{adj}(\bD, \rho)$ 
that is an unbiased estimator of $\theta$ given the true value of $\rho$:
\begin{equation*}
    \EX\left[\widehat{\theta}_{adj}(\bD, \rho)\mid \rho \right] = \theta.
\end{equation*}
However, the true values of $\rho$ are typically unknown and not identifiable from the data. Standard GSA compares estimates of $\widehat{\theta}_{adj}(\bD, \rho)$ (and possibly associated uncertainty quantification) across a fixed grid of $\rho$ values, treating each point in the grid as if it were the truth.
In contrast, PBA treats $\rho$ as a random variable governed by a probability distribution $p(\rho)$. This distribution reflects the researcher's prior knowledge or uncertainty regarding the true values and structure of the bias.  

The goal of PBA is to estimate the ``posterior" distribution of the parameter of interest $\theta$, accounting for both the uncertainty in $\rho$ and the sampling variability of the estimator. This distribution can be approximated via Monte Carlo simulation \citep{Greenland2005}.
First, we draw $M$ samples $\rho^{(1)}, \ldots, \rho^{(M)}$ from the prior $p(\rho)$. 
For each draw $\rho^{(m)}$, we calculate the adjusted point estimate $\widehat{\theta}_{adj}(\bD, \rho^{(m)})$ and its associated variance estimate, denoted $\widehat{V}(\bD, \rho^{(m)})$. 
To fully propagate the statistical uncertainty, we simulate a value $\theta^{(m)}$ from the estimator's approximate sampling distribution, typically assumed to be normal:
\begin{equation*}
    \theta^{(m)} \sim N\left(\widehat{\theta}_{adj}(\bD, \rho^{(m)}), \, \widehat{V}(\bD, \rho^{(m)}) \right).
\end{equation*}
The resulting values $\theta^{(1)}, \dots, \theta^{(M)}$ approximate the marginal ``posterior" distribution of the parameter of interest $\theta$, accounting for both the uncertainty regarding the sensitivity parameters and the statistical uncertainty arising from the data. This distribution can be summarized through standard summary statistics (e.g., the mean and percentiles). 

 In our context, the only randomness in the data is due to the randomization $\bZ$.
 Thus,
 $\widehat{\theta}$ corresponds to the naive estimators $\widehat{IE}$ or $\widehat{DE}$. The sensitivity $\rho$ consists of the edge probabilities 
    $\rho_{ij}^a$ or $\rho_{ij}^e$ and the interaction parameter $\kappa$.
    The adjusted estimators $\widehat{\theta}_{adj}$ correspond to
    $\widehat{IE}_{adj}$ or $\widehat{DE}_{adj}$.
    This formulation yields the PBA procedure described in Section~3.5.

\section{Simulations}
\label{apdx.sec:sim}
Code for replicating the simulation study and data analysis is available at \url{https://github.com/barwein/ENRT_SA}.

We generated three covariates for each unit.
\begin{equation*}
    \begin{aligned}
        X_{i1} &\sim \text{Bernoulli}(p_{1}), \\
        X_{i2} &\sim \text{Bernoulli}(p_{2}), \\
        X_{i3} &\sim \text{Normal}(0, 1),
    \end{aligned}
\end{equation*}
where $p_{1} = 0.6, p_{2} = 0.2$ for egos and $p_{1} = 0.5, p_{2} = 0.3$ for alters.
The covariates coefficients in the potential outcome models were set to
$\bbeta_e = (-0.5, -0.3, 0.2)$ for egos and $\bbeta_a = (-0.4, -0.2, 0.1)$ for alters.
Heterogeneous edge probabilities were generated with weights computed 
using the Euclidean distance $\lVert \cdot \rVert_2$ between the covariates of units.

Tables~\ref{apdx.tab:sim_ie_homo_aug} and \ref{apdx.tab:sim_de_homo_aug}
report the simulation results for the indirect and direct effects, respectively,
with homogeneous exposure probabilities and augmented estimators.
All results are qualitatively similar to those presented in the main text.

Table~\ref{apdx.tab:sim_de_misspec_kappa} shows the results of the direct effect estimators under heterogeneous exposure probabilities (with $m_e=150$) under misspecification of $\kappa$ values. That is, the true value is $\kappa=1.5$ while we tested how using $\kappa=0.5$ and $\kappa=1.1$ impacts the estimates. The table shows that using $\kappa=0.5$ severely affects the bias-corrected estimators $\widehat{DE}_{adj}$ and $\widehat{DE}_{adj}^{aug}$ by increasing the bias and reducing the coverage compared to the naive HT estimator $\widehat{DE}$. That is the case since using $\kappa < 1$ when the true value is larger than one, changes the direction of the bias correction in the weights used in \eqref{eq:DE_corrected}, resulting in inferior performance to the naive estimator. However, using a misspecified value closer to the true value $\kappa=1.1$ (which is still in the same direction $\kappa>1$ as the true value), negatively impacts the bias and coverage of the bias-corrected estimators, but still results in better performance than the naive estimator.

Table~\ref{apdx.tab:sim_de_viol_ass4} shows the results of the direct effect estimators under heterogeneous exposure probabilities (with $m_e=150$) with violation of Assumption~4. The violation was simulated by modifying the data-generating process of the potential outcomes. Specifically, we have added a heterogeneous interaction term such that the potential outcomes for egos $i \in \cR_e$ are now generated from
\begin{equation*}
   Y_i(z,f) = -0.5 + 2 z + 0.5 f + zf + \beta_4 zf X_{i1} + \bX_i^T \bbeta_e + \varepsilon_i.
\end{equation*}
Taking $\beta_4=0$ yields the original data-generating process used in the other setups. However, taking $\beta_4\neq 0$ results in heterogeneous treatment and exposure interactions, leading to violation of Assumption~4.
Table~\ref{apdx.tab:sim_de_viol_ass4} shows the results with $\beta_4=0.5$ and $\beta_4 =1.1$, while using the true $\kappa$ value in the bias-corrected estimators.
Larger $\beta_4$ leads to stronger violation of Assumption~4.
The table shows that the violation of Assumption~4 only slightly affects the bias-corrected estimators regardless of their specification and $\beta_4$ values. Specifically, the bias was minimal, but the augmented estimators tend to underestimate the variance ($SD/SE>1$).

Finally, Tables~\ref{apdx.tab:sim_ie_binary} and \ref{apdx.tab:sim_de_binary} show the results for the indirect and direct effects, respectively, with binary potential outcomes under homogeneous or heterogeneous exposure probabilities with $m_a=100$ and $m_e=150$. The binary potential outcomes were sampled from a logistic model with
\begin{equation*}
        \begin{aligned}
            \text{logit} \Pr(Y_i(z,f)=1) &= -0.5 + 2 z + 0.5 f + zf + \bX_i^T \bbeta_e,   & \text{for} \; i \in \cR_e,
            \\ 
            \text{logit} \Pr(Y_i(0,f)=1) &= -0.5 + 2 f + \bX_i^T \bbeta_a , \; &   \text{for} \; i \in \cR_a,        
        \end{aligned}
    \end{equation*}
using the same values for $\bbeta_e$ and $\bbeta_a$ as described above. We targeted the DE and IE on the difference scale, but the outcome-augmented estimators used the predicted probabilities from a logistic regression model. The results are qualitatively similar to those with continuous potential outcomes, showing that the bias-corrected estimators performed better than the naive estimators.

{\renewcommand{\arraystretch}{0.8} 

\begin{table}[H]

\caption{Simulation results for indirect effect with homogeneous exposure probabilities.
True effect is $IE = 2$.}
\centering
\footnotesize
\begin{tabular}[t]{lccccc}
\toprule
Scenario & Specification & Augmented & Bias & Coverage & SD/SE\\
\midrule
 &  & FALSE & 0.000 & 0.963 & 0.941\\
\cmidrule{3-6}
 & \multirow{-2}{*}{\centering\arraybackslash Heterogeneous} & TRUE & -0.001 & 0.954 & 0.986\\
\cmidrule{2-6}
 &  & FALSE & -0.001 & 0.962 & 0.942\\
\cmidrule{3-6}
 & \multirow{-2}{*}{\centering\arraybackslash Homogeneous} & TRUE & -0.002 & 0.954 & 0.986\\
\cmidrule{2-6}
 &  & FALSE & -0.236 & 0.506 & 0.942\\
\cmidrule{3-6}
\multirow{-6}{*}[1\dimexpr\aboverulesep+\belowrulesep+\cmidrulewidth]{\raggedright\arraybackslash $m^a=100$} & \multirow{-2}{*}{\centering\arraybackslash Naive} & TRUE & -0.237 & 0.400 & 0.986\\
\cmidrule{1-6}
 &  & FALSE & 0.008 & 0.972 & 0.881\\
\cmidrule{3-6}
 & \multirow{-2}{*}{\centering\arraybackslash Heterogeneous} & TRUE & 0.006 & 0.954 & 0.962\\
\cmidrule{2-6}
 &  & FALSE & 0.003 & 0.973 & 0.882\\
\cmidrule{3-6}
 & \multirow{-2}{*}{\centering\arraybackslash Homogeneous} & TRUE & 0.001 & 0.956 & 0.959\\
\cmidrule{2-6}
 &  & FALSE & -0.440 & 0.063 & 0.882\\
\cmidrule{3-6}
\multirow{-6}{*}[1\dimexpr\aboverulesep+\belowrulesep+\cmidrulewidth]{\raggedright\arraybackslash $m^a=200$} & \multirow{-2}{*}{\centering\arraybackslash Naive} & TRUE & -0.442 & 0.020 & 0.959\\
\bottomrule
\label{apdx.tab:sim_ie_homo_aug}
\end{tabular}
\end{table}
}

{\renewcommand{\arraystretch}{0.8} 

\begin{table}[H]

\caption{Simulation results for direct effect with homogeneous exposure probabilities.
True effect is $DE =2$.}
\centering
\footnotesize
\begin{tabular}[t]{lccccc}
\toprule
Scenario & Specification & Augmented & Bias & Coverage & SD/SE\\
\midrule
 &  & FALSE & 0.012 & 0.984 & 0.798\\
\cmidrule{3-6}
 & \multirow{-2}{*}{\centering\arraybackslash Heterogeneous} & TRUE & 0.008 & 0.956 & 0.961\\
\cmidrule{2-6}
 &  & FALSE & 0.000 & 0.982 & 0.810\\
\cmidrule{3-6}
 & \multirow{-2}{*}{\centering\arraybackslash Homogeneous} & TRUE & -0.003 & 0.955 & 0.969\\
\cmidrule{2-6}
 &  & FALSE & 0.529 & 0.247 & 1.079\\
\cmidrule{3-6}
\multirow{-6}{*}[1\dimexpr\aboverulesep+\belowrulesep+\cmidrulewidth]{\raggedright\arraybackslash $m^e=150, \kappa=1.5$} & \multirow{-2}{*}{\centering\arraybackslash Naive} & TRUE & 0.525 & 0.132 & 1.052\\
\cmidrule{1-6}
 &  & FALSE & 0.026 & 0.999 & 0.583\\
\cmidrule{3-6}
 & \multirow{-2}{*}{\centering\arraybackslash Heterogeneous} & TRUE & 0.023 & 0.980 & 0.827\\
\cmidrule{2-6}
 &  & FALSE & 0.007 & 0.999 & 0.600\\
\cmidrule{3-6}
 & \multirow{-2}{*}{\centering\arraybackslash Homogeneous} & TRUE & 0.004 & 0.978 & 0.843\\
\cmidrule{2-6}
 &  & FALSE & 0.724 & 0.076 & 1.070\\
\cmidrule{3-6}
\multirow{-6}{*}[1\dimexpr\aboverulesep+\belowrulesep+\cmidrulewidth]{\raggedright\arraybackslash $m^e=250, \kappa=1.5$} & \multirow{-2}{*}{\centering\arraybackslash Naive} & TRUE & 0.720 & 0.012 & 1.031\\
\bottomrule
\label{apdx.tab:sim_de_homo_aug}
\end{tabular}
\end{table}
}

\begin{table}[H]

\caption{Simulation results for direct effect with heterogeneous exposure probabilities and misspecified $\kappa$ values.
True effect is $DE =2$, true $\kappa = 1.5$, and $m_e = 150$.}
\centering
\footnotesize
\begin{tabular}[t]{lccccc}
\toprule
Scenario & Specification & Augmented & Bias & Coverage & SD/SE\\
\midrule
 &  & FALSE & 1.385 & 0.008 & 0.755\\
\cmidrule{3-6}
 & \multirow{-2}{*}{\centering\arraybackslash Heterogeneous} & TRUE & 1.376 & 0.000 & 0.951\\
\cmidrule{2-6}
 &  & FALSE & 1.417 & 0.006 & 0.772\\
\cmidrule{3-6}
 & \multirow{-2}{*}{\centering\arraybackslash Homogeneous} & TRUE &1.408 & 0.000 & 0.959\\
\cmidrule{2-6}
 &  & FALSE & 0.514 & 0.279 & 1.068\\
\cmidrule{3-6}
\multirow{-6}{*}[1\dimexpr\aboverulesep+\belowrulesep+\cmidrulewidth]{\raggedright\arraybackslash Using $\kappa=0.5$} & \multirow{-2}{*}{\centering\arraybackslash Naive} & TRUE & 0.508 & 0.175 & 1.045\\
\cmidrule{1-6}
 &  & FALSE & 0.392 & 0.716 & 0.791\\
\cmidrule{3-6}
 & \multirow{-2}{*}{\centering\arraybackslash Heterogeneous} & TRUE & 0.385 & 0.430 & 0.952\\
\cmidrule{2-6}
 &  & FALSE & 0.388 & 0.697 & 0.813\\
\cmidrule{3-6}
 & \multirow{-2}{*}{\centering\arraybackslash Homogeneous} & TRUE &0.382 & 0.429 & 0.960\\
\cmidrule{2-6}
 &  & FALSE & 0.514 & 0.279 & 1.068\\
\cmidrule{3-6}
\multirow{-6}{*}[1\dimexpr\aboverulesep+\belowrulesep+\cmidrulewidth]{\raggedright\arraybackslash Using $\kappa=1.1$} & \multirow{-2}{*}{\centering\arraybackslash Naive} & TRUE & 0.508 & 0.175 & 1.045\\
\bottomrule
\label{apdx.tab:sim_de_misspec_kappa}
\end{tabular}
\end{table}

\begin{table}[H]
\caption{Simulation results for direct effect with heterogeneous exposure probabilities and violation of Assumption~4.
True effect is $DE =2$ and $m_e = 150$.
The value $\beta_4$ controls the magnitude of violation, where larger values correspond to larger violations.
Values of $\kappa$ depend on $\beta_4$ and are computed from the potential outcomes.
}
\centering
\footnotesize
\begin{tabular}[t]{lccccc}
\toprule
Scenario & Specification & Augmented & Bias & Coverage & SD/SE\\
\midrule
 &  & FALSE & 0.003 & 0.984 & 0.792\\
\cmidrule{3-6}
 & \multirow{-2}{*}{\centering\arraybackslash Heterogeneous} & TRUE &-0.002 & 0.955 & 0.969\\
\cmidrule{2-6}
 &  & FALSE & -0.010 & 0.981 & 0.815\\
\cmidrule{3-6}
 & \multirow{-2}{*}{\centering\arraybackslash Homogeneous} & TRUE & -0.016 & 0.952 & 0.977\\
\cmidrule{2-6}
 &  & FALSE & 0.681 & 0.122 & 1.094\\
\cmidrule{3-6}
\multirow{-6}{*}[1\dimexpr\aboverulesep+\belowrulesep+\cmidrulewidth]{\raggedright\arraybackslash $\beta_4=0.5$ and $\kappa=1.68$} & \multirow{-2}{*}{\centering\arraybackslash Naive} & TRUE & 0.673 & 0.049 & 1.063\\
\cmidrule{1-6}
 &  & FALSE & 0.009 & 0.980 & 0.814\\
\cmidrule{3-6}
 & \multirow{-2}{*}{\centering\arraybackslash Heterogeneous} & TRUE & 0.003 & 0.950 & 0.993\\
\cmidrule{2-6}
 &  & FALSE & -0.010 & 0.976 & 0.840\\
\cmidrule{3-6}
 & \multirow{-2}{*}{\centering\arraybackslash Homogeneous} & TRUE & -0.015 & 0.949 & 1.000\\
\cmidrule{2-6}
 &  & FALSE & 1.013 & 0.020 & 1.134 \\
\cmidrule{3-6}
\multirow{-6}{*}[1\dimexpr\aboverulesep+\belowrulesep+\cmidrulewidth]{\raggedright\arraybackslash $\beta_4=1.5$ and $\kappa=1.97$} & \multirow{-2}{*}{\centering\arraybackslash Naive} & TRUE & 1.004 & 0.005 & 1.087\\
\bottomrule
\label{apdx.tab:sim_de_viol_ass4}
\end{tabular}
\end{table}

\begin{table}[H]
\caption{Simulation results for indirect effect with binary potential outcomes under homogeneous or heterogeneous exposure probabilities with $m_a=100$.
Augmented estimators used a logistic regression model for the augmentation. True effect is $IE = 0.4775$.}
\centering
\footnotesize
\begin{tabular}[t]{lccccc}
\toprule
Scenario & Specification & Augmented & Bias & Coverage & SD/SE\\
\midrule
 &  & FALSE & 0.001 & 0.968 & 0.913\\
\cmidrule{3-6}
 & \multirow{-2}{*}{\centering\arraybackslash Heterogeneous} & TRUE & 0.001 & 0.984 & 0.789\\
\cmidrule{2-6}
 &  & FALSE & 0.001 & 0.967 & 0.913\\
\cmidrule{3-6}
 & \multirow{-2}{*}{\centering\arraybackslash Homogeneous} & TRUE & 0.001 & 0.985 & 0.788\\
\cmidrule{2-6}
 &  & FALSE & -0.055 & 0.928 & 0.913\\
\cmidrule{3-6}
\multirow{-6}{*}[1\dimexpr\aboverulesep+\belowrulesep+\cmidrulewidth]{\raggedright\arraybackslash Heterogeneous} & \multirow{-2}{*}{\centering\arraybackslash Naive} & TRUE & -0.055 & 0.834 & 0.788\\
\cmidrule{1-6}
 &  & FALSE & -0.003 & 0.968 & 0.912\\
\cmidrule{3-6}
 & \multirow{-2}{*}{\centering\arraybackslash Heterogeneous} & TRUE & -0.003 & 0.987 & 0.786\\
\cmidrule{2-6}
 &  & FALSE & -0.003 & 0.967 & 0.912\\
\cmidrule{3-6}
 & \multirow{-2}{*}{\centering\arraybackslash Homogeneous} & TRUE & -0.003 & 0.986 & 0.785\\
\cmidrule{2-6}
 &  & FALSE & -0.059 & 0.920 & 0.912 \\
\cmidrule{3-6}
\multirow{-6}{*}[1\dimexpr\aboverulesep+\belowrulesep+\cmidrulewidth]{\raggedright\arraybackslash Homogeneous} & \multirow{-2}{*}{\centering\arraybackslash Naive} & TRUE & -0.059 & 0.804 & 0.785\\
\bottomrule
\label{apdx.tab:sim_ie_binary}
\end{tabular}
\end{table}

\begin{table}[H]
\caption{Simulation results for direct effect with binary potential outcomes under homogeneous or heterogeneous exposure probabilities with $m_e=150$.
Augmented estimators used a logistic regression model for the augmentation. True effect is $DE = 0.52$ and $\kappa = 0.932$.}
\centering
\footnotesize
\begin{tabular}[t]{lccccc}
\toprule
Scenario & Specification & Augmented & Bias & Coverage & SD/SE\\
\midrule
 &  & FALSE & -0.005 & 0.994 & 0.674\\
\cmidrule{3-6}
 & \multirow{-2}{*}{\centering\arraybackslash Heterogeneous} & TRUE & -0.006 & 0.982 & 0.812\\
\cmidrule{2-6}
 &  & FALSE & -0.004 & 0.993 & 0.692\\
\cmidrule{3-6}
 & \multirow{-2}{*}{\centering\arraybackslash Homogeneous} & TRUE & -0.005 & 0.981 & 0.818\\
\cmidrule{2-6}
 &  & FALSE & -0.023 & 0.954 & 0.939\\
\cmidrule{3-6}
\multirow{-6}{*}[1\dimexpr\aboverulesep+\belowrulesep+\cmidrulewidth]{\raggedright\arraybackslash Heterogeneous} & \multirow{-2}{*}{\centering\arraybackslash Naive} & TRUE & -0.024 & 0.963 & 0.886\\
\cmidrule{1-6}
 &  & FALSE & -0.012 & 0.996 & 0.675\\
\cmidrule{3-6}
 & \multirow{-2}{*}{\centering\arraybackslash Heterogeneous} & TRUE & -0.012 & 0.987 & 0.772\\
\cmidrule{2-6}
 &  & FALSE & -0.011 & 0.994 & 0.689\\
\cmidrule{3-6}
 & \multirow{-2}{*}{\centering\arraybackslash Homogeneous} & TRUE & -0.012 & 0.986 & 0.778\\
\cmidrule{2-6}
 &  & FALSE & -0.029 & 0.953 & 0.934 \\
\cmidrule{3-6}
\multirow{-6}{*}[1\dimexpr\aboverulesep+\belowrulesep+\cmidrulewidth]{\raggedright\arraybackslash Homogeneous} & \multirow{-2}{*}{\centering\arraybackslash Naive} & TRUE & -0.030 & 0.964 & 0.845\\
\bottomrule
\label{apdx.tab:sim_de_binary}
\end{tabular}
\end{table}
\section{Data analysis}
\label{apdx.sec:data}
\begin{figure}[H]
    \centering
    \includegraphics[width=0.4\linewidth]{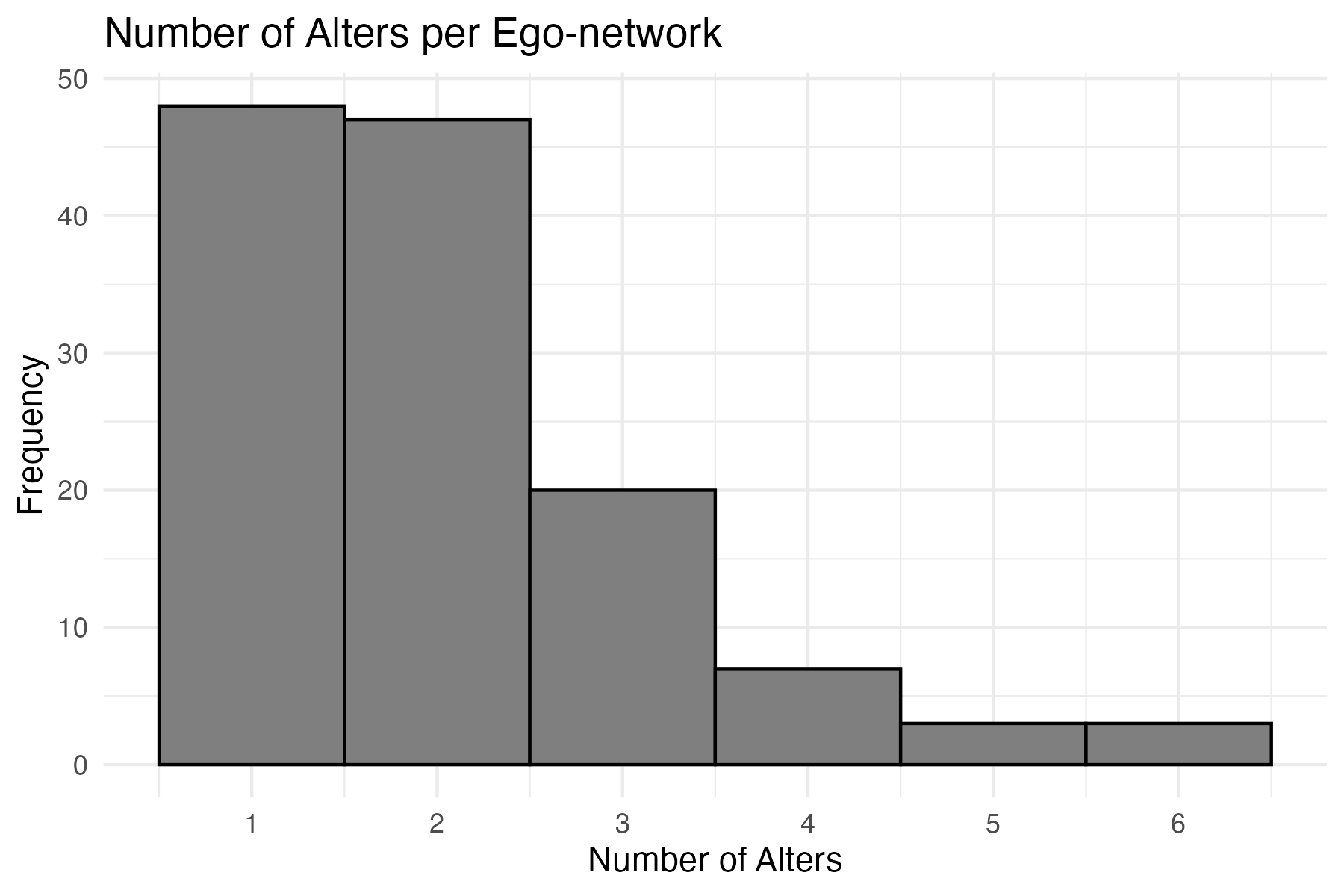}
    \caption{Distribution of number of alters per ego-network.}
    \label{fig:hptn_degree}
\end{figure}

\begin{figure}[H]
    \centering
    \includegraphics[width=0.5\linewidth]{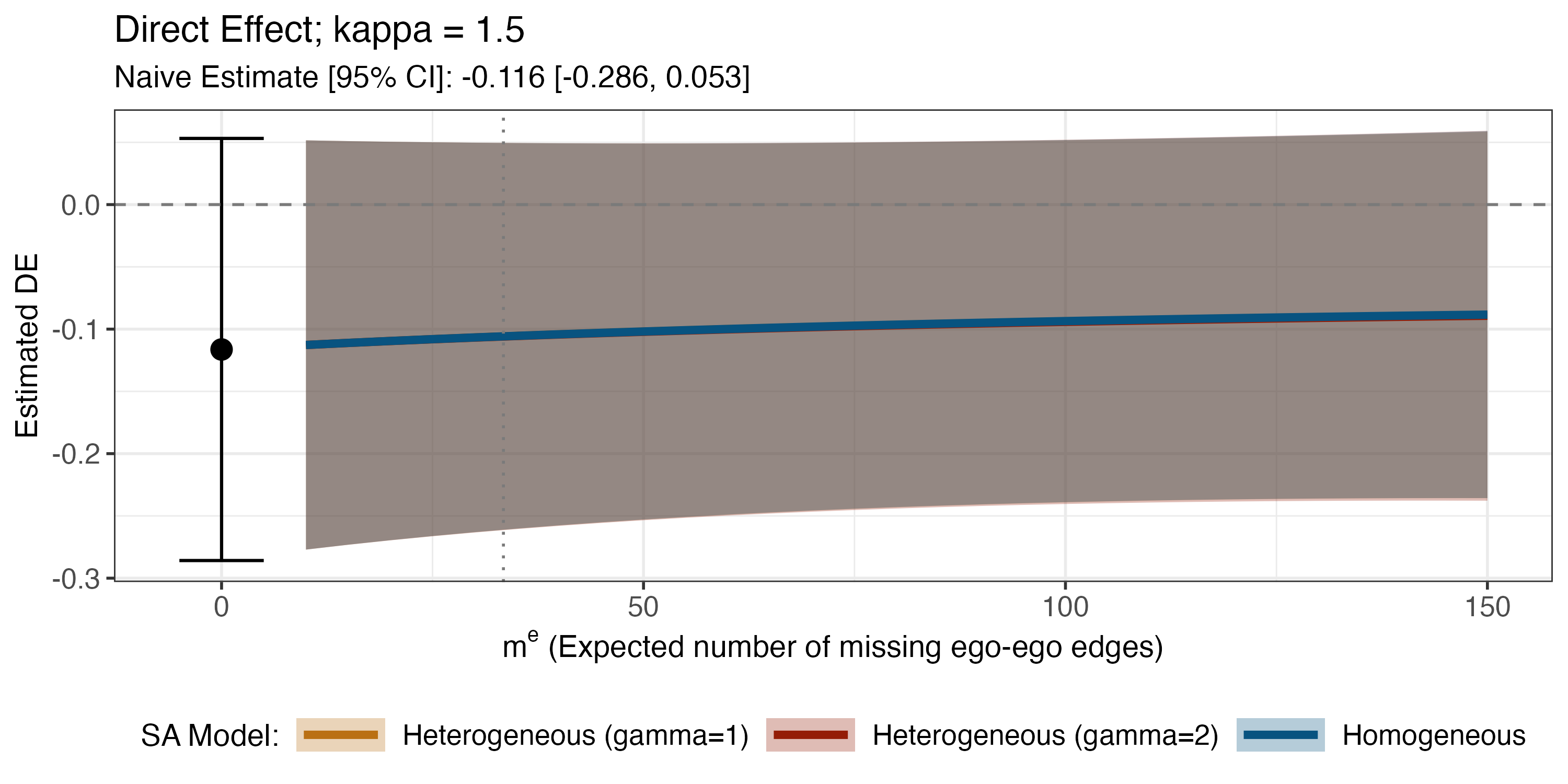}
    \caption{GSA for direct effect with $\kappa=1.5$ using the augmented estimators $\widehat{DE}_{adj}^{aug}$. The dashed vertical line at $33.4$ is the estimate of $m^e$ from the internal validation data.}
    \label{fig:sa_de_kappa2}
\end{figure}

\begin{figure}[H]
    \centering
    \includegraphics[width=0.7\linewidth]{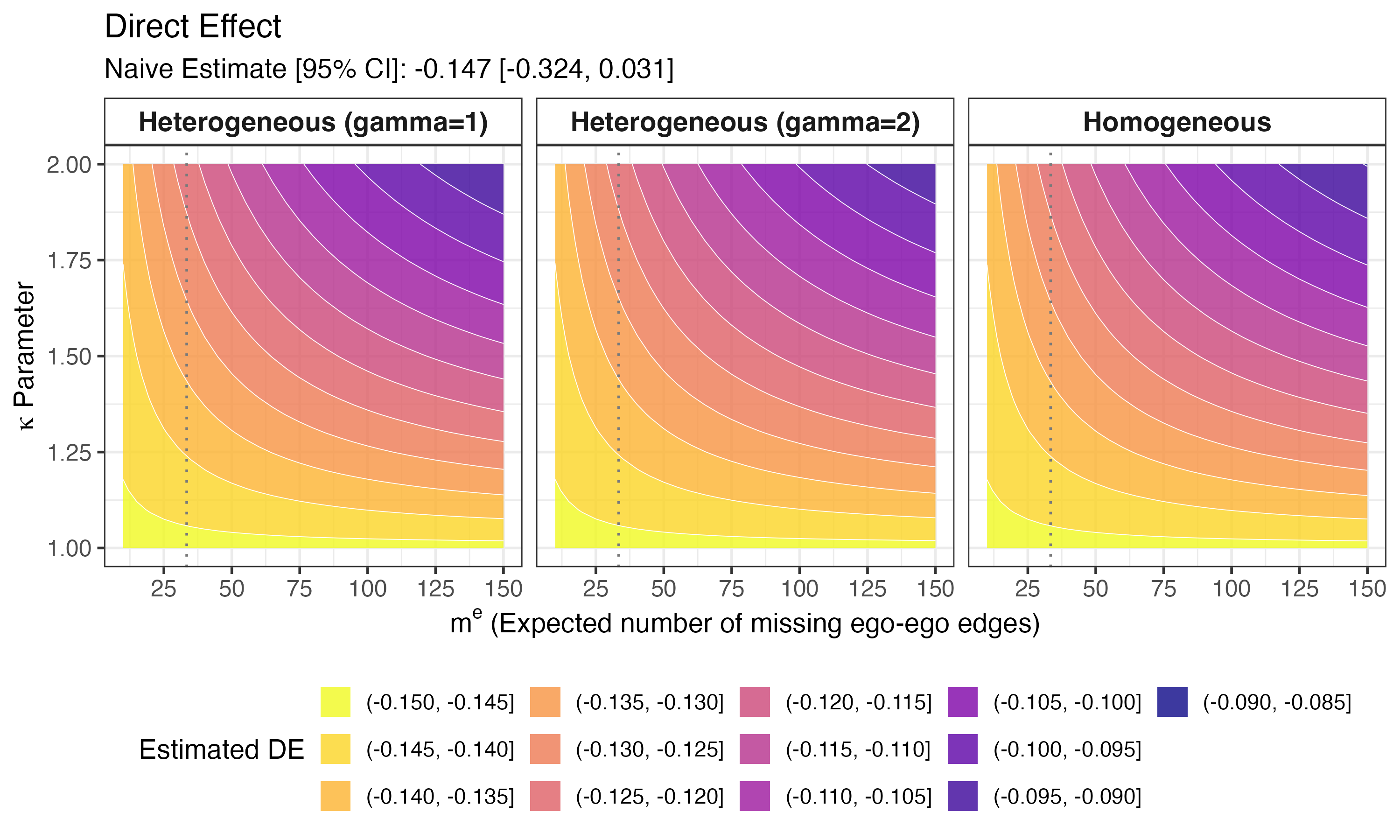}
    \caption{GSA for direct effect with non-augmented estimators $\widehat{DE}_{adj}$. The dashed vertical line at $33.4$ is the estimate of $m^e$ from the internal validation data.}
    \label{fig:sa_de_not_aug}
\end{figure}

\begin{figure}[H]
    \centering
    \includegraphics[width=0.5\linewidth]{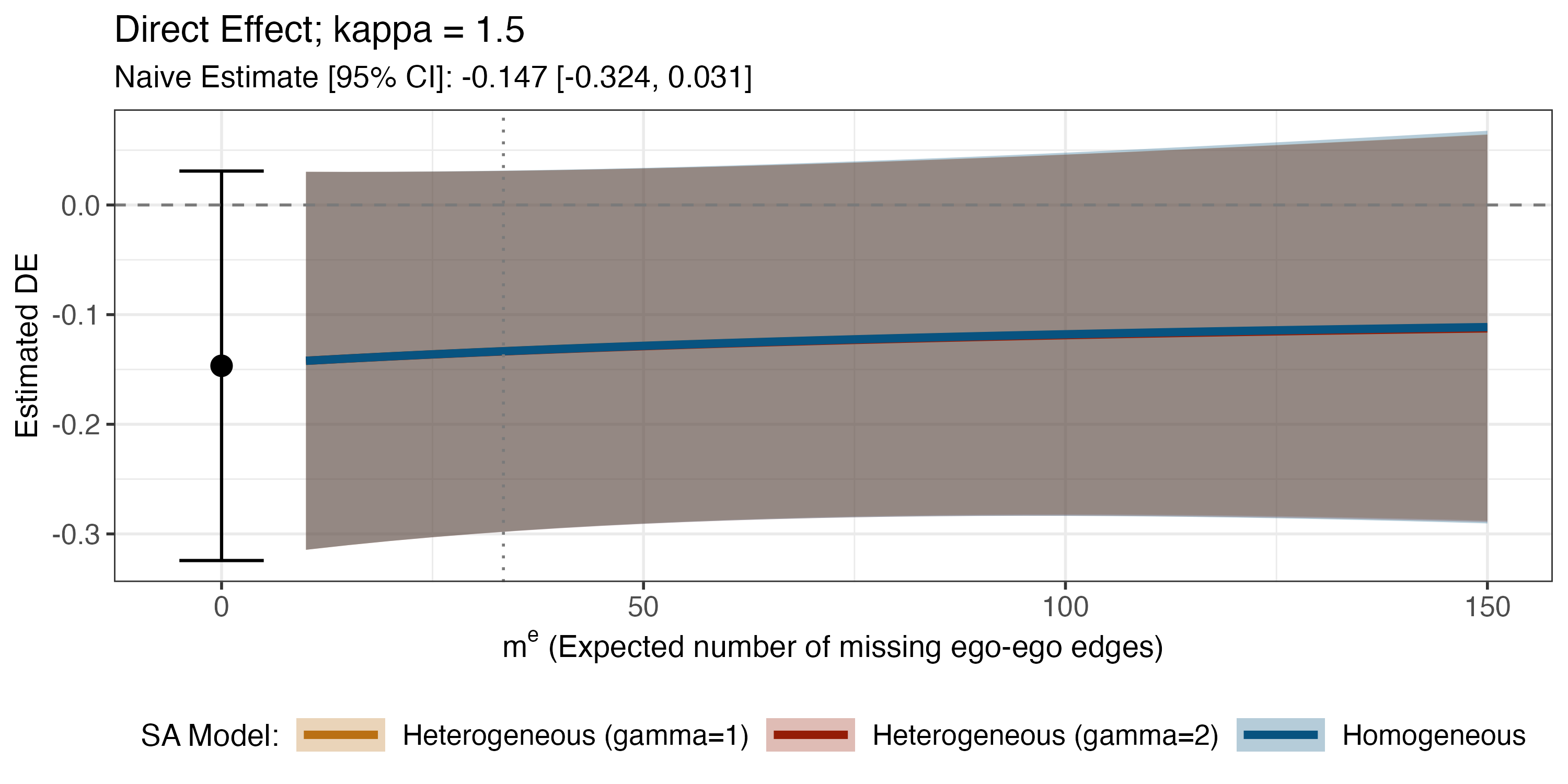}
    \caption{GSA for direct effect with $\kappa=1.5$.
    Not-augmented estimator. The dashed vertical line at $33.4$ is the estimate of $m^e$ from the internal validation data.}
    \label{fig:sa_de_kappa2_not_aug}
\end{figure}

\begin{figure}[H]
    \centering
    \includegraphics[width=0.5\linewidth]{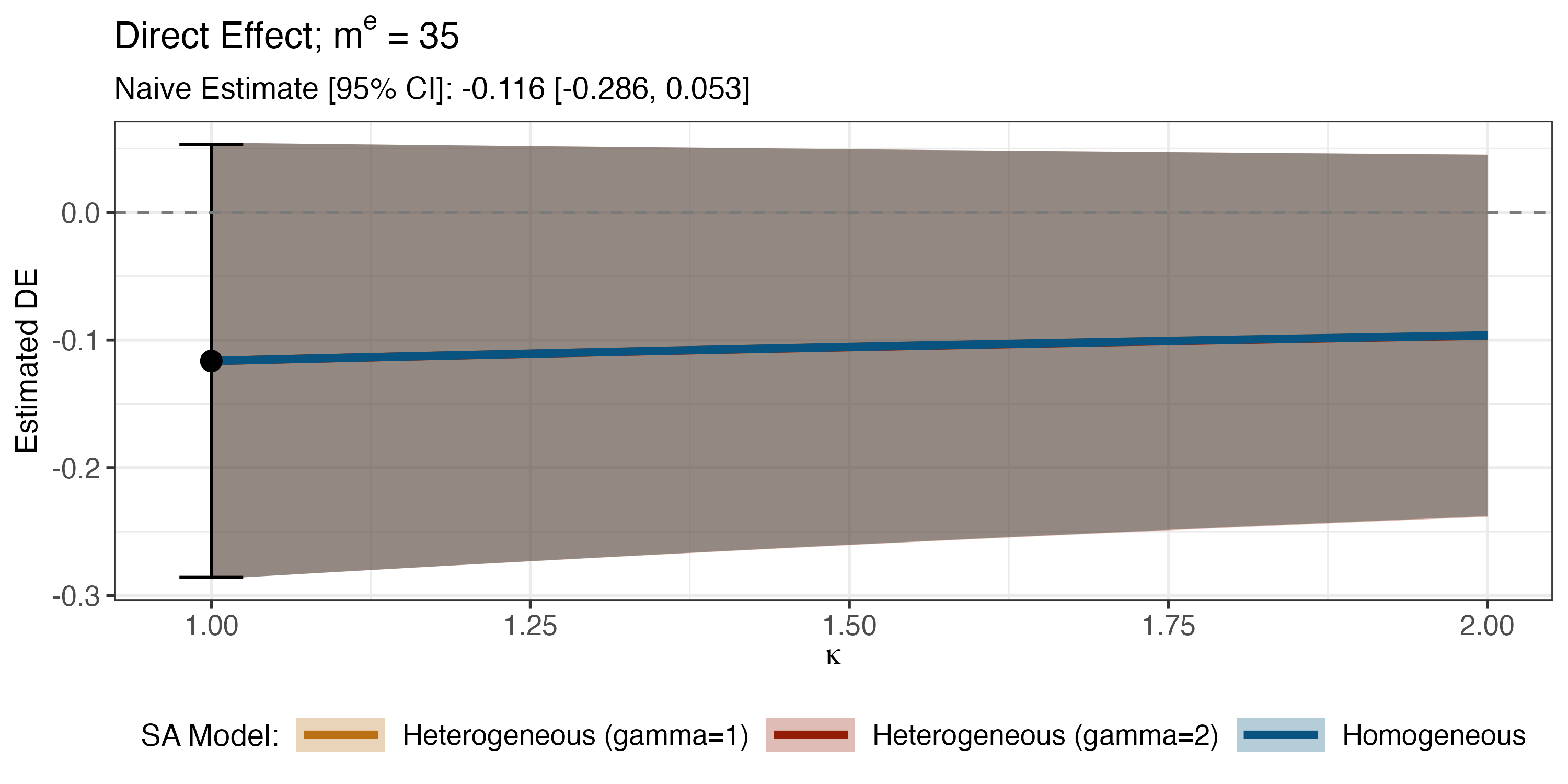}
    \caption{GSA for direct effect with augmented estimators using $m^e=35$ to approximate the bias-corrected estimates obtained using the the estimated value $\widehat{m}^e=33.4$ from the internal validation data.}
    \label{fig:sa_de_m_e_esti}
\end{figure}

Figure~\ref{fig:hptn_degree} shows the distribution of the number of alters per ego-network in the 
data. The average and median number is $2$, the range is from $1$ to $6$.
Figure~\ref{fig:sa_de_kappa2} complements Figure~\ref{fig:hptn_sa_de_augmented} and shows the 
the GSA results (Section~5.1)
for the direct effect using the augmented estimators given that $\kappa=1.5$. 
We also included the results using $\gamma^e=2$, representing a stronger influence of covariates' similarity on the exposure probabilities $\pi_i^e$.
This presentation, similar to that of the indirect effect, enables us 
to see how changing the expected number of missing ego--ego edges ($m^e$) impacts 
the estimated direct effect. Clearly, for all specifications, the 
estimated effects were almost identical and approached zero as the contamination level increased
Furthermore, Figure~\ref{fig:sa_de_not_aug} shows the results of GSA with the non-augmented estimators $\widehat{DE}_{adj}$, also including $\gamma^e=2$. The figure shows that under all three specifications, the bias-corrected estimates are similar. Figure~\ref{fig:sa_de_kappa2_not_aug} complements Figure~\ref{fig:sa_de_kappa2} by showing the bias-corrected estimates with the non-augmented estimators for $\kappa=1.5$. This figure highlights the similarity of the results across the different edge probability specifications.
Finally, Figure~\ref{fig:sa_de_m_e_esti} shows the augmented bias-corrected $DE$ estimates as a function of $\kappa$ given the value $\widehat{m}^e=33.4$ estimated from the internal validation data, showing that increasing $\kappa$ yields bias-corrected estimates closer to zero.

\begin{figure}[H]
    \centering
    \includegraphics[width=0.5\linewidth]{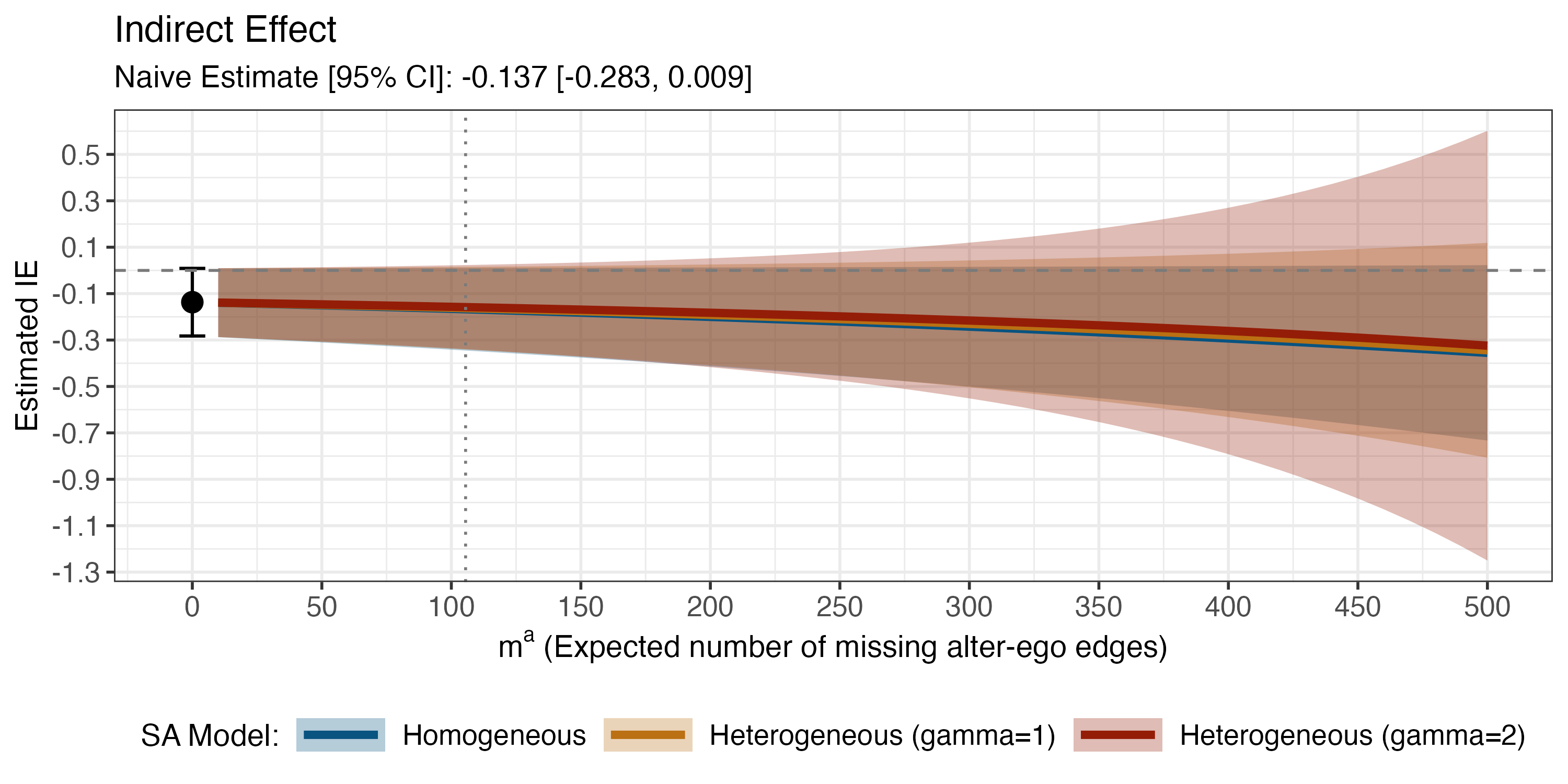}
    \caption{GSA for indirect effect with non-augmented estimators $\widehat{IE}_{adj}$. The dashed vertical line at $105.5$ is the estimate of $m^a$ from the internal validation data.}
    \label{fig:sa_ie_not_aug}
\end{figure}

Figure~\ref{fig:sa_ie_not_aug} shows the results of the GSA
for indirect effect using the non-augmented estimators $\widehat{IE}_{adj}$.
The structure is similar to that of the non-augmented estimator
presented in the main text (Figure~\ref{fig:hptn_sa_ie_augmented}). 
However, the augmented estimator had values closer to zero
for all levels of contamination (including no contamination) and edge probability specifications. Taking $\gamma^e=2$ results in higher variance estimates for large levels of missing alter--ego edges ($m^a\geq 250$). 

\begin{figure}[H]
    \centering
    \includegraphics[width=0.5\linewidth]{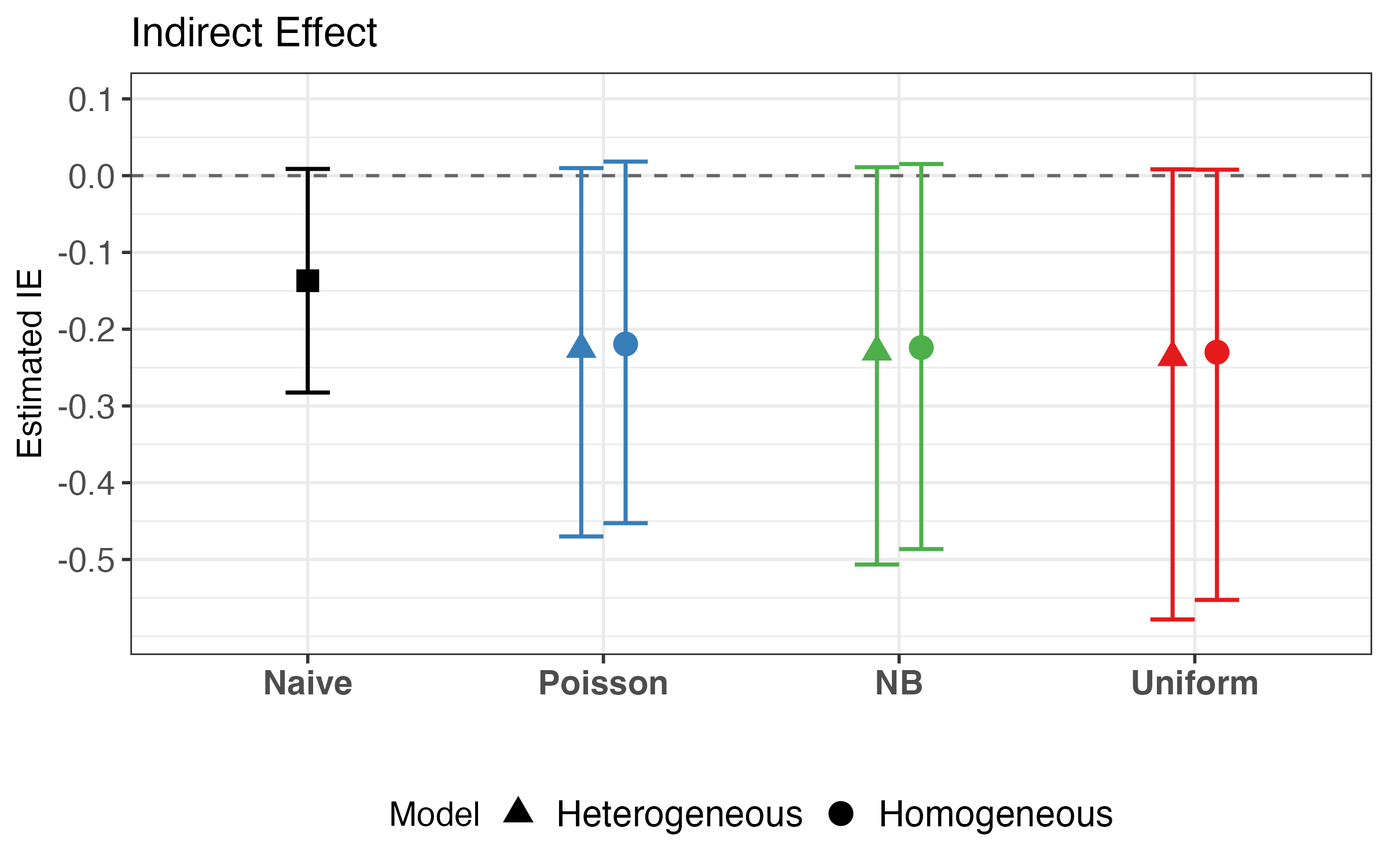}
    \caption{PBA for indirect effect with non-augmented estimators $\widehat{IE}_{adj}$.
    Results are shown as mean ($95\%$ intervals) of bias-corrected estimates across $B=10^4$ Monte Carlo samples. The results account for both bias and statistical uncertainty.}
    \label{fig:pba_ie_not_augmented}
\end{figure}

\begin{figure}[H]
    \centering
    \includegraphics[width=0.5\linewidth]{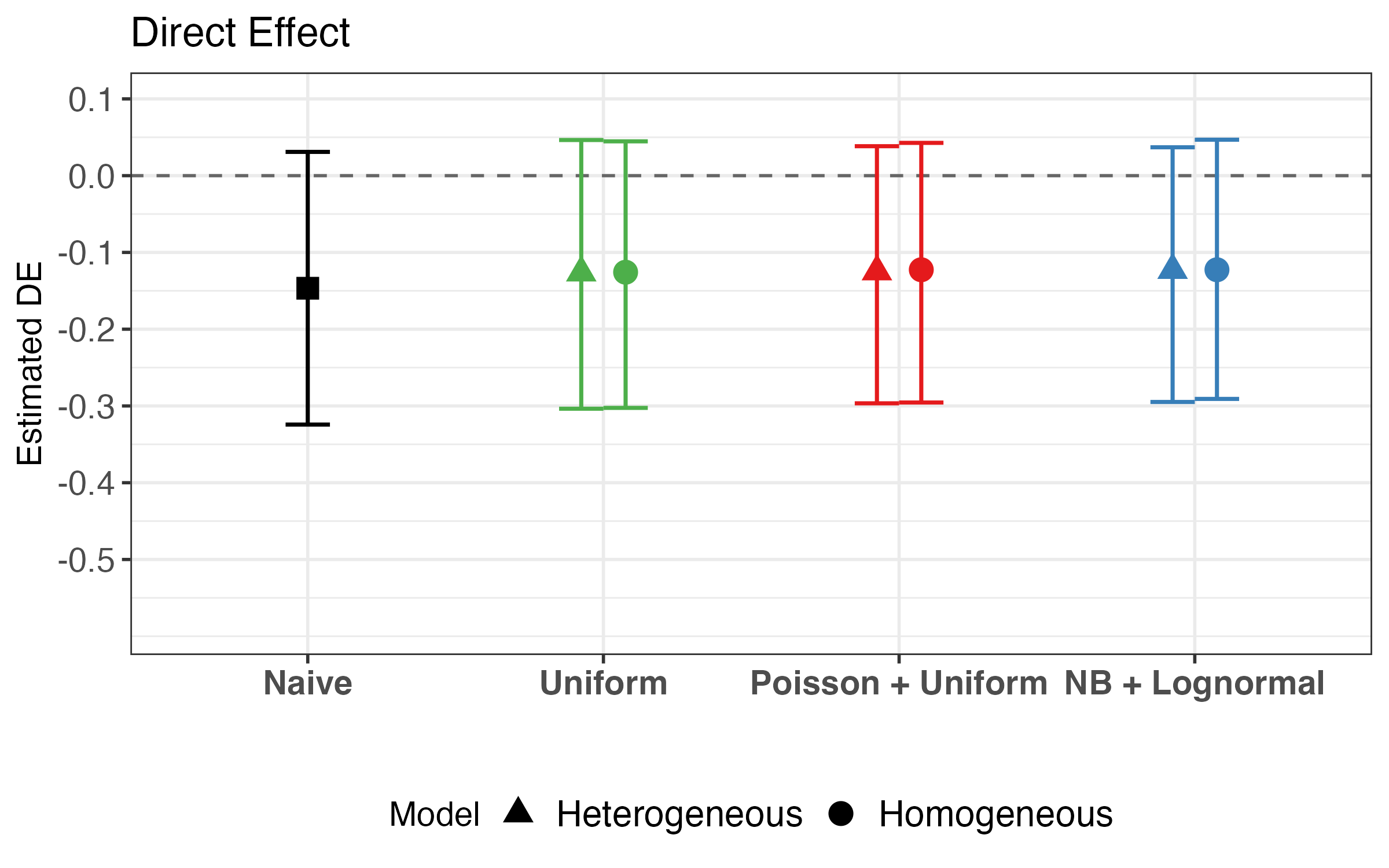}
    \caption{Probabilistic bias analysis for direct effect with non-augmented estimators 
    $\widehat{DE}_{adj}$.
    Results are shown as mean ($95\%$ intervals) of bias-corrected estimates across $B=10^4$ Monte Carlo samples. The results account for both bias and statistical uncertainty.}
    \label{fig:pba_de_not_augmented}
\end{figure}

Figures~\ref{fig:pba_ie_not_augmented} and \ref{fig:pba_de_not_augmented} show the results of the PBA
for indirect and direct effect using the non-augmented estimators $\widehat{IE}_{adj}$ and $\widehat{DE}_{adj}$.
The results are similar to those of the non-augmented estimators
presented in the main text (Figure~\ref{fig:hptn_pba_augmented}).

\section{Results of the example analysis code}
Analysis code applied to a simulated dataset similar to the design of HPTN 037 can be found at \url{https://github.com/barwein/ENRT_SA}.
To replicate it, run the script \texttt{ENRT\_example\_code.R}. 
We provide here the results from running the code, in the order of their appearance in the script.

\begin{figure}[H]
    \centering
    \includegraphics[width=0.45\linewidth]{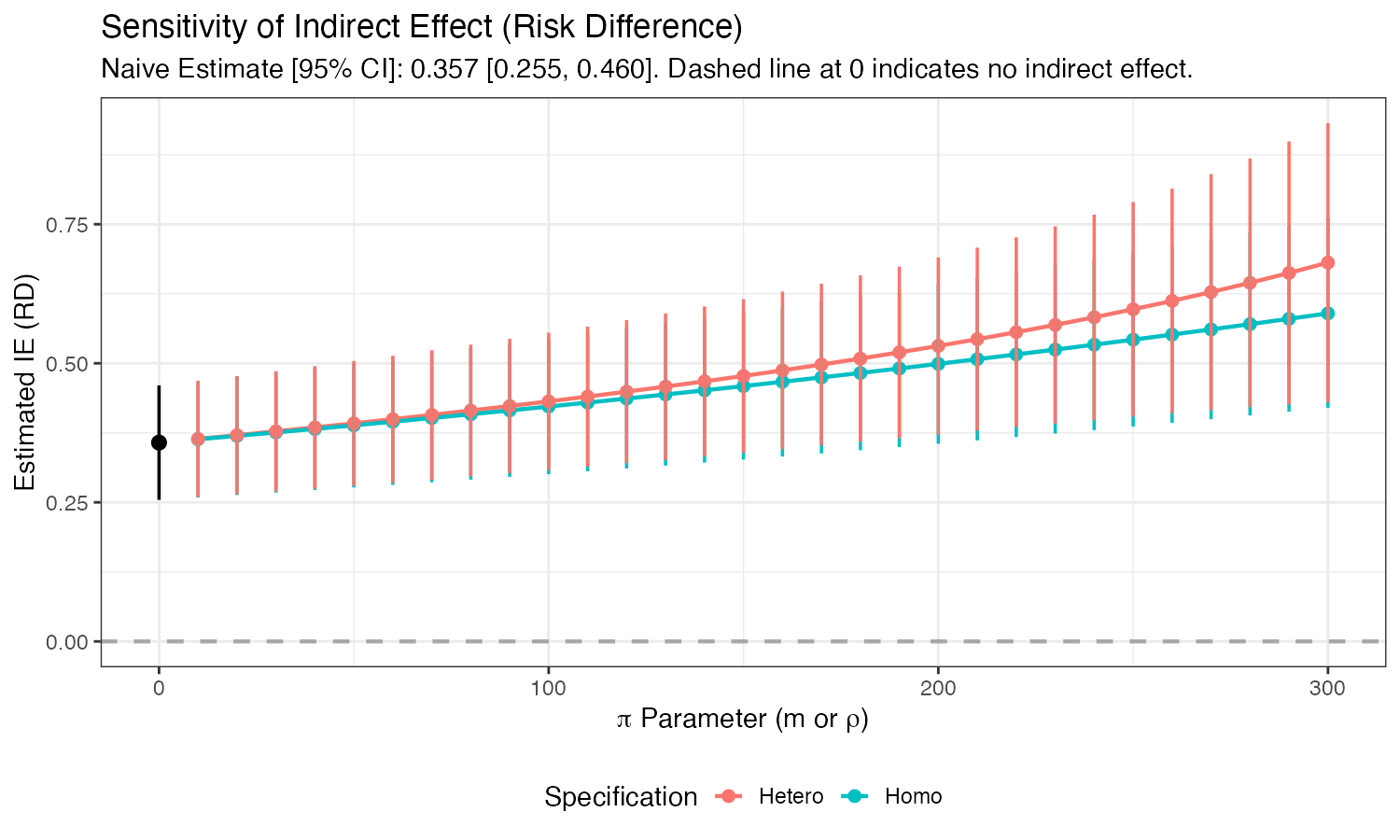}
    \caption{Results for $IE$ estimand with GSA in the ENRT example code.}
\end{figure}

\begin{figure}[H]
    \centering
    \includegraphics[width=0.55\linewidth]{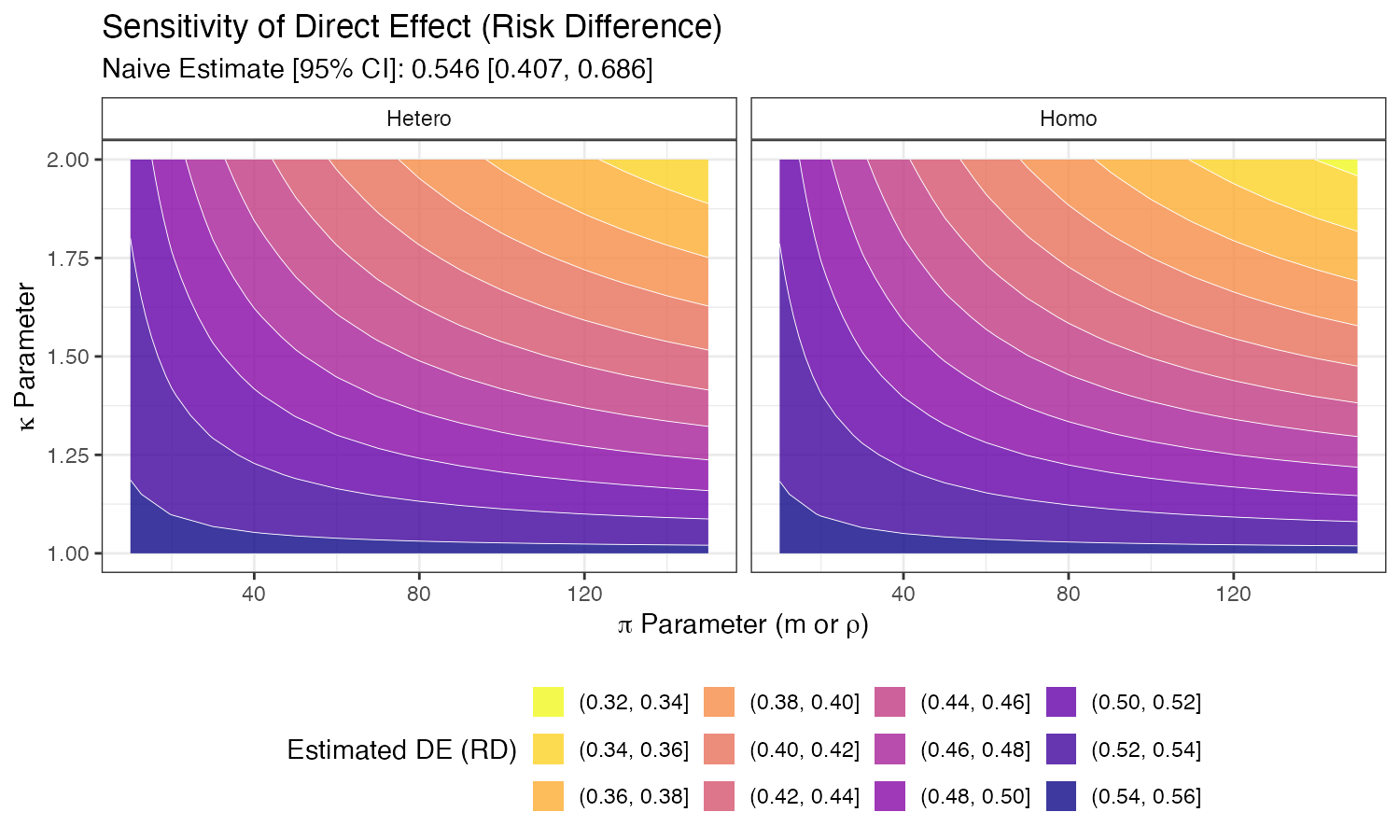}
    \caption{Results for $DE$ estimand with GSA in the ENRT example code.}
\end{figure}

\begin{figure}[H]
    \centering
    \includegraphics[width=0.45\linewidth]{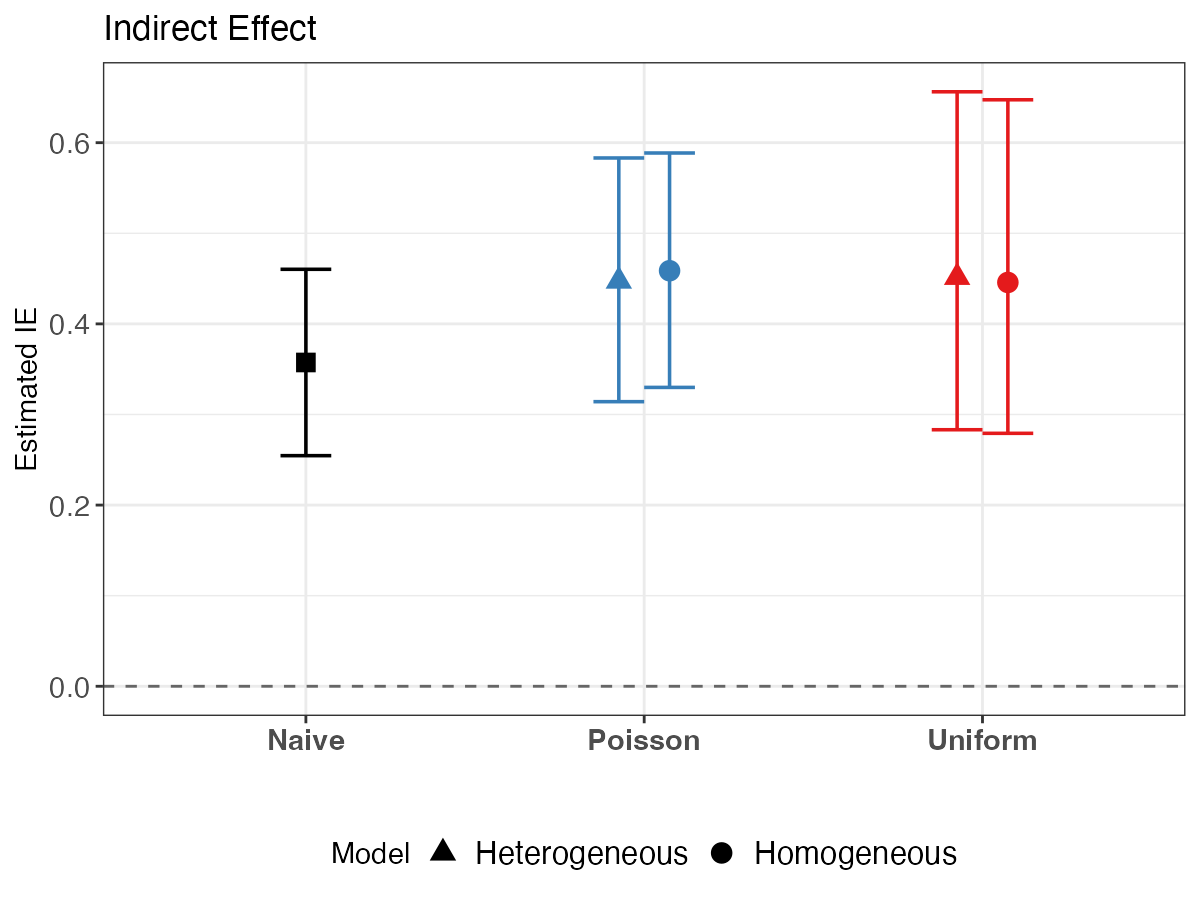}
    \caption{Results for $IE$ estimand with PBA in the ENRT example code.}
\end{figure}

\begin{figure}[H]
    \centering
    \includegraphics[width=0.45\linewidth]{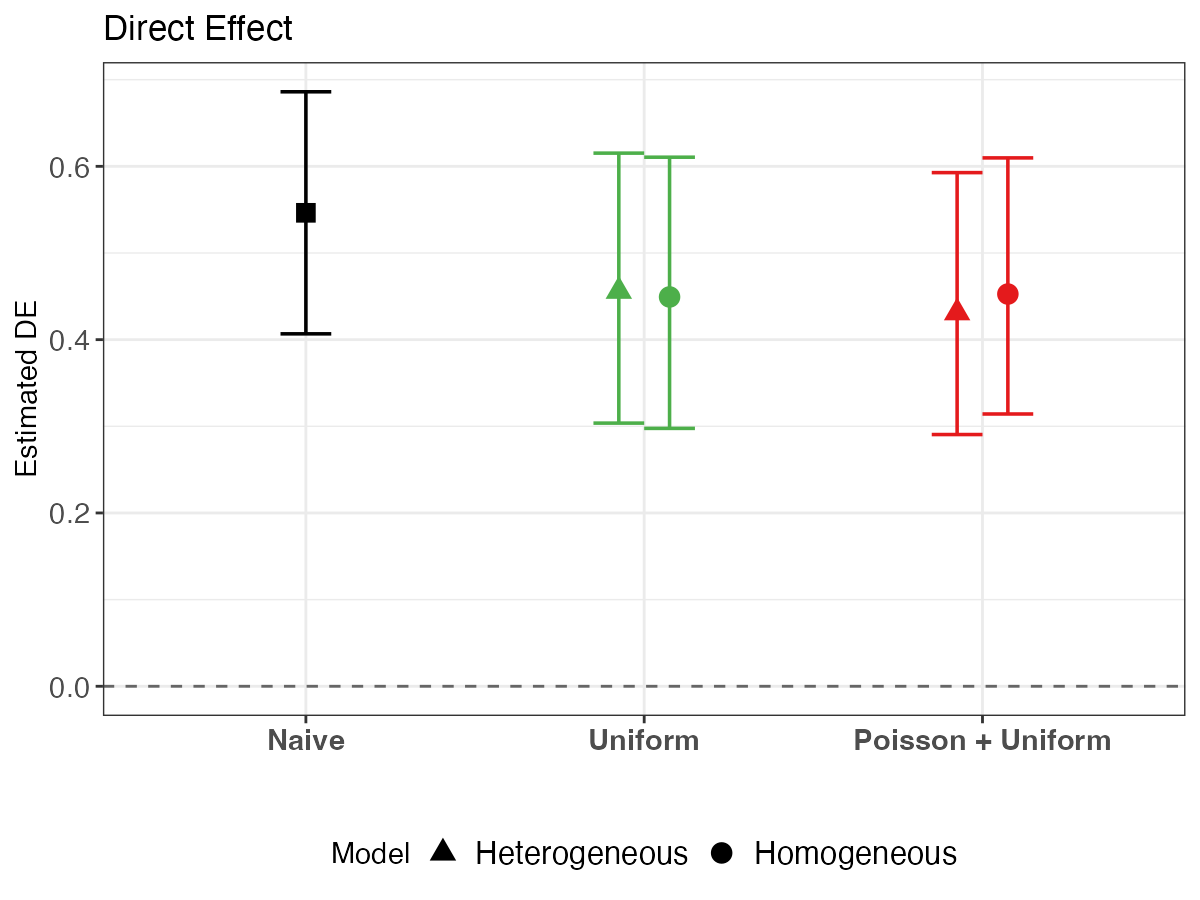}
    \caption{Results for $DE$ estimand with PBA in the ENRT example code.}
\end{figure}

\end{document}